\documentclass[preprint,10pt,authoryear]{elsarticle}
\usepackage{amssymb}
\usepackage{amsthm}
\usepackage{amsmath}
\usepackage{amsfonts}
\usepackage{mathrsfs}
\usepackage{bm}
\usepackage{lineno}
\usepackage{hyperref}
\usepackage{caption}
\usepackage{subfig}
\usepackage{multirow}
\usepackage{placeins}
\usepackage{chngcntr}
\usepackage{float}
\usepackage{makecell}
\usepackage{rotating}
\usepackage{longtable}
\usepackage{booktabs}
\usepackage{xcolor}
\usepackage{orcidlink}

\newcommand{\blue}[1]{\textcolor{black}{#1}}

\journal{arXiv}

\usepackage[a4paper, total={6.25in, 8.75in}]{geometry}

\newcommand{\bdelta}{\boldsymbol{\delta}}
\newcommand{\E}{\text{E}}

\makeatletter
\gdef\emailauthor#1#2{\stepcounter{ead}%
\g@addto@macro\@elseads{\raggedright%
\let\corref\@gobble\let\orcidlink\@gobble\def\@@tmp{#1}%
\eadsep{\ttfamily\expandafter\strip@prefix\meaning\@@tmp}
(#2)\def\eadsep{\unskip,\space}}%
}
\makeatother

\begin{document}

\begin{frontmatter}

\title{Modelling and detecting mild and gross anomalies in circular data via double-contaminated models}

\author[1]{Antonio Punzo \orcidlink{0000-0001-7742-1821}}
\author[2,3]{Andri\"ette Bekker \orcidlink{0000-0003-4793-5674}}
\author[2]{Arnoldus F. Otto\corref{cor1} \orcidlink{0000-0002-6565-2910}}
\ead{arno.otto@up.ac.za}
\author[4,3]{Priyanka Nagar \orcidlink{0000-0002-1775-8905}}
\author[5]{Cristina Tortora \orcidlink{0000-0001-8351-3730}}

\cortext[cor1]{Corresponding author}

\affiliation[1]{organization={Department of Economics and Business, University of Catania},
            city={Catania},
            postcode={55 - 95129},
            country={Italy}}

\affiliation[2]{organization={Department of Statistics, University of Pretoria},
            city={Pretoria},
            postcode={0028},
            country={South Africa}}

\affiliation[3]{organization={National Institute of Theoretical and Computational Sciences (NITheCS)},
            country={South Africa}}

\affiliation[4]{organization={Department of Statistics and Actuarial Science, Stellenbosch University},
            city={Stellenbosch},
            postcode={7602},
            country={South Africa}}

\affiliation[5]{organization={Department of Mathematics and Statistics, San Jose State University},
            city={San Jose},
            state={California},
            postcode={95192},
            country={United States of America}}

\begin{abstract}
In this paper, we propose a model-based framework to robustify inference for circular data in the presence of anomalous observations, distinguishing between mild and gross anomalies.
Starting from a unimodal and symmetric reference model on $[0,2\pi)$, parametrized by a mean direction and concentration, we construct a family of finite mixtures: a gross-anomaly model obtained by adding a circular uniform component; a mild-anomaly (contaminated) model obtained by mixing the reference distribution with a less concentrated version sharing the same mean direction; and a general three-component specification combining both models, the double-contaminated model. Posterior component probabilities provide an automatic classification of observations without ad hoc thresholds, while the mixing weights yield interpretable measures of anomaly prevalence and dispersion inflation.
For illustration, we consider two classical circular reference distributions, the wrapped normal and von Mises.
\blue{The methodology is evaluated through an extensive simulation study and three real-data applications involving animal movement directions and wind directions.} The results indicate that jointly modelling mild and gross departures improves model fit and yields an informative decomposition of the directional data, demonstrating that mixture-based robustness is valuable not only for anomaly detection but also for the interpretation and the identification of latent structure in directional data.
\end{abstract}


\end{frontmatter}

\section{Introduction} 
\label{sec:Intro}
\blue{Angular measurements arise naturally across a broad range of environmental and ecological contexts. Slope aspect governs vegetation distribution, habitat suitability, solar radiation receipt, and wildfire behaviour (Stage, 2003; Pierce et al., 2005); wind direction and ocean currents structure atmospheric and marine transport processes (Carta et al., 2008); and animal movement bearings reflect behavioural responses to landscape features and resource gradients (Morales et al., 2004; Codling et al., 2008). Because such variables are intrinsically periodic or angular, they are termed circular data, for which standard linear statistical methods are generally inappropriate because they ignore the intrinsic circular topology of the sample space \citep{mardia2009directional,jammalamadaka2001topics}.}
 
\blue{To properly account for this periodicity, probability distributions defined on the circle are required. A wide range of parametric circular distributions has been developed for this purpose, including the von Mises distribution \citep{abe2012circular}, the wrapped normal distribution \citep{jammalamadaka2001topics,mardia2009directional}, and related families; see also \citet{pewsey2013circular}, \citet{kato2015tractable}, and \citet{ley2017modern}. Directional statistics enable inference on directional trends while respecting the continuity of the sample space, and have been applied successfully to wind energy assessment (Carta et al., 2008), animal navigation (Codling et al., 2008), and landscape ecology (Pierce et al., 2005), among other domains.}
 
\blue{Regardless of the specific distributional framework adopted, parameter estimation may be strongly influenced by anomalous observations or outliers---hereafter referred to simply as anomalies---and is typically sensitive to departures from model assumptions, particularly with respect to distributional shape \citep{huber2011robust}. In environmental and ecological datasets, such anomalies may originate from sensor malfunctions, data transmission errors, extreme meteorological events, unusual ecological behaviours, or rare environmental processes (Hodge and Austin, 2004; Chandola et al., 2009). Because the geometry of circular data renders conventional outlier detection methods inappropriate and potentially misleading, anomalous observations that go undetected can bias parameter estimates, distort inferred directional patterns, and compromise subsequent modelling and decision-making. This concern is compounded by the fact that environmental monitoring systems increasingly generate large volumes of high-frequency, directional data (Lark, 2014), placing increasing demands on robust methods that are both statistically principled and computationally tractable. Detecting anomalies and constructing estimation procedures that are resilient to them and to model misspecification, therefore, remain fundamental open challenges in directional statistics, with significant implications for environmental and ecological research.}
 
From a model-based perspective, anomalies are observations that deviate from a reference (also referred to as target or baseline) model (\citealp{Agga:Outl:2013} and \citealp{Hawk:Iden:2013}), which, for simplicity, is assumed here to be any absolutely continuous, unimodal, and
symmetric circular distribution.
For a discussion on the notion of a reference model, refer to \citet{davies1993identification} and \citet{tomarchio2020dichotomous}.
In turn, anomalies are often broadly categorized as either ``mild'' or ``gross'' \citep[cf.][pp.~79--80]{Ritt:Robu:2015}.
 
The first contribution of this paper is the adaptation of anomaly definitions from the linear domain to the directional domain.
Mild anomalies correspond to observations that still share the main structural features of the reference model—such as the same mean direction—but arise from a secondary component characterized by lower concentration (i.e., increased dispersion). 
In directional applications, these may correspond, for instance, to occasional wind directions that remain centred around the prevailing flow yet exhibit greater variability. 
Such observations typically reflect limitations in model specification and motivate the adoption of a distribution flexible enough to accommodate moderate departures from the assumed concentration structure.
Gross anomalies, instead, represent observations that do not follow any concentrated directional pattern and may be viewed as arising from a diffuse background mechanism. 
These may be caused by recording errors or sporadic random orientations unrelated to the underlying process. 
A natural probabilistic representation of such a diffuse mechanism is provided by the circular uniform distribution, which assigns equal probability to all directions on the circle. 
In their presence, it is generally preferable not to force them into the structural component of the model, but rather to account for them through a separate probabilistic mechanism that leaves to the user the possibility to detect and suppress them if needed.
 
The model-based framework we propose in Section~\ref{sec:proposed models} aims to protect the reference model for the regular observations against the occurrence of mild and gross anomalies, in accordance with the definitions introduced above. 
The underlying idea is to represent the overall distribution of the data as a finite mixture (double-contaminated model), where each component corresponds to a specific data type (regular observations, mild anomalies, or gross anomalies).
Owing to the mixture structure, the model assigns to each point on the circle \emph{a posteriori} probability of belonging to each component. 
This feature provides the user with the option of automatically detecting and classifying observations according to the maximum \emph{a posteriori} probability rule, without requiring the specification of any subjective threshold. 
As a consequence of this classification rule, the circle is partitioned into regions associated with the data types under consideration.
 
In more detail, mild anomalies are handled by augmenting the reference model with a contaminant component; see \citet{punzo2016parsimonious,punzo2017robust} and \citet{otto2025contaminated,otto2026modeling} for examples of contaminated models in the non-directional setting. 
Specifically, the contaminant component shares the same mean direction as the reference component but incorporates an attenuation parameter that deflates the concentration parameter. 
Consequently, while the mean direction remains unchanged, the contaminant component exhibits lower concentration (i.e., greater dispersion), enabling it to capture moderately anomalous observations.
To the best of our knowledge, the earliest reference to a two-component von Mises mixture sharing a common mean direction appears in \citet{jones2012inverse}, where it was introduced in the context of a reduced model for data fitting. 
Similarly, albeit on a different manifold, \citet{zhang2023model} proposed a contaminated von Mises–Fisher distribution on the sphere, extending the classical directional model by incorporating a contamination component to accommodate mild anomalies and thereby enhance robustness. A further extension to the Kent distribution was then explored by \citet{dong2024contaminated}.
 
To address gross anomalies, the model is further extended by incorporating a circular uniform component, as naturally suggested by the considerations above. 
Because the circular uniform distribution does not exhibit any preferred direction, its inclusion weakens the overall directional structure of the mixture. 
In particular, it reduces the resultant length and flattens the distribution without introducing spurious directional patterns.
 
Overall, this construction leads to a flexible family of mixture models: a general specification that jointly accounts for regular observations, mild anomalies, and gross anomalies (Section~\ref{sec:mild and gross anomalies}), together with two simpler sub-models tailored to settings where only one type of anomaly is of interest (Sections~\ref{sec:Mild anomalies} and \ref{sec:Gross anomalies}).
An additional strength of the proposed framework lies in the direct interpretability of its parameters, a feature of fundamental importance in parametric statistical modelling. 
In particular, the mixing proportions quantify the prevalence of mild and gross anomalies in the data, while the attenuation parameter measures the extent of dispersion inflation relative to the reference model. 
The remaining parameters retain their usual interpretation as measures of mean direction and concentration. 
Moreover, the interpretability of the parameters enables practitioners to translate model estimates into meaningful statements about directional persistence, variability, and anomaly prevalence.
 
This is, of course, not the first contribution addressing "anomalies" on the circle. 
Beyond the works already cited above, a number of additional studies have mainly focused on anomaly-detection procedures, which play a crucial role in ensuring reliable model fitting and in preventing masking or swamping effects that may bias inference. 
Recently, \citet{Demni2023anomaly} addressed anomaly detection by fitting a von Mises distribution to the data, estimating its centre and concentration using robust statistics, and flagging observations whose circular distance from the estimated centre exceeds a pre-specified threshold.
Earlier approaches primarily relied on deletion-based strategies, whereby suspected observations are removed and their influence subsequently assessed through changes in summary statistics—such as the mean resultant length—or via discordancy tests \citep{ko1992robust}.
These methods are typically suited to small samples, often focus on detecting a single anomaly, and are vulnerable to masking effects. 
A seminal contribution in the circular setting is due to \citet{collett1980outliers}, who investigated discordancy tests through Monte Carlo simulation.
Other proposed procedures were based on circular distances, cumulative distance measures, or spacing theory. 
Although these methods represent methodological progress, they often remain limited to single-anomaly detection, rely on simulation-based critical values under specific distributional assumptions, or require multiple anomalies to be sufficiently separated in order to be identifiable (see \citealp{abuzaid2009new,abuzaid2012statistics, mohamed2016new, mahmood2017detection}). 
 
The main contributions of this paper can be summarized as follows:
\begin{itemize}
    \item[(i)] We formalize definitions of mild and gross anomalies within the directional setting.
    \item[(ii)] We propose a model-based framework that safeguards inference on the assumed reference model against the influence of anomalies.
    \item[(iii)] The proposed double‑contaminated model yields a five‑parameter formulation with interpretable parameters related to mean direction, concentration, and contamination.
    \item[(iv)] The proposed model enables automatic detection of mild and gross anomalies through outputs that remain directly interpretable in practical scenarios. 
    
\end{itemize}

The remainder of the paper is organized as follows. 
Section~\ref{sec:proposed models} introduces the proposed models and highlights their key features. 
Section~\ref{sec:illustrative examples} briefly presents illustrative choices of reference distributions; these should be regarded as examples only, since the framework is not restricted to them and can accommodate other circular models, provided that they satisfy the properties required of a reference distribution, namely symmetry and unimodality.  
Section~\ref{sec: maximum likelihood estimation} addresses model identifiability and describes parameter estimation via maximum likelihood, implemented through the expectation-maximization algorithm. 
Section~\ref{Sec: simulation study} reports the results of a simulation study assessing the performance of the proposed framework and providing further insight into its behaviour. 
Finally, Section~\ref{sec: data application} illustrates its effectiveness through three real-data applications, followed by concluding remarks.
 
\section{Proposed framework}
\label{sec:proposed models}
 
Let $\Theta$ be a circular random variable defined on $[0,2\pi)$ with an absolutely continuous, unimodal, and symmetric distribution.
Let $f_{\text{R}}(\theta;\mu,\kappa)$ denote the probability density function (PDF) assumed for $\Theta$, parametrized such that $\mu \in [0,2\pi)$ represents the mean direction and $\kappa>0$ is a concentration parameter, with larger values of $\kappa$ corresponding to higher concentration of the distribution around $\mu$.
If $\Theta$ follows this distribution, we write $\Theta \sim \mathcal{R}(\mu,\kappa)$.
The distribution $\mathcal{R}(\mu,\kappa)$ will serve as the reference model for the regular observations---also referred to as ``good'' in the nomenclature of \citet{aitkin1980mixture}.
 
To protect the reference model against the presence of anomalous observations—whether mild or gross, as discussed in Section~\ref{sec:Intro}—and to enable their probabilistic detection, we construct suitable mixture extensions of the reference distribution.
The specific form of the mixture depends on the type of contamination affecting the data.
 
More precisely, in Section~\ref{sec:Gross anomalies} we consider a mixture of the reference model with the circular uniform distribution to accommodate gross anomalies.
In Section~\ref{sec:Mild anomalies}, we instead mix the reference model with a less concentrated (i.e., more dispersed) version of itself to account for mild anomalies.
Finally, in Section~\ref{sec:mild and gross anomalies}, we combine these ideas into a three-component mixture model that simultaneously captures regular observations, mild anomalies, and gross anomalies which we term as the double-contaminated model.
 
\subsection{Gross anomalies}
\label{sec:Gross anomalies}
 
A general framework for protecting the reference model against the occurrence of gross anomalies (GA), and allowing for their detection, can be constructed by mixing the reference PDF with the PDF $f_{\text{U}}(\theta)=1/(2\pi)$ of the circular uniform distribution, denoted as $\text{Uniform}(0,2\pi)$.
In this case, the parametric formulation of the PDF of $\Theta$ becomes the following two-component, three-parameter mixture:
\begin{align}\label{pdf gross anomaly model}
    f_{\text{GA}}(\theta;\mu,\kappa,\delta_{\text{U}}) 
    = (1-\delta_{\text{U}})\underbrace{f_{\text{R}}(\theta;\mu,\kappa)}_{\text{reference}}
    + \delta_{\text{U}}\underbrace{\frac{1}{2\pi}}_{\text{uniform}},
\end{align}
where the contamination parameter $\delta_{\text{U}}\in(0,1)$ represents the proportion of observations arising from the uniform component, that is, the proportion of gross anomalies.
If $\Theta$ has the PDF in \eqref{pdf gross anomaly model}, then we write $\Theta\sim \mathcal{GA}(\mu,\kappa,\delta_{\text{U}})$.
Clearly, as $\delta_{\text{U}} \to 0^+$, 
the model $\mathcal{GA}(\mu,\kappa,\delta_{\text{U}})$ 
converges to the reference distribution $\mathcal{R}(\mu,\kappa)$. 
This corresponds to a boundary (limiting) case of the parameter space.
 
Once the parameters $\mu$, $\kappa$, and $\delta_{\text{U}}$ are estimated, say $\hat\mu$, $\hat\kappa$, and $\hat\delta_{\text{U}}$, respectively, it is possible to determine whether a generic observation $\theta$ is a gross anomaly with respect to the reference distribution via the \emph{a posteriori} probability
\begin{equation}
\label{eq a posteriori probability gross outlier}
    P\!\left(\theta \text{ arises from the uniform component}\mid \hat\mu, \hat\kappa,\hat\delta_{\text{U}}\right)
    = \frac{\hat\delta_{\text{U}}}{2\pi f_{\text{GA}}(\theta;\hat\mu,\hat\kappa,\hat\delta_{\text{U}})}.
\end{equation}
It is natural to classify $\theta$ as a gross anomaly if the probability in \eqref{eq a posteriori probability gross outlier} exceeds $0.5$, and as a regular observation otherwise.
 
\subsection{Mild anomalies}
\label{sec:Mild anomalies}
 
A general framework for protecting the reference model against the occurrence of mild anomalies (MA), and allowing for their detection, can be constructed by mixing the reference PDF with a contaminant component having a lower concentration.
In this case, the parametric formulation of the PDF of $\Theta$ becomes the following two-component, four-parameter mixture:
\begin{align}\label{pdf mild anomaly model}
    f_{\text{MA}}(\theta;\mu,\kappa,\delta_{\text{C}},\eta)
    =(1-\delta_{\text{C}})
    \underbrace{f_{\text{R}}(\theta;\mu,\kappa)}_{\text{reference}}
    +\delta_{\text{C}}
    \underbrace{f_{\text{R}}(\theta;\mu,\kappa/\eta)}_{\text{contaminant}},
\end{align}
where $\delta_{\text{C}}\in(0,1)$ represents the proportion of observations arising from the contaminant component and $\eta>1$ denotes the degree of contamination.
If $\Theta$ has the PDF in \eqref{pdf mild anomaly model}, we write 
$\Theta\sim \mathcal{MA}(\mu,\kappa,\delta_{\text{C}},\eta)$.
Since $\eta>1$, it acts as an attenuation parameter: it reduces the concentration (and consequently increases the variability) of observations originating from the contaminant component relative to the reference distribution.
 
As either $\delta_{\text{C}} \to 0^+$ or $\eta \to 1^+$, the model $\mathcal{MA}(\mu,\kappa,\delta_{\text{C}},\eta)$ converges to the reference distribution $\mathcal{R}(\mu,\kappa)$, corresponding to a boundary configuration of the parameter space.
Moreover, as $\eta \to \infty$ (so that $\kappa/\eta \to 0^+$), the contaminant PDF $f_{\text{R}}(\theta;\mu,\kappa/\eta)$ converges to the circular uniform PDF, and the model approaches the GA framework introduced in Section~\ref{sec:Gross anomalies}.
Hence, the formulation in \eqref{pdf mild anomaly model} encompasses both mild anomalies and---at a specific boundary configuration of the parameter space (namely, $\eta \to \infty$)---gross anomalies.
 
Once the parameters $\mu$, $\kappa$, $\delta_{\text{C}}$, and $\eta$ are estimated, say $\hat\mu$, $\hat\kappa$, $\hat\delta_{\text{C}}$, and $\hat\eta$, respectively, it is possible to determine whether a generic observation $\theta$ is a mild anomaly with respect to the reference distribution via the \emph{a posteriori} probability
\begin{align}\label{eq a posteriori probability mild or gross outlier}
    P\!\left(\theta \text{ arises from the contaminant component} 
    \mid \hat\mu, \hat\kappa,\hat\delta_{\text{C}},\hat\eta\right)
    = \frac{\hat\delta_{\text{C}}
    f_{\text{R}}(\theta;\hat\mu,\hat\kappa/\hat\eta)}
    {f_{\text{MA}}(\theta;\hat\mu,\hat\kappa,\hat\delta_{\text{C}},\hat\eta)}.
\end{align}
It is natural to classify $\theta$ as a mild anomaly if the probability in \eqref{eq a posteriori probability mild or gross outlier} exceeds $0.5$, and as a regular observation otherwise.
 
\subsection{Mild and gross anomalies}
\label{sec:mild and gross anomalies}
 
Putting together the ideas developed in Sections~\ref{sec:Gross anomalies} and \ref{sec:Mild anomalies}, a general framework for protecting the reference model against the occurrence of both mild and gross anomalies (MGA) is obtained by mixing three components: a reference component, a contaminant component with lower concentration to capture mild anomalies, and a uniform component to account for gross anomalies.
In this case, the parametric formulation of the PDF of $\Theta$ becomes the following three-component, five-parameter mixture:
\begin{equation}\label{pdf mild and gross anomaly model}
    f_{\text{MGA}}(\theta;\mu,\kappa,\delta_{\text{U}},\delta_{\text{C}},\eta)
    =
    (1-\delta_{\text{U}}-\delta_{\text{C}})
    \underbrace{f_{\text{R}}(\theta;\mu,\kappa)}_{\text{reference}}
    + \delta_{\text{C}}
    \underbrace{f_{\text{R}}(\theta;\mu,\kappa/\eta)}_{\text{contaminant}}
    + \delta_{\text{U}}
    \underbrace{\frac{1}{2\pi}}_{\text{uniform}},
\end{equation}
where $\delta_{\text{U}}\in(0,1)$ and $\delta_{\text{C}}\in(0,1)$ satisfy $\delta_{\text{U}}+\delta_{\text{C}}<1$, and $\eta>1$.
If $\Theta$ has the PDF in \eqref{pdf mild and gross anomaly model}, we write
$\Theta\sim \mathcal{MGA}(\mu,\kappa,\delta_{\text{U}},\delta_{\text{C}},\eta)$.
Here, $\delta_{\text{U}}$ represents the proportion of gross anomalies arising from the uniform component, $\delta_{\text{C}}$ denotes the proportion of mild anomalies arising from the contaminant component, and $\eta$ is the degree of contamination controlling the reduction in concentration of the contaminant component relative to the reference distribution.  
The model in \eqref{pdf mild and gross anomaly model} encompasses the previously introduced specifications (i.e., MA and GA) as limiting cases under suitable boundary configurations of the parameter space.
In particular, as $\delta_{\text{U}}\to 0^+$, the formulation converges to $\mathcal{MA}(\mu,\kappa,\delta_{\text{C}},\eta)$.
Conversely, as $\delta_{\text{C}}\to 0^+$ or $\eta\to 1^+$, 
\eqref{pdf mild and gross anomaly model} converges to the GA model $\mathcal{GA}(\mu,\kappa,\delta_{\text{U}})$.
Finally, when both contamination mechanisms vanish---either through $\delta_{\text{U}}\to 0^+$ and $\delta_{\text{C}}\to 0^+$, or equivalently through $\delta_{\text{U}}\to 0^+$ and $\eta\to 1^+$---the specification converges to the reference distribution $\mathcal{R}(\mu,\kappa)$.
 
Once the parameters $\mu$, $\kappa$, $\delta_{\text{U}}$, $\delta_{\text{C}}$, and $\eta$ are estimated, say $\hat\mu$, $\hat\kappa$, $\hat\delta_{\text{U}}$, $\hat\delta_{\text{C}}$, and $\hat\eta$, respectively, the \emph{a posteriori} probabilities for component membership can be computed.
Specifically, the \emph{a posteriori} probability that an observation $\theta$ arises from the contaminant component is
\begin{align}\label{eq a posteriori probability contaminated 3}
    P\!\left(\theta \text{ arises from the contaminant component}
    \mid \hat\mu, \hat\kappa,\hat\delta_{\text{U}},\hat\delta_{\text{C}},\hat\eta\right)
    =
    \frac{\hat\delta_{\text{C}}
    f_{\text{R}}(\theta;\hat\mu,\hat\kappa/\hat\eta)}
    {f_{\text{MGA}}(\theta;\hat\mu,\hat\kappa,\hat\delta_{\text{U}},\hat\delta_{\text{C}},\hat\eta)}
    =: \hat p_{\text{C}},
\end{align}
while the \emph{a posteriori} probability that $\theta$ arises from the uniform component is
\begin{align}\label{eq a posteriori probability uniform 3}
    P\!\left(\theta \text{ arises from the uniform component}
    \mid \hat\mu, \hat\kappa,\hat\delta_{\text{U}},\hat\delta_{\text{C}},\hat\eta\right)
    =
    \frac{\hat\delta_{\text{U}}\frac{1}{2\pi}}
    {f_{\text{MGA}}(\theta;\hat\mu,\hat\kappa,\hat\delta_{\text{U}},\hat\delta_{\text{C}},\hat\eta)}
    =: \hat p_{\text{U}}.
\end{align}
The \emph{a posteriori} probability that $\theta$ arises from the reference component is therefore $\hat p_{\text{R}} = 1 - \hat p_{\text{C}} - \hat p_{\text{U}}$.
An observation is classified as a regular point, mild anomaly, or gross anomaly according to the component with the largest \emph{a posteriori} probability, that is,
$\operatorname{argmax}\{\hat p_{\text{R}}, \hat p_{\text{C}}, \hat p_{\text{U}}\}$.
 
\subsection{Hierarchical representation}
\label{subsec:Hierarchical representation}
 
It is worth noting that all the proposed models admit a hierarchical representation.
In the most general case, when $\Theta\sim \mathcal{MGA}(\mu,\kappa,\delta_{\text{U}},\delta_{\text{C}},\eta)$, we have
\begin{equation}\label{eq hierarchical representation}
\begin{aligned}
W &\sim \mathcal{D}_{\mathcal{S}_{\text{MGA}}}(\bdelta_{\text{MGA}}),\\
\Theta \mid W = w &\sim \mathcal{R}(\mu, w\kappa),
\end{aligned}
\end{equation}
where $\mathcal{S}_{\text{MGA}} = \left\{1,1/\eta,0\right\}$ and $\bdelta_{\text{MGA}} = \left(1-\delta_{\text{U}}-\delta_{\text{C}}, \delta_{\text{C}}, \delta_{\text{U}}\right)$, and where $\mathcal{D}_{\mathcal{S}}(\boldsymbol{\delta})$ denotes a finite discrete distribution supported on the set $\mathcal{S}=\{s_1,\ldots,s_K\}$ with probability mass function (PMF) $P(W=s_k)=\delta_k$, $k=1,\ldots,K$,
where $\bdelta=(\delta_1,\ldots,\delta_K)$ satisfies 
$\delta_k > 0$ and $\sum_{k=1}^K\delta_k=1$.
Thus, in the MGA case,
\[
W=
\begin{cases}
1 & \text{with probability } 1-\delta_{\text{U}}-\delta_{\text{C}},\\[4pt]
\dfrac{1}{\eta} & \text{with probability } \delta_{\text{C}},\\[6pt]
\epsilon & \text{with probability } \delta_{\text{U}},
\end{cases}
\]
where $\epsilon>0$ is a value close to zero.
Under the assumption that
\[
\mathcal{R}(\mu,\kappa) \xrightarrow[\kappa \to 0^+]{} \text{Uniform}(0,2\pi),
\]
the hierarchical representation in \eqref{eq hierarchical representation}
is equivalent to the mixture formulation in
\eqref{pdf mild and gross anomaly model}.
 
Analogous hierarchical representations, as the one given in \eqref{eq hierarchical representation}, arise for the GA and MA models.
Specifically, in the GA model, $\mathcal{S}_{\text{GA}} = \left\{1,0\right\}$ and $\bdelta_{\text{GA}} = \left(1-\delta_{\text{U}}, \delta_{\text{U}}\right)$, while in the MA model, $\mathcal{S}_{\text{MA}} = \left\{1,1/\eta\right\}$ and $\bdelta_{\text{MA}} = \left(1-\delta_{\text{C}}, \delta_{\text{C}}\right)$.
Hence, all proposed models can be interpreted as scale mixtures of the reference distribution, where the latent variable $W$ rescales the concentration parameter and governs the degree and type of contamination.

\subsection{Further properties}
\label{subsec:Circular moments and flexibility}
 
In order to better understand the structural properties of the proposed models, we briefly discuss their trigonometric moments (see, e.g., \citet{mardia2009directional} and \citet{pewsey2004large} for further details). The mean resultant length, $\rho$, measures the concentration of $\Theta$ around $\mu$, and is defined as
 
\begin{align*}
 \rho & = \E\!\left[\cos(\Theta-\mu)\right] \in [0,1].
\end{align*}
For a finite mixture, the mean resultant length is the convex combination of the component mean resultant lengths.
Denoting by $\rho_{\text{R}}(\kappa)$ the mean resultant length of the reference model $\mathcal{R}(\mu,\kappa)$, and recalling that the circular uniform distribution has mean resultant length equal to zero, we obtain
\begin{align*}
\rho_{\text{GA}}
& = (1-\delta_{\text{U}})\rho_{\text{R}}(\kappa),\\
\rho_{\text{MA}}
& = (1-\delta_{\text{C}})\rho_{\text{R}}(\kappa)
  + \delta_{\text{C}}\rho_{\text{R}}(\kappa/\eta),\\
\rho_{\text{MGA}}
& = (1-\delta_{\text{U}}-\delta_{\text{C}})\rho_{\text{R}}(\kappa)
  + \delta_{\text{C}}\rho_{\text{R}}(\kappa/\eta).
\end{align*}
Since $\rho_{\text{R}}(\kappa/\eta) < \rho_{\text{R}}(\kappa)$ for $\eta>1$, the mixture models allow for a broader range of variability than the reference distribution alone, and variability increases as $\delta_{\text{U}}$ or $\delta_{\text{C}}$ increase \citep{jammalamadaka2001topics}.
 
Since all components of the GA, MA and MGA models are symmetric about the same mean direction $\mu$, the mean direction remains equal to $\mu$ whenever the mean resultant length is strictly positive; hence contamination does not alter the mean direction. 
Moreover, because the reference distribution is assumed to be unimodal and symmetric about $\mu$, and all mixture components share the same centre $\mu$ (with the circular uniform component being constant over the circle), the resulting mixtures preserve symmetry and, under the stated assumptions, retain a unique mode at $\mu$.
In particular, the MA model is a convex combination of unimodal symmetric densities with common mode and therefore remains unimodal, while the GA and MGA models, although including a uniform component, do not introduce additional local maxima since the uniform PDF is constant over the circle. 
Thus, the proposed extensions preserve both symmetry and unimodality.
 
\section{Illustrative reference models}
\label{sec:illustrative examples}
 
To illustrate the proposed framework, we consider two classical choices for the reference distribution, namely the wrapped normal and the von Mises distributions. Detailed discussions of these models can be found in \citet{mardia2009directional} and \citet{jammalamadaka2001topics}.
 
\subsection{Wrapped normal}
\label{subsec:WN}
 
A circular random variable $\Theta$ is said to follow a wrapped normal (WN) distribution if its PDF is
\begin{equation}\label{pdf wrapped normal}
    f_{\text{WN}}(\theta;\mu,\kappa)
    =\frac{1}{2\pi}\left\{1+2\sum_{p=1}^\infty 
    \exp\left(-\frac{p^2}{2\kappa}\right)
    \cos\bigl(p(\theta-\mu)\bigr)\right\},
    \quad 0\leq\theta<2\pi,
\end{equation}
where $\mu \in [0,2\pi)$ is the mean direction and $\kappa>0$ is a concentration parameter, with larger values of $\kappa$ corresponding to stronger concentration around $\mu$. 
If $\Theta$ has the PDF in \eqref{pdf wrapped normal}, we denote it as $\Theta\sim\mathcal{WN}(\mu,\kappa)$.
 
The parametrization in \eqref{pdf wrapped normal} is equivalent to the classical variance-based representation of the WN distribution (see, e.g., \citealp{mardia2009directional,jammalamadaka2001topics}), obtained through the reparametrization $\kappa = 1/\sigma^2$, where $\sigma^2$ denotes the variance of the underlying linear normal distribution. 
We adopt this concentration-based formulation to ensure coherence with the contamination mechanism described in Section~\ref{sec:proposed models}.
 
\begin{figure}[!h]
\centering
\subfloat[Effect of varying $\delta_{\text{U}}$ of the cuWN distribution with fixed $\delta_{\text{C}} = 0.2$ and $\eta = 2$.]{\includegraphics[width=0.48\textwidth]{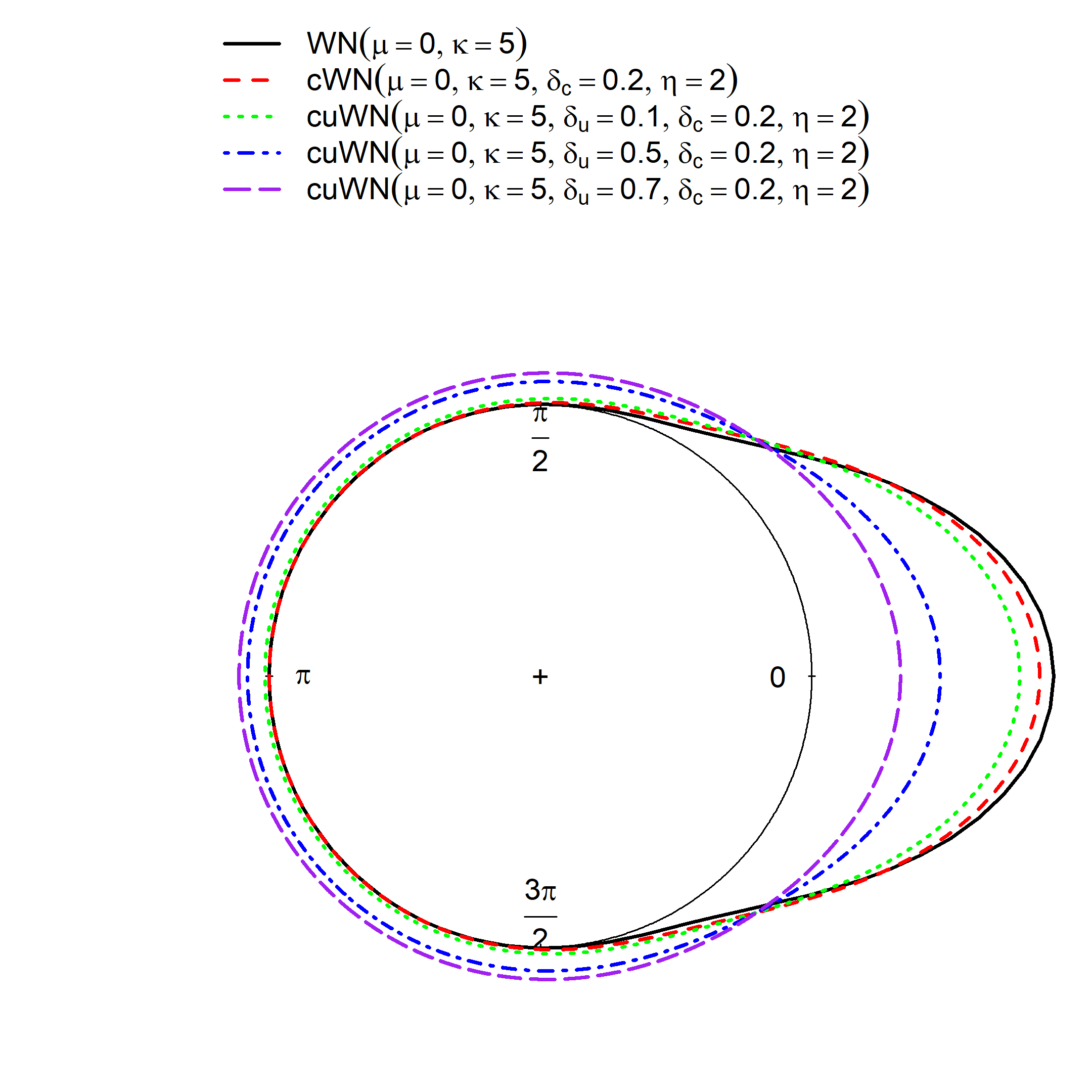}}
\hfill
\subfloat[Effect of varying $\delta_{\text{C}}$ of the cuWN distribution with fixed $\delta_{\text{U}} = 0.5$ and $\eta = 2$.]{\includegraphics[width=0.48\textwidth]{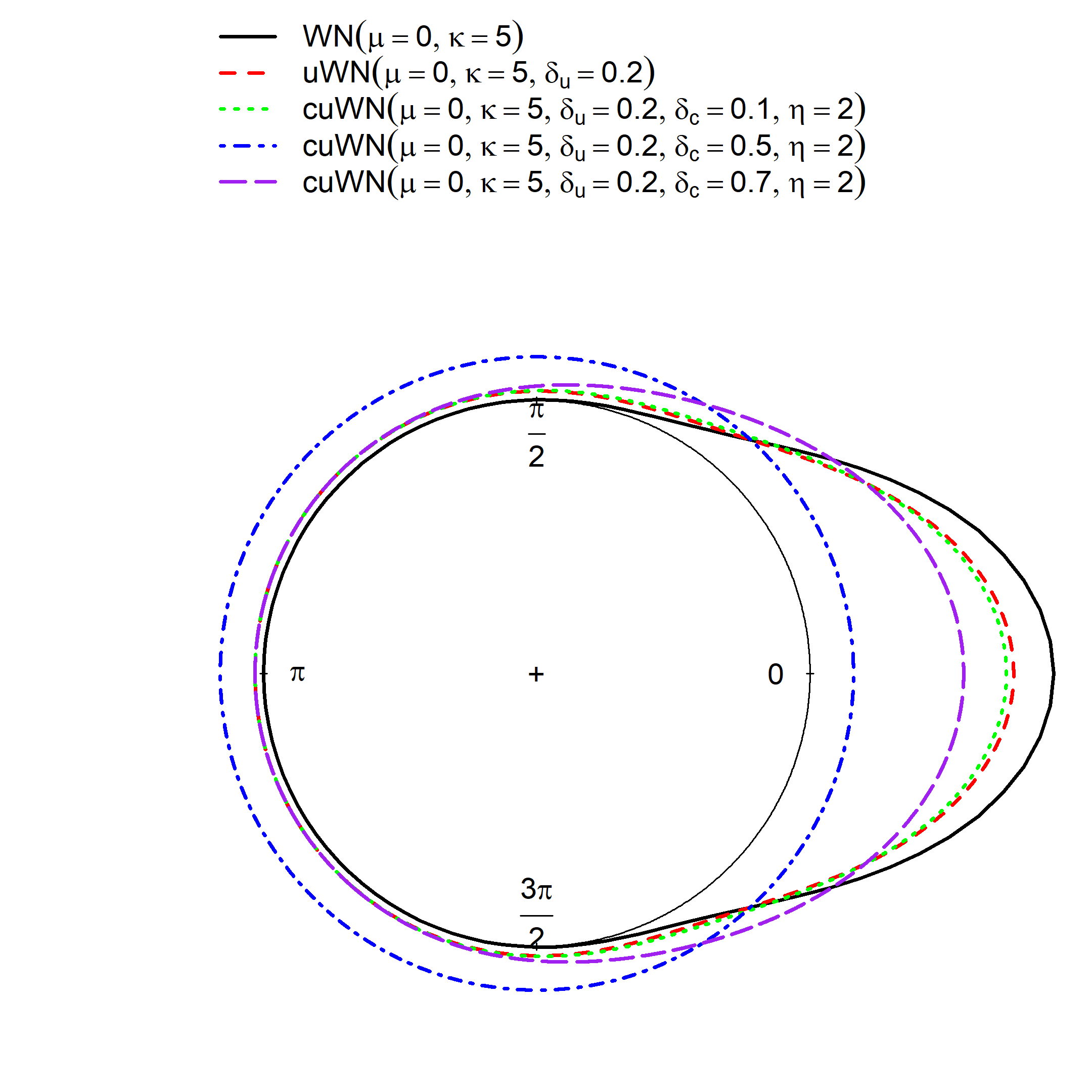}}
\hfill
\subfloat[Effect of varying $\eta$ on the cuWN distribution with fixed $\delta_{\text{U}} = \delta_{\text{C}} = 0.2$.]{\includegraphics[width=0.48\textwidth]{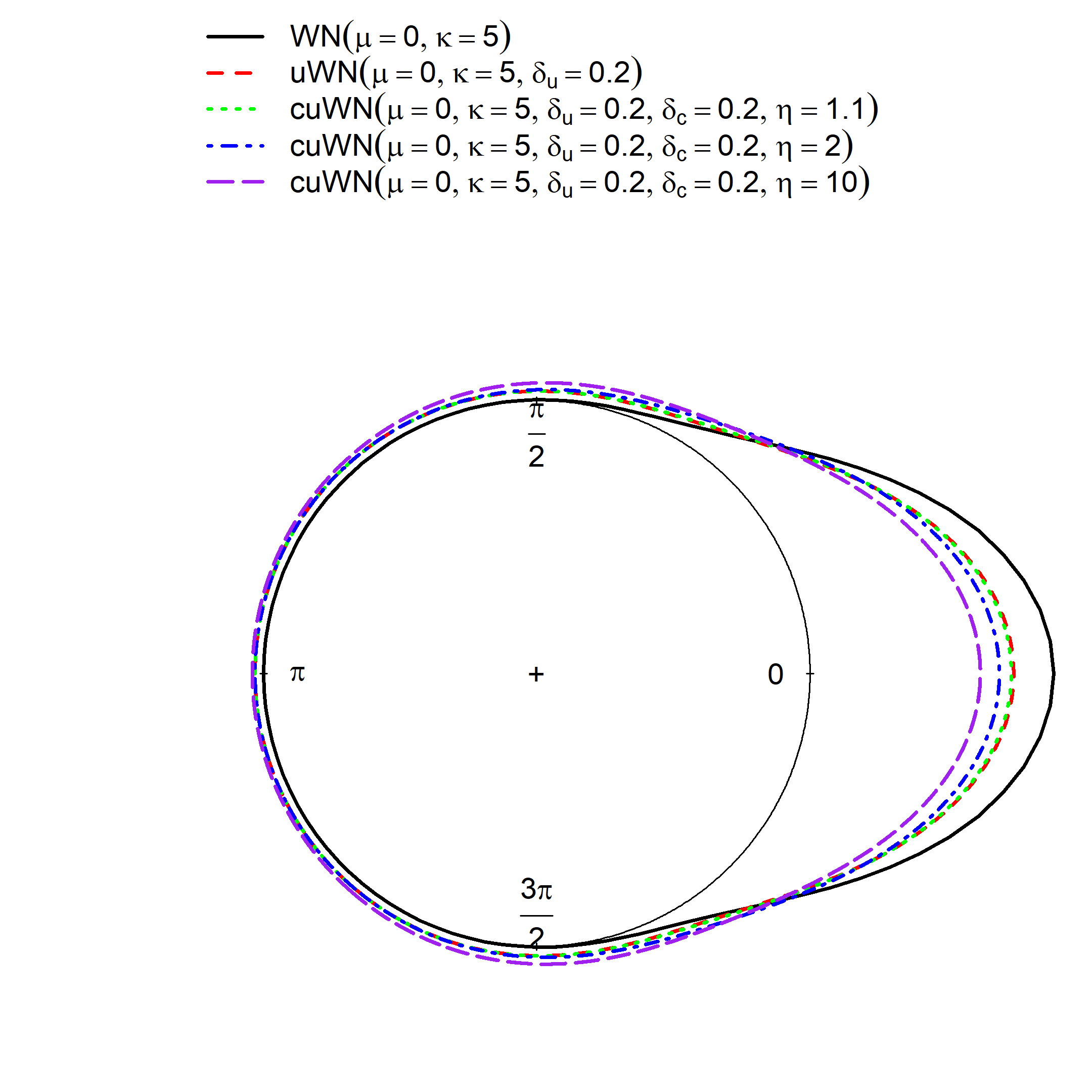}}
\caption{Probability density functions of WN-based mixture models for anomaly detection. The reference WN distribution (black solid line) is shown together with selected limiting cases (red line) and the effect of varying contamination parameters within the MGA framework.}
\label{fig:wn_mixtures}
\end{figure}

When the WN distribution is adopted as the reference distribution in the general anomaly detection framework presented in Section~\ref{sec:proposed models}, three specialized models emerge naturally. 
First, substituting $f_{\text{R}}$ with $f_{\text{WN}}$ in \eqref{pdf gross anomaly model} yields the uniform WN (uWN) distribution, capable of detecting gross anomalies. 
Second, using $f_{\text{WN}}$ in \eqref{pdf mild anomaly model} produces the contaminated WN (cWN) distribution, which can identify mild anomalies through a contaminant component with attenuated concentration $\kappa/\eta$. 
Finally, the framework in \eqref{pdf mild and gross anomaly model} leads to the contaminated-uniform WN (cuWN) distribution, a three-component mixture that simultaneously detects both mild and gross anomalies by combining the reference WN, a contaminated WN with reduced concentration, and a circular uniform component. 
\figurename~\ref{fig:wn_mixtures} illustrates the behaviour of these mixture models for varying parameter values, demonstrating how the distributional shape changes as the contamination parameters are adjusted.
 
\subsection{Von Mises distribution}
\label{subsec:VMM}
 
A random variable $\Theta$ is said to follow a von Mises (VM) distribution if its PDF is given by
\begin{align}\label{pdf von mises}
    f_{\text{VM}}(\theta;\mu,\kappa)
    = \frac{1}{2\pi I_0(\kappa)} \exp\bigl(\kappa \cos(\theta-\mu)\bigr),
    \quad 0 \leq \theta < 2\pi,
\end{align}
where $\mu \in [0,2\pi)$ is the mean direction, $\kappa > 0$ is the concentration parameter, and $I_0(\cdot)$ denotes the modified Bessel function of the first kind of order zero. Note that we restrict $\kappa >0$ for arguments specified in Section \ref{sec:identifiability}.
If $\Theta$ has PDF \eqref{pdf von mises}, we write $\Theta \sim \mathcal{VM}(\mu,\kappa)$.
 
When the VM distribution is used as the reference model in the general framework, three corresponding models arise: the uniform VM (uVM) for gross anomaly detection via \eqref{pdf gross anomaly model}, the contaminated VM (cVM) for mild anomalies via \eqref{pdf mild anomaly model}, and the contaminated-uniform VM (cuVM) for the simultaneous detection of mild and gross anomalies under \eqref{pdf mild and gross anomaly model}.
 
\section{Maximum likelihood estimation}
\label{sec: maximum likelihood estimation}
 
In Section~\ref{sec:identifiability}, we first discuss the identifiability of the proposed anomaly detection framework.
We then present an expectation–maximization (EM) algorithm for maximum likelihood (ML) estimation (Section~\ref{sec:em algorithm}) of the parameters of the general MGA model introduced in Section~\ref{sec:mild and gross anomalies}. 
This is followed by a discussion of the initialization strategy for the considered EM algorithm (Section~\ref{sec:initialization}).
 
\subsection{Identifiability}
\label{sec:identifiability}
 
Identifiability is a fundamental prerequisite for statistical inference, ensuring that distinct parameter values correspond to distinct probability distributions. This property underlies the consistency and asymptotic normality of ML estimators. 
For finite mixtures of directional distributions, identifiability results have been established by \citet{kent1983identifiability} and \citet{holzmann2004identifiability}, providing the theoretical foundation for mixture-based modelling on the circle.
 
The proposed MGA model in \eqref{pdf mild and gross anomaly model} can be written as a three-component mixture consisting of: 
(i) a reference distribution with parameters $(\mu,\kappa)$, 
(ii) a contaminated component sharing the same mean direction $\mu$ and having reduced concentration $\kappa/\eta$, and 
(iii) a circular uniform component. 
The corresponding mixing proportions are functions of $\delta_{\text{C}}$ and $\delta_{\text{U}}$.
 
Because the MGA model is a finite mixture, two classical sources of non-identifiability must be addressed: (i) label-switching between components, and (ii) overfitting due to degenerate or vanishing components (see \citealp{fruhwirth2006finite}, Chapter~1.3). 
We show that the structural constraints imposed in the MGA specification effectively prevent both issues.
 
The structure of the MGA model closely parallels contamination constructions developed in the non-directional setting. For instance, in the multivariate contaminated normal model, identifiability can fail due to label-switching unless suitable structural constraints are imposed; see \citet{melnykov2025contaminated} and the Supplementary Material of \citet{lim2025heckmanCN}. The key insight is that one must prevent components from being interchangeable via reciprocal reparameterizations of scale and weight. A similar reasoning applies here.
 
First, the constraint $\eta>1$ guarantees that $\kappa \neq \kappa/\eta$ whenever $\kappa>0$, thereby ensuring that the reference and contaminated components correspond to distinct circular densities and cannot be interchanged without violating the admissible parameter space. This constraint induces a strict concentration hierarchy,
$\kappa > \kappa/\eta > 0$, which assigns a clear interpretation to the reference and contaminated components. 
Consequently, the usual label-switching ambiguity between these two concentration-based components does not arise.
Second, since $\kappa>0$, the circular uniform PDF cannot be represented as a special case of either the reference or contaminated components. 
Although the circular uniform PDF may arise as a limiting case when $\kappa \to 0$, it is not contained in the admissible parameter space, and this places the circular uniform at the bottom of the concentration hierarchy. 
Hence, it is structurally distinct from the other two components and cannot be permuted with them.
Third, by assuming $\delta_{\text{C}} \in (0,1)$ and $\delta_{\text{U}} \in (0,1)$, degenerate submodels in which one or more components vanish are excluded. 
This prevents overfitting-type identifiability issues.
 
Taken together, the concentration ordering induced by $\eta>1$ and the structural distinctness of the uniform component ensure that no nontrivial permutation of the mixture components is admissible. Under these conditions, the parameter vector $(\mu,\kappa,\delta_{\text{C}},\delta_{\text{U}},\eta)$ associated with the MGA model is uniquely determined by the distribution, providing a rigorous basis for estimation and inference. 
The same arguments apply to the MA and GA models.
 
\subsection{EM algorithm}
\label{sec:em algorithm}
 
Let $\theta_1, \dots, \theta_n$ be the observed sample from the MGA model \eqref{pdf mild and gross anomaly model}.  
To simplify ML estimation of the parameters $(\mu,\kappa,\delta_{\text{C}},\delta_{\text{U}},\eta)$, we consider the EM algorithm, which is standard for models admitting a mixture representation.  
 
For the application of the EM algorithm, the observed data are viewed as incomplete. The incompleteness stems from the fact that we do not know whether a generic observation $\theta_i$ originates from the reference, contaminant, or uniform component. To formalize this, we introduce two latent indicator variables. The first one is the MA-indicator vector $\mathbf{w}_{\text{C}}=(w_{\text{C}_1},\dots,w_{\text{C}_n})$, where $w_{\text{C}_i}=1$ if $\theta_i$ comes from the contaminant component and $w_{\text{C}_i}=0$ otherwise. The second one is the GA-indicator vector $\mathbf{w}_{\text{U}}=(w_{\text{U}_1},\dots,w_{\text{U}_n})$, where $w_{\text{U}_i}=1$ if $\theta_i$ comes from the uniform component and $w_{\text{U}_i}=0$ otherwise. The reference component corresponds to $1-w_{\text{C}_i}-w_{\text{U}_i}$. For each $i$, at most one of $w_{\text{C}_i}$ or $w_{\text{U}_i}$ equals 1, ensuring that each observation arises from exactly one component.
 
The complete data are thus given by $(\theta_i,w_{\text{C}_i},w_{\text{U}_i})$, $i=1,\dots,n$. 
From \eqref{pdf mild and gross anomaly model}, the complete-data likelihood function is
\begin{align}
L(\mu,\kappa,\delta_{\text{U}},\delta_{\text{C}},\eta)
=\prod_{i=1}^n
\left[(1-\delta_{\text{U}}-\delta_{\text{C}})f_{\text{R}}(\theta_i;\mu,\kappa)\right]^{1-w_{\text{C}_i}-w_{\text{U}_i}}
\left[\delta_{\text{C}}f_{\text{R}}(\theta_i;\mu,\kappa/\eta)\right]^{w_{\text{C}_i}}
\left(\frac{\delta_{\text{U}}}{2\pi}\right)^{w_{\text{U}_i}}.
\end{align}
The complete-data log-likelihood can be decomposed as
\begin{align}\label{eq complete loglikelihood}
\ell_c(\mu,\kappa,\delta_{\text{U}},\delta_{\text{C}},\eta)
=\ell_{c_1}(\delta_{\text{U}},\delta_{\text{C}})
+\ell_{c_2}(\mu,\kappa,\eta),
\end{align}
where
\begin{align}
\ell_{c_1}(\delta_{\text{U}},\delta_{\text{C}})
&=\sum_{i=1}^n
\Big[
(1-w_{\text{C}_i}-w_{\text{U}_i})\log(1-\delta_{\text{U}}-\delta_{\text{C}})
+w_{\text{C}_i}\log(\delta_{\text{C}})
+w_{\text{U}_i}\log(\delta_{\text{U}})
\Big],
\end{align}
and
\begin{align}
\ell_{c_2}(\mu,\kappa,\eta)
=\sum_{i=1}^n
\Big[
(1-w_{\text{C}_i}-w_{\text{U}_i})\log f_{\text{R}}(\theta_i;\mu,\kappa)\\ \nonumber
+ w_{\text{C}_i}\log f_{\text{R}}(\theta_i;\mu,\kappa/\eta)
-w_{\text{U}_i}\log\left(2\pi\right)
\Big].
\end{align}
The algorithm alternates between the E-step and M-step until convergence.
These steps, for the $(k+1)$th iteration of the algorithm, are detailed below.
 
\subsubsection*{E-step}
 
In the E-step, the conditional expectation of the complete-data log-likelihood function is computed as 
\begin{align*}
Q\left(\mu,\kappa,\delta_{\text{U}},\delta_{\text{C}},\eta|\Psi^{(k)}\right) &= Q_1\left(\delta_{\text{U}},\delta_{\text{C}}|\Psi^{(k)}\right)+Q_2\left(\mu,\kappa,\eta|\Psi^{(k)}\right),
\end{align*}
for the $(k+1)$-th iteration, which is in the same order as \eqref{eq complete loglikelihood}, with $\Psi^{(k)}=\{\mu^{(k)},\kappa^{(k)},\delta_{\text{U}}^{(k)},\delta_{\text{C}}^{(k)},\eta^{(k)}\}$.
The conditional expectation of the complete-data log-likelihood is obtained by replacing $w_{\text{C}_i}$ and $w_{\text{U}_i}$ with their \emph{a posteriori} probabilities:
\begin{align*}
w_{\text{C}_i}^{(k+1)}
&=\E(W_{\text{C}_i}\mid \theta_i;\Psi^{(k)})
=\frac{\delta_{\text{C}}^{(k)}f_{\text{R}}(\theta_i;\mu^{(k)},\kappa^{(k)}/\eta^{(k)})}
{f_{\text{MGA}}(\theta_i;\mu^{(k)},\kappa^{(k)},\delta_{\text{U}}^{(k)},\delta_{\text{C}}^{(k)},\eta^{(k)})},
\\
w_{\text{U}_i}^{(k+1)}
&=\E(W_{\text{U}_i}\mid \theta_i;\Psi^{(k)})
=\frac{\delta_{\text{U}}^{(k)}/(2\pi)}
{f_{\text{MGA}}(\theta_i;\mu^{(k)},\kappa^{(k)},\delta_{\text{U}}^{(k)},\delta_{\text{C}}^{(k)},\eta^{(k)})}.
\end{align*}
 
\subsubsection*{M-step}
 
The updates for the mixing proportions are obtained by maximizing $Q_1$ subject to the parameter constraints, yielding
\begin{align}
\delta_{\text{C}}^{(k+1)}=\frac{1}{n}\sum_{i=1}^n w_{\text{C}_i}^{(k+1)}
\quad\text{and}\quad
\delta_{\text{U}}^{(k+1)}=\frac{1}{n}\sum_{i=1}^n w_{\text{U}_i}^{(k+1)}.
\end{align}
The updates for $\mu$, $\kappa$, and $\eta$ are obtained by maximizing
\begin{align*}
Q_2(\mu,\kappa,\eta|\Psi^{(k)})
= &\sum_{i=1}^n
\Big[
(1-w_{\text{C}_i}^{(k+1)}-w_{\text{U}_i}^{(k+1)})\log f_{\text{R}}(\theta_i;\mu,\kappa)\\
+ & w_{\text{C}_i}^{(k+1)}\log f_{\text{R}}(\theta_i;\mu,\kappa/\eta)
-w_{\text{U}_i}^{(k+1)}\log(2\pi)
\Big].
\end{align*}
This maximization can be performed in \textsf{R} (R Core Team, \citeyear{r2020r}) using the \texttt{optim()} function from the \textbf{stats} package. The Nelder-Mead or BFGS algorithms can be employed via the \texttt{method} argument. 
Although these procedures involve unconstrained optimization, parameter constraints can be enforced through suitable transformations---for example using $\log(\kappa)$ for $\kappa$ and $\log(\eta-1)$ for $\eta$---under a transformation/back-transformation approach.
 
The EM algorithms for the MA and GA models arise naturally as special cases of the above procedure.
 
\subsection{Initialization strategy}
\label{sec:initialization}
 
Starting values play a critical role in EM-based algorithms and can substantially affect the accuracy and stability of the resulting estimates \citep{biernacki2003choosing}. We propose an initialization strategy that exploits the structure of the models through a hierarchical approach.
 
For the reference model $\mathcal{R}(\mu,\kappa)$, sample circular moments are used to obtain the starting values.  For the implementation of $\mathcal{GA}(\mu,\kappa,\delta_{\text{U}})$, we first fit the corresponding reference model and use its parameter estimates as starting values for the shared parameters $(\mu,\kappa)$. The proportion of observations arising from the uniform component, $\delta_{\text{U}}$, is initialized at a conservative value of $0.01$. This choice is consistent with the contamination literature \citep{punzo2016parsimonious,otto2026modeling}, where contamination parameters are typically initialized so that the model remains close to the reference distribution.
 
Similarly, the $\mathcal{MA}(\mu,\kappa,\delta_{\text{C}},\eta)$ model is initialized using the parameter estimates from the reference distribution for $(\mu,\kappa)$, with additional starting values $\delta_{\text{C}}=0.01$ and $\eta=1.01$ for the contamination parameters. Initializing $\eta$ slightly above 1 ensures compatibility with the admissible parameter space while maintaining proximity to the reference model.
 
Finally, the full $\mathcal{MGA}(\mu,\kappa,\delta_{\text{U}},\delta_{\text{C}},\eta)$ model is initialized adaptively. We perform model selection among the three limiting models using the Akaike information criterion (AIC; \citealp{akaike1974new}) and adopt the parameter estimates from the selected model as starting values for the shared parameters. 
Any remaining parameters are initialized as described above. 
This hierarchical initialization strategy aims to position the EM algorithm close to the global optimum and to reduce the risk of convergence to local maxima.

\section{Simulation study}
\label{Sec: simulation study}
 
To illustrate the practical usefulness of the proposed framework, we conducted a comprehensive simulation study in \textsf{R}, the design and results of which are described in this section. 
For simplicity, we focus on the wrapped normal (WN) distribution as the reference model, with the aim of highlighting the scenarios in which each model variant (reference, mild-only, gross-only, and full MGA) is required to adequately capture different contamination structures. 
The primary objective of the simulation is to examine the anomaly-detection behaviour of the proposed family of models.
 
One hundred samples of size $n = 100$ were generated from a WN distribution with $\mu = 0$ and $\kappa \in \{100,10\}$. 
To systematically explore the full range of possible anomaly configurations, we adopted a grid-based design in which two additional points, say $(\theta_1, \theta_2)$, were appended to each simulated dataset. 
These points were placed at all possible combinations of 32 equally spaced angular positions over the entire circle, with increments of $\pi/16$. 
This design yields $32 \times 32 = 1{,}024$ distinct two-point placement configurations. 
The purpose of spanning the added points over the entire circle is to generate scenarios in which, depending on their location relative to the bulk of the WN-generated data, each added point may be classified as a regular observation, a mild anomaly, or a gross anomaly under the proposed framework.

The MGA model in \eqref{pdf mild and gross anomaly model} was subsequently fitted, and the \emph{a posteriori} probabilities in \eqref{eq a posteriori probability contaminated 3} and \eqref{eq a posteriori probability uniform 3} were used to automatically classify each of the two added points $\theta_1$ and $\theta_2$ into one of three categories: regular (or ``good'') observations, mild anomalies, or gross anomalies. 
Based on the joint classification of $\theta_1$ and $\theta_2$, the anomaly-detection outcome for each configuration was assigned to one of the following four cases:
\begin{enumerate}
    \item {WN}, if both points were classified as regular,
    \item {uWN}, if at least one point was classified as a gross anomaly and neither was classified as mild,
    \item {cWN}, if at least one point was classified as a mild anomaly and neither was classified as gross,
    \item {cuWN}, if one point was classified as mild and the other as gross.
\end{enumerate}
 
The primary output of this simulation is a detection map displayed as a $32 \times 32$ heatmap, where each cell corresponds to one of the 1,024 configurations $(\theta_1, \theta_2)$. 
For each configuration, we determined the modal model across the 100 replications, that is, the model most frequently selected by the automatic detection procedure. 
The color scheme distinguishes the four possible outcomes: green for configurations where no anomalies were detected (WN), orange for gross anomaly detection (uWN), sky blue for mild anomaly detection (cWN), and red for the simultaneous detection of mild and gross anomalies (cuWN).
 
\begin{figure}[!htbp]
\centering
\subfloat[Sample WN data, with added points in red.]{\includegraphics[width=0.48\textwidth]{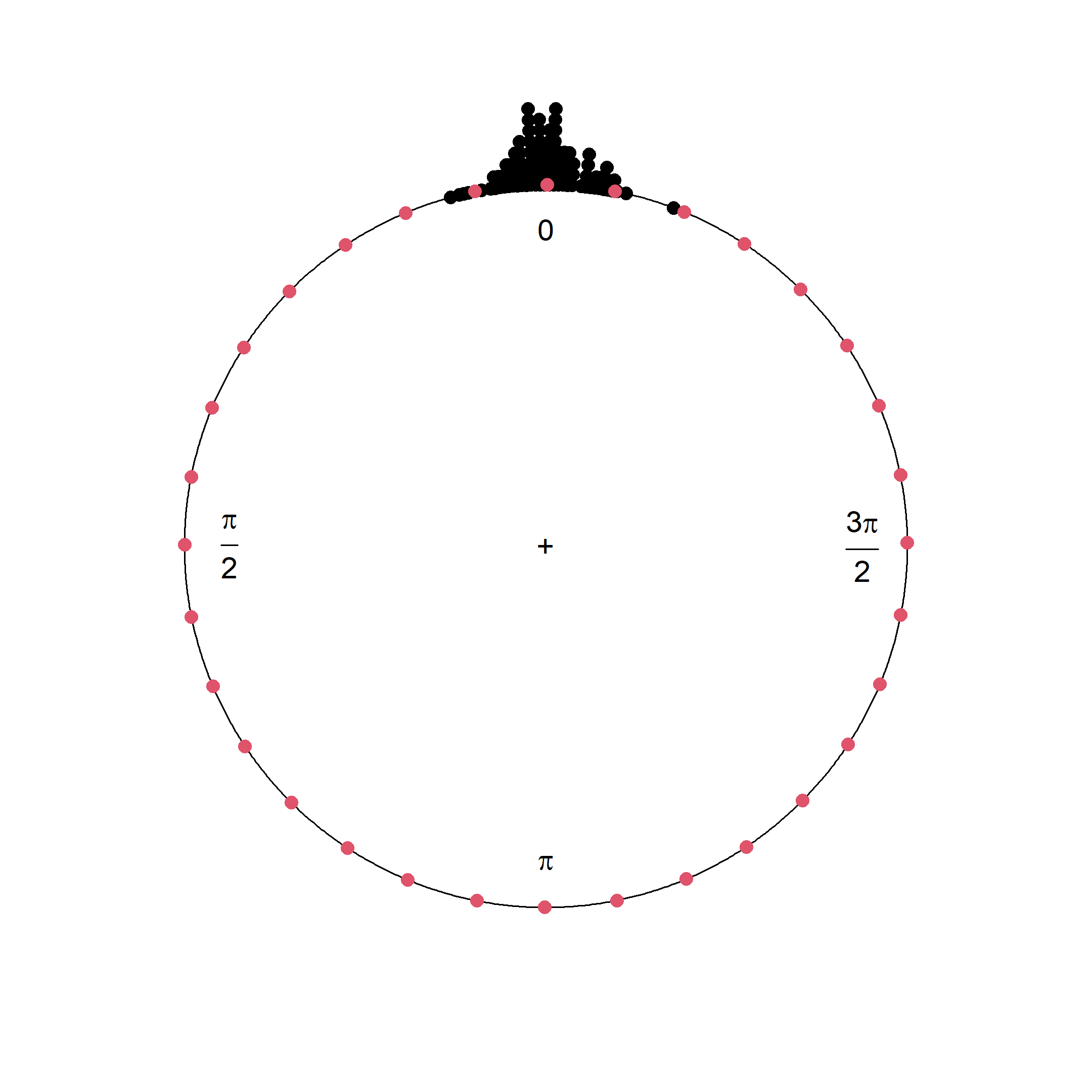}}
\hfill
\subfloat[Anomaly detection map.]{\includegraphics[width=0.48\textwidth]{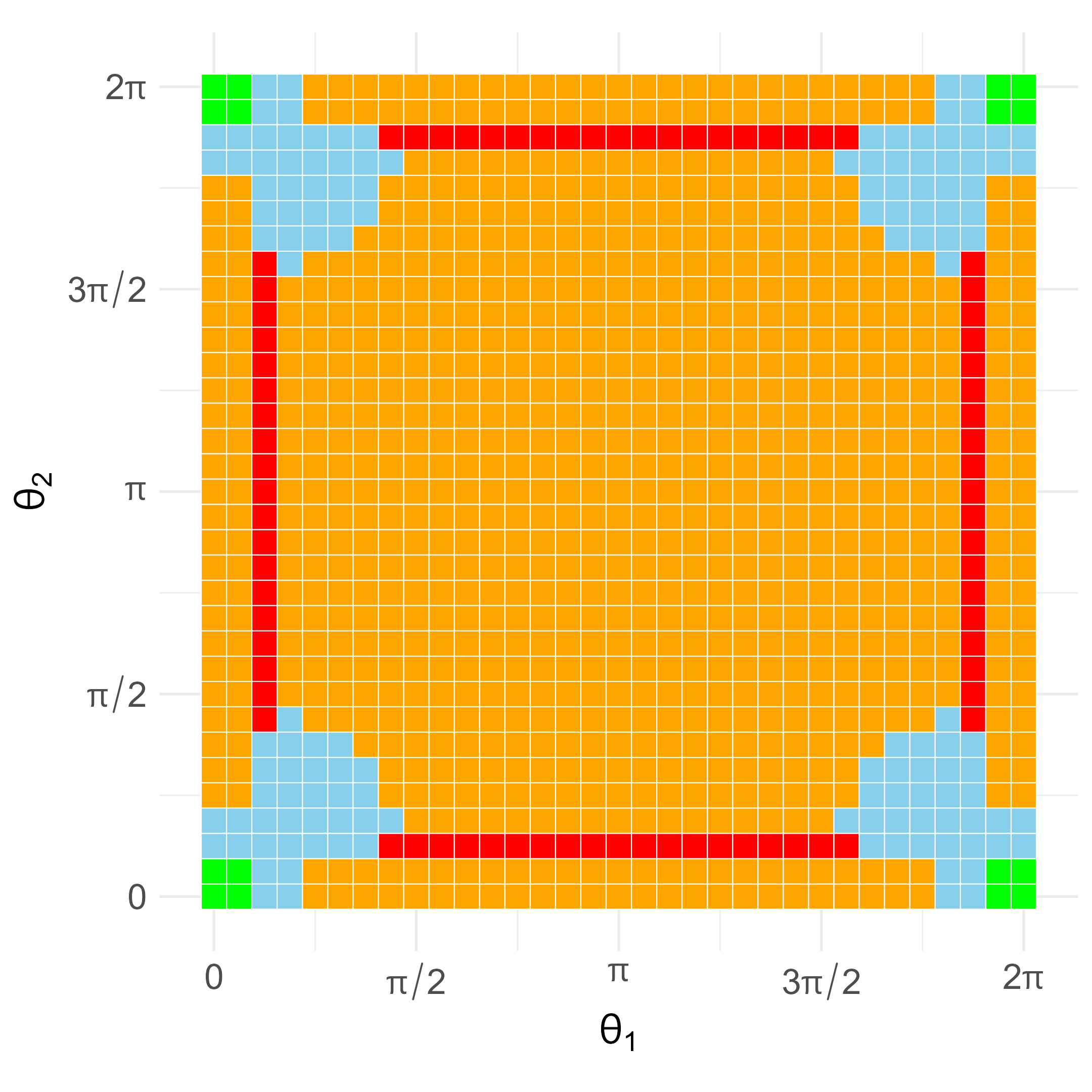}}
\caption{Anomaly detection map for $n = 100$ and $\kappa = 100$. 
(a) Example of simulated WN data, with red points indicating the 32 equally spaced locations at which additional points may be placed on the circle. 
(b) Detection heatmap where each cell corresponds to the modal classification (across 100 replications) for two added points located at positions $(\theta_1, \theta_2)$. 
Green: no anomalies detected (WN); orange: gross anomaly only (uWN); sky blue: mild anomaly only (cWN); red: simultaneous detection of mild and gross anomalies (cuWN).}
\label{fig:detection_n100_sigma01}
\end{figure}

\begin{figure}[!htbp]
\centering
\subfloat[Sample WN data, with added points in red.]{\includegraphics[width=0.48\textwidth]{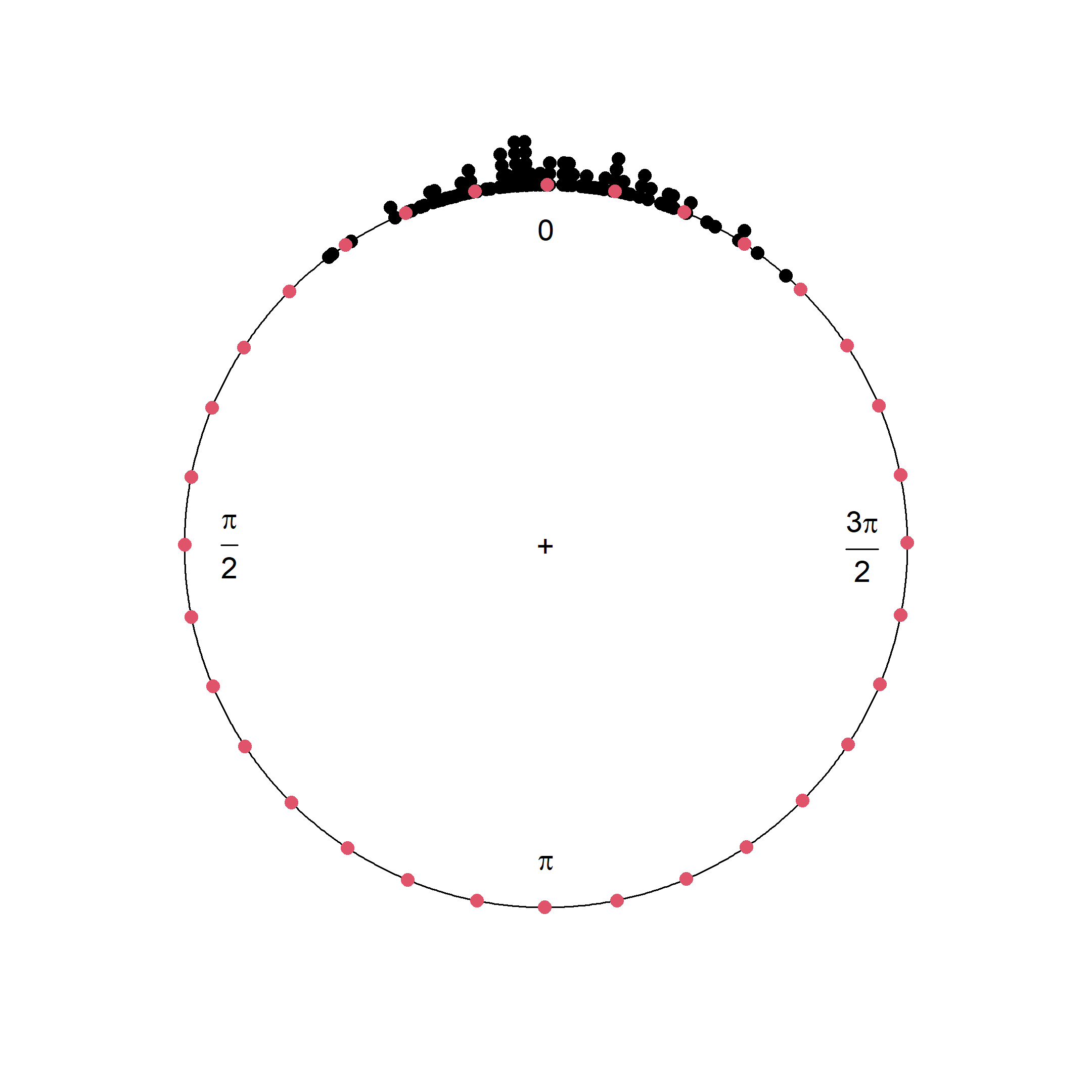}}
\hfill
\subfloat[Anomaly detection map.]{\includegraphics[width=0.48\textwidth]{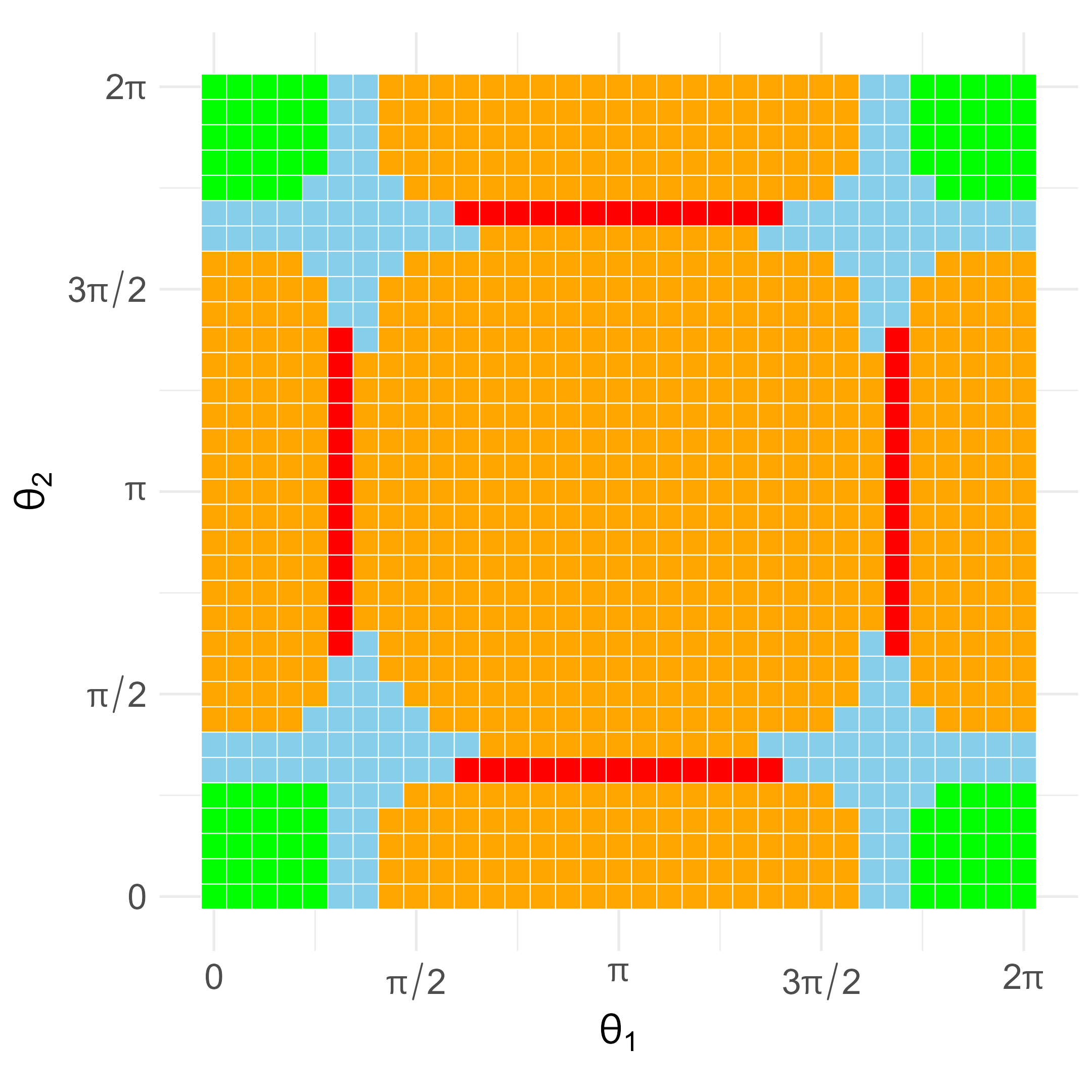}}
\caption{Anomaly detection map for $n = 100$ and $\kappa = 10$. 
(a) Example of simulated WN data, with red points indicating the 32 equally spaced locations at which additional points may be placed on the circle. 
(b) Detection heatmap where each cell corresponds to the modal classification (across 100 replications) for two added points located at positions $(\theta_1, \theta_2)$. 
Green: no anomalies detected (WN); orange: gross anomaly only (uWN); sky blue: mild anomaly only (cWN); red: simultaneous detection of mild and gross anomalies (cuWN).}
\label{fig:detection_n100_sigma03}
\end{figure}

The detection maps in \figurename~\ref{fig:detection_n100_sigma01} and \ref{fig:detection_n100_sigma03}, corresponding to $\kappa = 100$ and $\kappa = 10$, respectively, clearly delineate regions of the $(\theta_1, \theta_2)$ configuration space in which different types of anomalies are identified, confirming that the automatic selection procedure is capable of distinguishing between mild anomalies, gross anomalies, and scenarios exhibiting both. 
The resulting pattern exhibits a clear symmetry.
The green corners correspond to configurations in which the added points lie near the concentration centre (around $\theta_1, \theta_2 \approx 0$) and are therefore predominantly classified as regular observations, since such points are consistent with the generating WN model. 
Moving away from these regions, blue areas (cWN) emerge, indicating configurations where one (the other remaining regular) or both points are classified as mild anomalies. 
These represent moderate deviations that can be accommodated by the contaminated component without invoking the uniform component. 
The dominant orange central region (uWN) corresponds to configurations in which both added points are far from the mean direction, leading to their classification as gross anomalies. 
The surrounding orange bands represent cases where only one of the two points is classified as gross, while the other remains regular. 
Finally, the red boundary regions indicate configurations in which one point is classified as mild and the other as gross, highlighting the framework’s ability to simultaneously detect different types of anomalous behaviour within the same dataset.
 
Comparing \figurename~\ref{fig:detection_n100_sigma01} and \ref{fig:detection_n100_sigma03}, we observe that as $\kappa$ decreases from 100 to 10, the red region corresponding to the simultaneous detection of mild and gross anomalies (cuWN) moves toward the antimode. 
This behaviour is intuitively reasonable: as $\kappa$ decreases, the reference distribution becomes more diffuse and spread across the circle. 
Consequently, added points must be located closer to the antimode before they are considered sufficiently extreme to be classified as gross anomalies. 
In this more dispersed setting, the transition region---where one point is identified as a mild anomaly and the other as a gross anomaly---naturally shifts toward more extreme positions.
 
In addition to the detection maps, we assessed estimation performance by computing the mean squared error (MSE) for both $\mu$ and $\kappa$ across all 1,024 configurations. 
The MSE was computed using circular distance, defined as the minimum between the absolute angular difference and its complement with respect to $2\pi$. 
Separate MSE heatmaps are presented for the WN and cuWN models, illustrating how the presence and location of anomalies influence parameter estimation accuracy under different modelling assumptions.
 
\begin{figure}[!htbp]
\subfloat[$\mu$, WN model]{\includegraphics[width=0.48\textwidth]{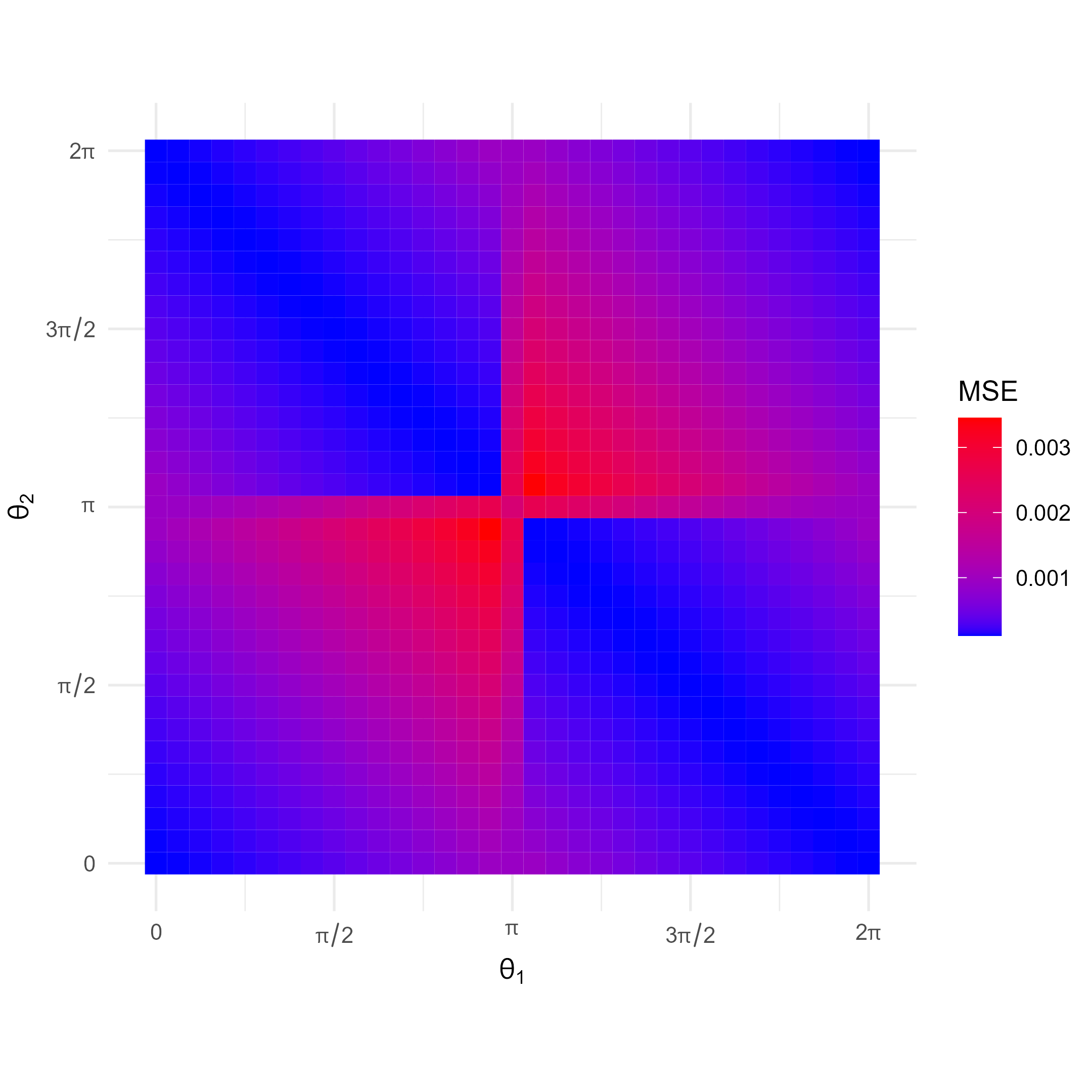}}
\hfill
\subfloat[$\mu$, cuWN model]{\includegraphics[width=0.48\textwidth]{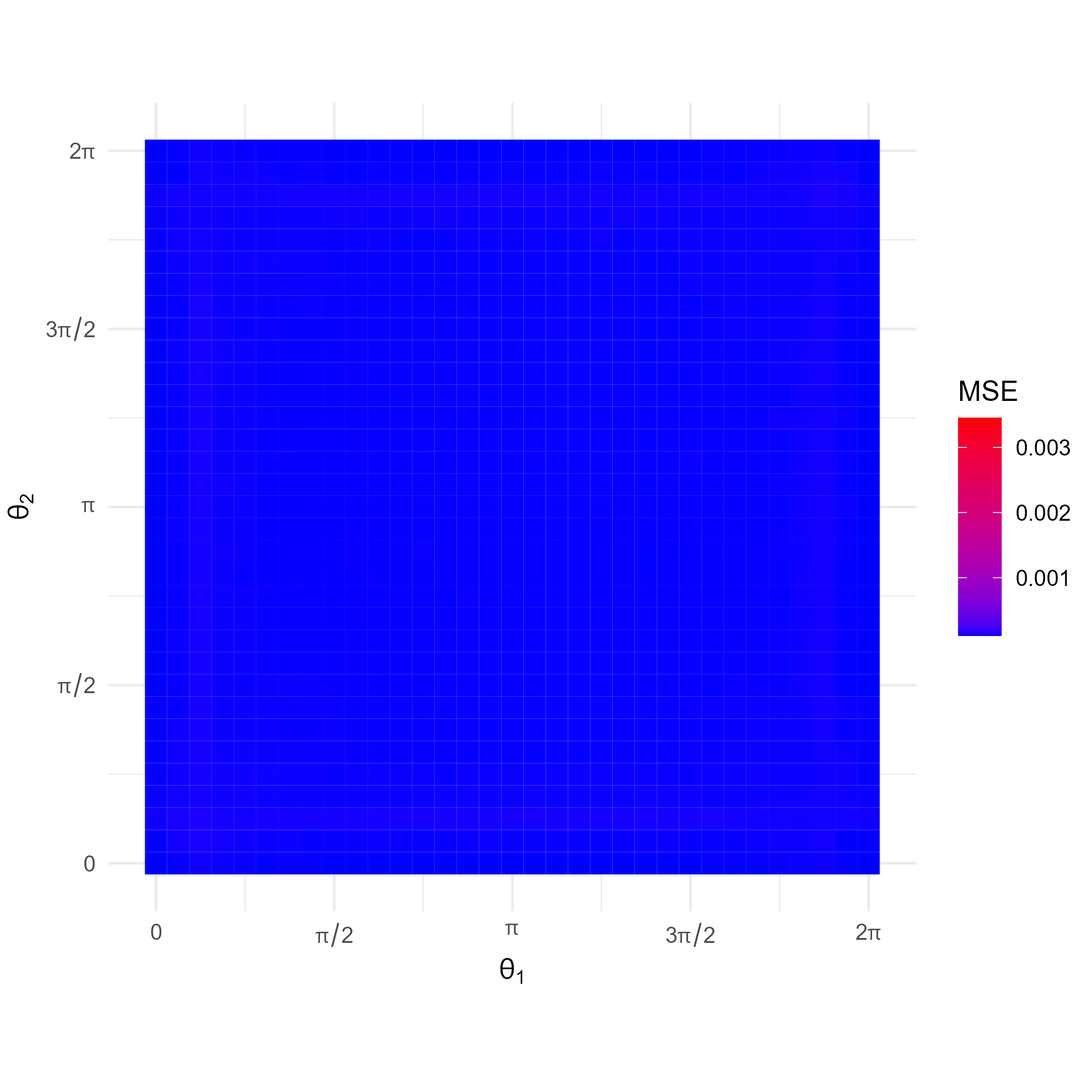}}
\hfill
\subfloat[$\kappa$, WN model]{\includegraphics[width=0.48\textwidth]{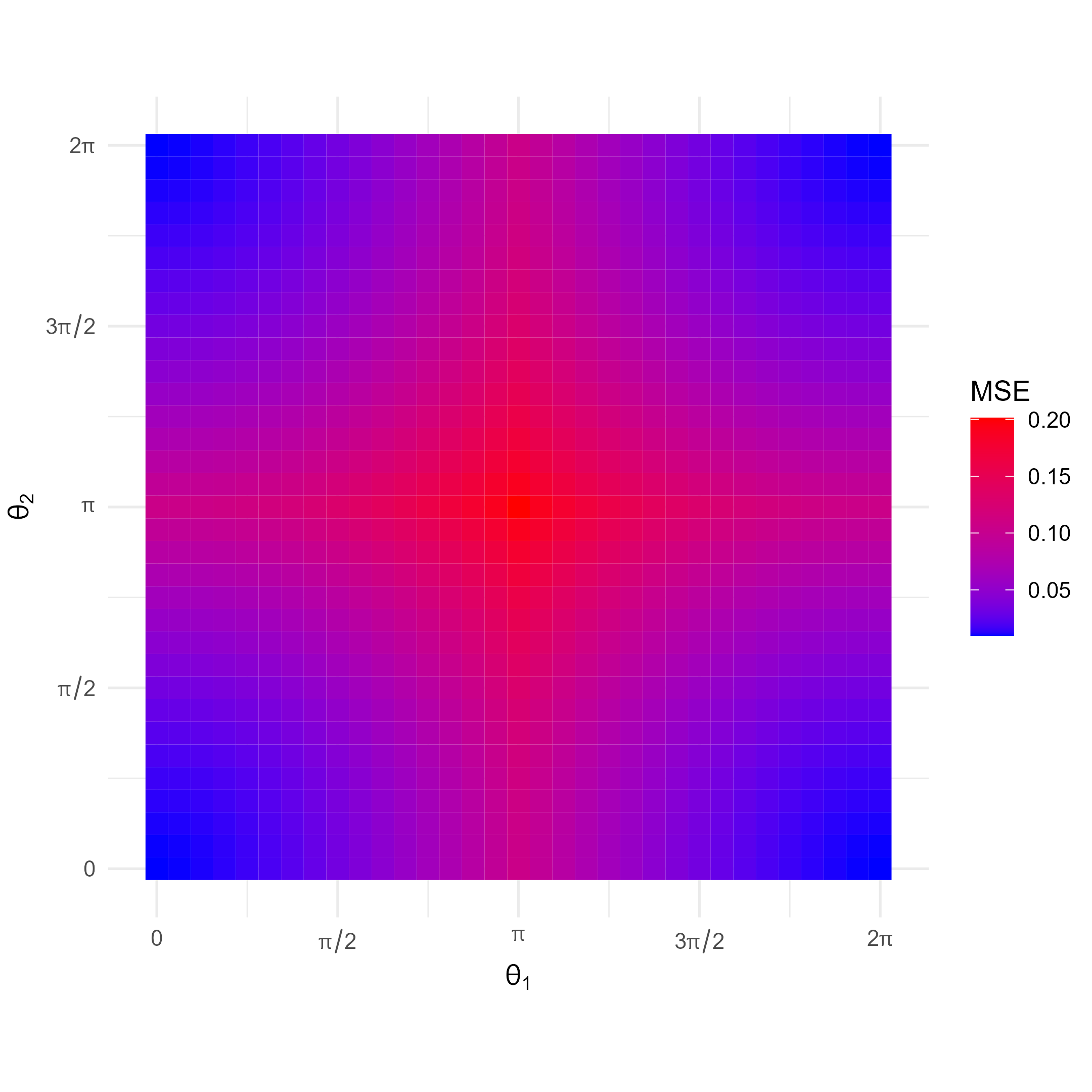}}
\hfill
\subfloat[$\kappa$, cuWN model]{\includegraphics[width=0.48\textwidth]{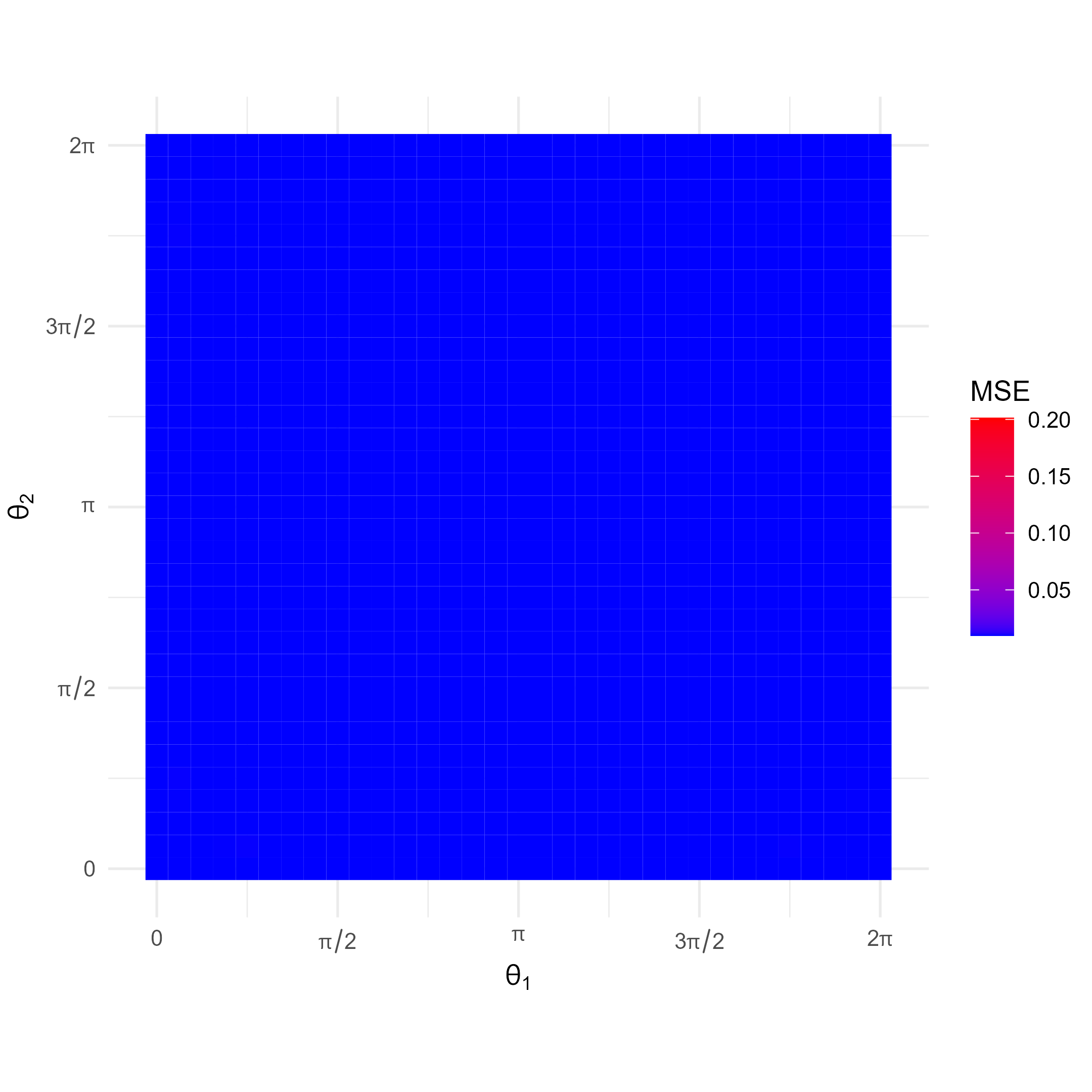}}
\caption{Mean squared error (MSE) for estimating $\mu$ (top row) and $\kappa$ (bottom row) with $n = 100$ and $\kappa = 100$. The left column displays results under the WN model, while the right column corresponds to the cuWN model. }
\label{fig:mse_n100_sigma01}
\end{figure}

 In \figurename~\ref{fig:mse_n100_sigma01}(a), the MSE of $\hat\mu$ under the WN model is shown. 
 High MSE (red regions) occur when the two added points are positioned close to each other on the same side of the circle. 
 In contrast, lower MSE values arise when points are placed on opposite sides of the circle. 
 This makes intuitive sense: when points lie on the same side (e.g., when both are near $\pi/2$, or both near $3\pi/2$ they exert a consistent, unidirectional pull on the estimated mean, creating substantial bias in $\hat\mu$ by shifting the estimate away from the true value of zero in a single direction. 
 Conversely, when the points are placed on opposite sides of the circle (e.g., one near one $\pi/2$ and one near $3\pi/2$), they pull the mean in opposing directions, effectively ``cancelling'' each other out to some degree, reducing the bias and thus the MSE. 
 In \figurename~\ref{fig:mse_n100_sigma01}(b), the cuWN model yields uniformly lower MSE values across all the configurations of $(\theta_1,\theta_2)$, suggesting that it effectively mitigates the effect of these anomalies on $\mu$.
 
A similar pattern emerges for the concentration parameter $\kappa$, as shown in \figurename~\ref{fig:mse_n100_sigma01}(c) and \figurename~\ref{fig:mse_n100_sigma01}(d) for the WN and cuWN models. 
Here, the MSE of $\hat\kappa$ increases as the added points are positioned farther from the mean direction toward the antimode at $\pi$. 
This occurs because points near the antimode represent extreme departures from the concentrated reference distribution, inflating the apparent dispersion when not properly accounted for as anomalies. 
Once again, the cuWN model consistently achieves lower MSE values for $\hat\kappa$ compared with the standard WN model, confirming its robustness in preserving estimation accuracy even in the presence of contamination. 
 
\begin{figure}[!htbp]
\subfloat[$\mu$, WN model]{\includegraphics[width=0.48\textwidth]{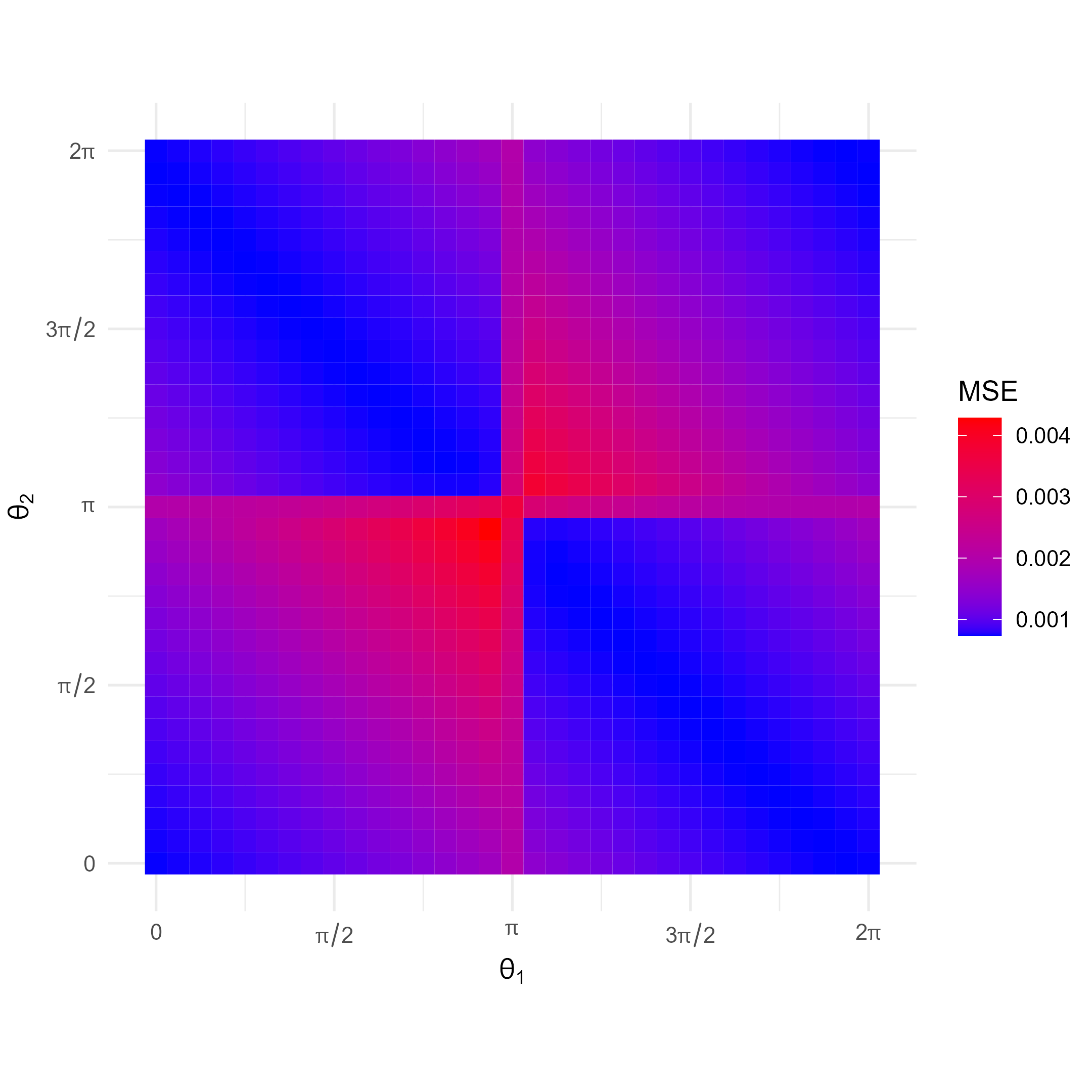}}
\hfill
\subfloat[$\mu$, cuWN model]{\includegraphics[width=0.48\textwidth]{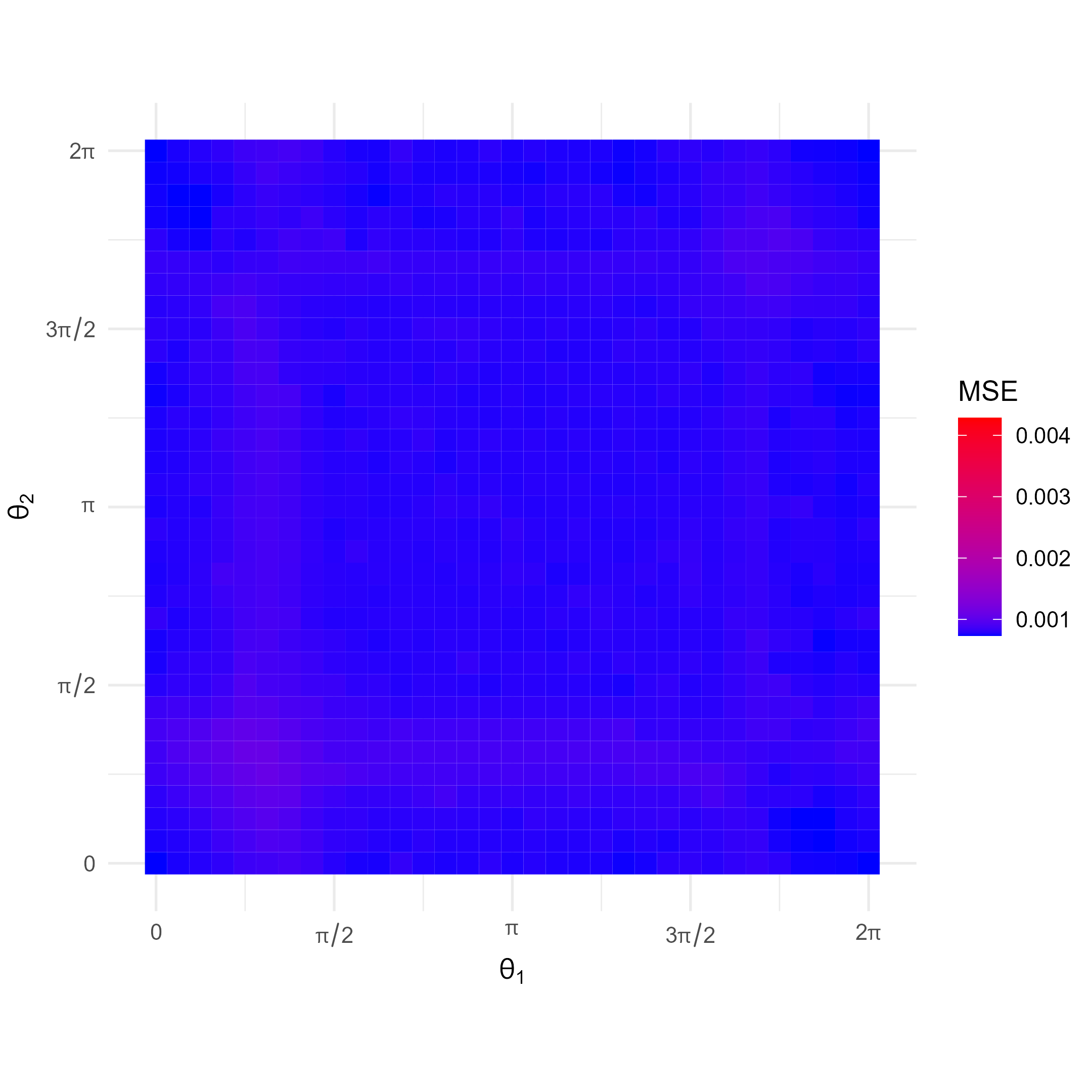}}
\hfill
\subfloat[$\kappa$, WN model]{\includegraphics[width=0.48\textwidth]{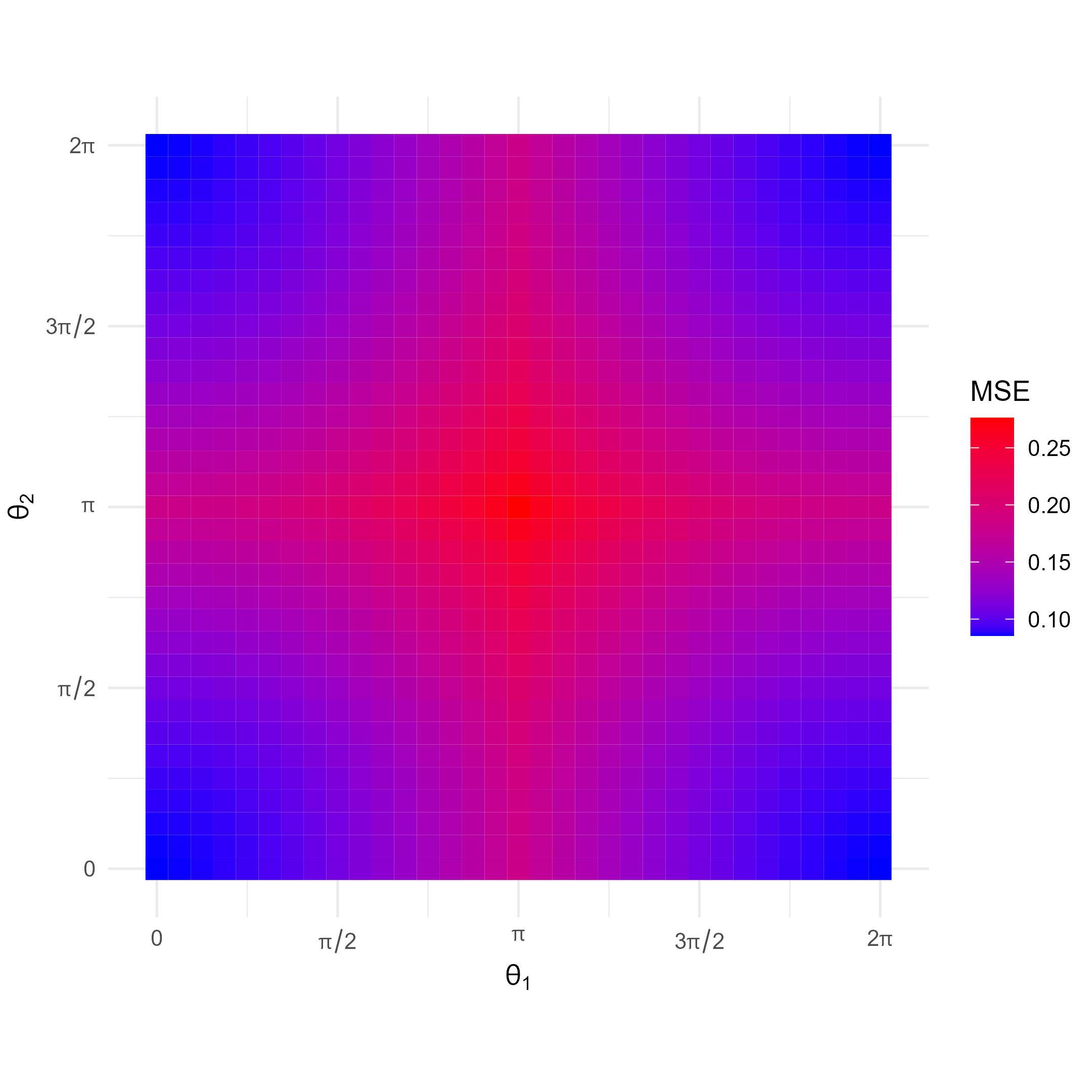}}
\hfill
\subfloat[$\kappa$, cuWN model]{\includegraphics[width=0.48\textwidth]{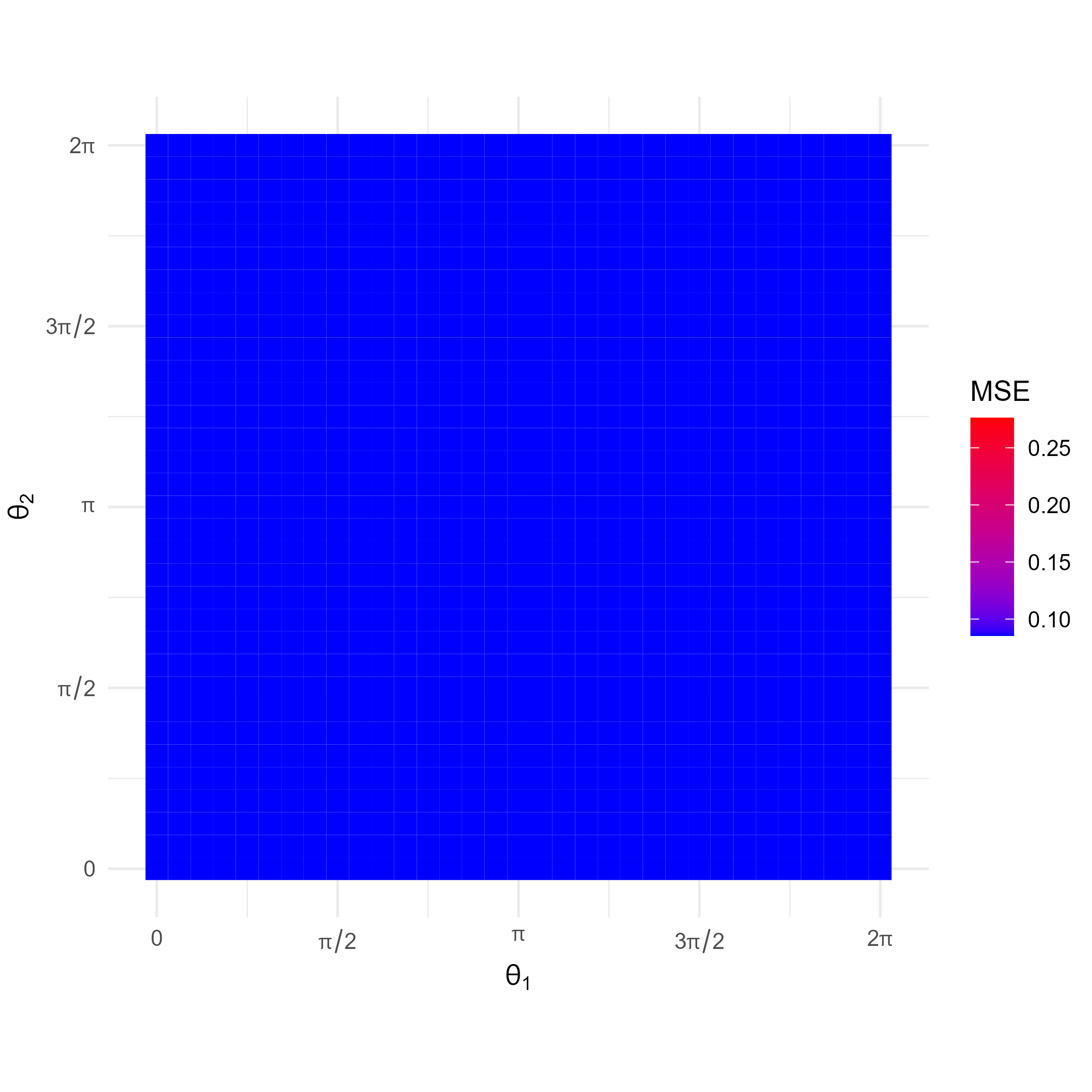}}
\caption{Mean squared error (MSE) for estimating $\mu$ (top row) and $\kappa$ (bottom row) with $n = 100$ and $\kappa = 10$. The left column displays results under the WN model, while the right column corresponds to the cuWN model. }
\label{fig:mse_n100_sigma03}
\end{figure}

A comparable structure is observed in \figurename~\ref{fig:mse_n100_sigma03}, corresponding to the less concentrated scenario with $\kappa = 10$.
For the WN model (left column), the MSE pattern for $\hat{\mu}$ remains qualitatively similar to that observed when $\kappa = 100$, with larger errors arising when the two added points lie on the same side of the circle and reinforce each other’s directional pull. 
However, the overall magnitude of the MSE is slightly reduced, reflecting the fact that, under a more diffuse reference distribution, additional points exert a comparatively smaller relative influence on the estimated mean.
For the concentration parameter, the MSE of $\hat{\kappa}$ under the WN model again increases as the added points approach the antimode. 
Nevertheless, because the reference distribution is already less concentrated, the distortion induced by extreme points is less abrupt than in the highly concentrated case.
In contrast, the cuWN model (right column) continues to display markedly lower MSE values for both $\hat{\mu}$ and $\hat{\kappa}$ across almost all configurations. 
The advantage of the anomaly-aware specification remains evident, although the gain relative to the WN model is slightly attenuated compared with the $\kappa = 100$ scenario. 
This behaviour is coherent with the detection maps: when the reference distribution is more dispersed, fewer configurations produce extreme distortions, and the need for strong corrective adjustments becomes less pronounced.
 
Overall, these results justify the necessity of the proposed framework, demonstrating that a model capable of handling both mild and gross anomalies simultaneously is essential. 
Results for a larger sample size scenario, with $n=500$, are provided in \ref{App: estimation}, where these patterns persist and lead to a reduced overall MSE in $\mu$ and $\kappa$.
 
\section{Data application}
\label{sec: data application}
 
In this section, we apply the proposed framework to three real-world datasets: the Marion Island dataset (Section \ref{sec: Marion island}), the larva dataset (Section \ref{sec: larva dataset}), and the Col de la Roa wind direction dataset (Section \ref{sec: wind data}). 
The model performance is ranked via the AIC, and the Bayesian information criterion (BIC; \citealp{schwarz1978estimating}).
 
\subsection{Marion Island dataset} \label{sec: Marion island}

\blue{The Marion Island wind dataset is drawn from a directional analysis of vegetation patterning on the island's volcanic scoria cones, for which surface-level wind direction was estimated using a Computational Fluid Dynamics (CFD) model \citep{goddard2022investigation}. Measurements correspond to 133 locations spanning the four aspects (NE, SE, SW, NW) of 35 scoria cones; aspect strongly differentiates the wind regime experienced at each location, since the cones obstruct and deflect the prevailing flow. On the southern aspects (SE, SW), the deflected flow generates substantial directional heterogeneity, with wind direction varying considerably across locations; such multimodal variability is better addressed through models accommodating multiple dominant directions than through an anomaly-based framework. On the northern aspects (NE, NW), by contrast, wind direction is closely aligned with a single dominant direction at most locations, with a small number of locations deviating noticeably from this norm. We therefore focus on the $n=44$ measurements recorded on the northern aspects, where this pattern of a dominant direction with occasional marked departures provides a natural setting for the proposed framework.}

\blue{Tables~\ref{tab:model_comparison_Marion} and \ref{tab:model_par marion} summarize the results of fitting the WN and VM distributions to the Marion Island dataset. 
Both the AIC and BIC unanimously select the uWN as the best-fitting model. 
The uVM ranks a close second under both criteria, confirming that the WN-based specification is slightly preferred but that the two distributional families yield broadly consistent conclusions. Notably, the cWN and cuWN models rank third and sixth by AIC, and fourth and seventh by BIC, respectively, indicating that the parsimony penalty outweighs any marginal gain in log-likelihood from adding a mild-anomaly component. The reference WN and VM models rank last or near-last, confirming that a small but non-negligible proportion of gross anomalies is present.}

\begin{figure}[!htbp]
\subfloat[uWN distribution]{\includegraphics[width=0.38\textwidth]{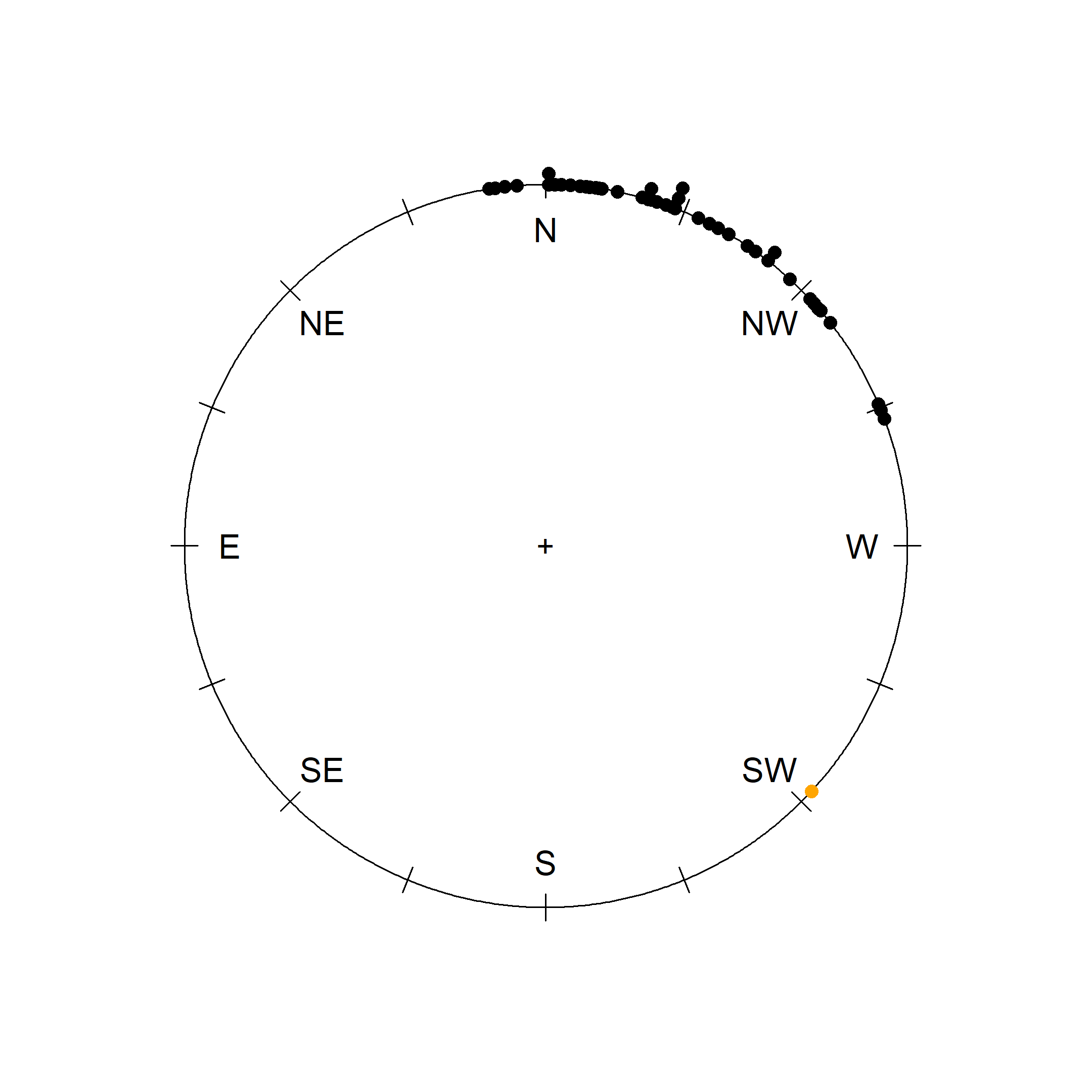}}
\hfill
\subfloat[uWN distribution]{\includegraphics[width=0.38\textwidth]{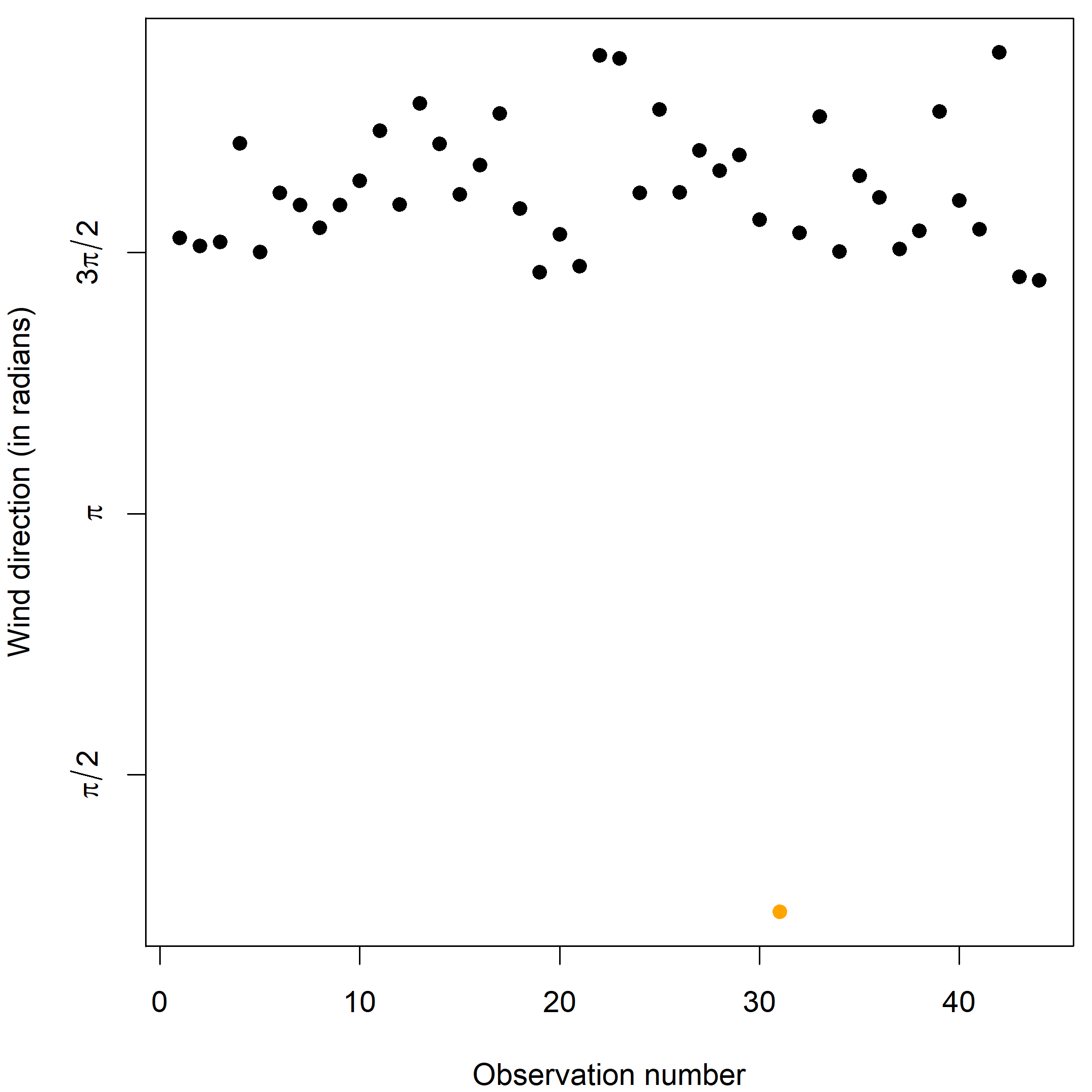}}
\caption{Anomaly detection results for the Marion Island dataset under the uWN model. Observations assigned to the reference component are shown in black, and gross anomalies (uniform component) in orange.}
\label{fig: marion island detection}
\end{figure}

\blue{Figure~\ref{fig: marion island detection} presents the anomaly detection results under the uWN model. The circular plot shows that the vast majority of observations cluster tightly around the north-northwest direction, with the single gross anomaly (indicated in orange) following a southwest wind direction. The estimated contamination proportion $\hat{\delta}_{\text{U}} = 0.033
$ implies that roughly 3.3\% of observations are attributed to the uniform component, which is broadly consistent with the 2\% of observations classified as gross anomalies via the \emph{a posteriori} probabilities in \eqref{eq a posteriori probability gross outlier}. }
 
\begin{table}[!htbp]
\centering
\caption{Model comparison for the Marion Island dataset based on log-likelihood, AIC, and BIC values, with corresponding rankings and number of parameters (\#par).} 
\begin{tabular}{lrrrrrr}
  \toprule
Model & \#par & loglike & AIC & rank & BIC & rank \\ 
  \midrule
  WN & 2 & -28.117 & 60.235 & 8 & 63.803 & 6 \\ 
  uWN & 3 & -23.508 & 53.017 & 1 & 58.369 & 1 \\ 
  cWN & 4 & -23.482 & 54.965 & 3 & 62.102 & 4 \\ 
  cuWN & 5 & -23.482 & 56.965 & 6 & 65.886 & 7 \\ 
  VM & 2 & -26.444 & 56.889 & 5 & 60.457 & 3 \\ 
  uVM & 3 & -23.615 & 53.231 & 2 & 58.584 & 2 \\ 
  cVM & 4 & -23.610 & 55.219 & 4 & 62.356 & 5 \\ 
  cuVM & 5 & -23.610 & 57.219 & 7 & 66.140 & 8 \\ 
   \bottomrule
\end{tabular}
\label{tab:model_comparison_Marion}
\end{table}

\begin{table}[!htbp]
\caption{Maximum likelihood estimates of model parameters for the Marion Island dataset.} 
\centering
\begin{tabular}{lrrrrr}
  \toprule
Model & $\hat\mu$ & $\hat\kappa$ & $\hat\delta_{\text{U}}$ & $\hat\delta_{\text{C}}$ & $\hat\eta$ \\ 
  \midrule
WN & 5.146 & 4.758 &  &  &  \\ 
  uWN & 5.100 & 7.762 & 0.033 &  &  \\ 
  cWN & 5.100 & 7.864 &  & 0.044 & 18.919 \\ 
  cuWN & 5.100 & 7.864 & 0.001 & 0.044 & 18.918 \\ 
  VM & 5.122 & 5.686 &  &  &  \\ 
  uVM & 5.095 & 8.254 & 0.033 &  &  \\ 
  cVM & 5.096 & 8.288 &  & 0.037 & 36.357 \\ 
  cuVM & 5.095 & 8.288 & 0.001 & 0.037 & 37.086 \\ 
   \bottomrule
\end{tabular}
\label{tab:model_par marion}
\end{table}
 
\subsection{Larva dataset}
\label{sec: larva dataset}
 
The larva dataset, originally analyzed by \citet{jones2012inverse}, records the directional changes of a fruit fly larva. The larva was observed for three minutes as it wriggled across a flat surface, and its direction of movement was noted every second. This dataset provides an example of short-term directional behavior in biological studies, where small deviations in movement direction may reflect exploratory behavior, environmental perturbations, or random fluctuations in motion.
 
The GA, MA, and MGA models were fitted using both the WN and VM as reference models. \tablename~\ref{tab:model_comparison} presents the model comparison using AIC and BIC criteria. According to the AIC, the cuVM model provides the best fit, while it ranks second according to the BIC, which slightly favors the more parsimonious cVM model. 
Importantly, both information criteria clearly indicate that models accounting for anomalies outperform the reference models (WN and VM). This suggests that the larval movement cannot be adequately described by a single unimodal symmetric distribution and that incorporating contamination components leads to a more realistic representation of the observed directional behavior. Within the WN family, the cuWN model provides the best fit, confirming that allowing for both mild and gross anomalies improves explanatory power.
 
\begin{table}[!htbp]
\centering
\caption{Model comparison for the larva dataset based on log-likelihood, AIC, and BIC values, with corresponding rankings and number of parameters (\#par).} 
\begin{tabular}{lrrrrrr}
  \toprule
Model & \#par & loglike & AIC & rank & BIC & rank \\ 
  \midrule
WN & 2 & -176.664 & 357.328 & 8 & 363.714 & 8 \\ 
  uWN & 3 & -125.712 & 257.424 & 6 & 267.003 & 6 \\ 
  cWN & 4 & -120.487 & 248.973 & 4 & 261.745 & 4 \\ 
  cuWN & 5 & -117.799 & 245.599 & 2 & 261.563 & 3 \\ 
  VM & 2 & -155.289 & 314.578 & 7 & 320.964 & 7 \\ 
  uVM & 3 & -125.448 & 256.897 & 5 & 266.476 & 5 \\ 
  cVM & 4 & -119.114 & 246.229 & 3 & 259.001 & 1 \\ 
  cuVM & 5 & -117.500 & 245.000 & 1 & 260.965 & 2 \\ 
   \bottomrule
\end{tabular}
\label{tab:model_comparison}
\end{table}
 
\begin{figure}[!htbp]
\subfloat[uWN distribution]{\includegraphics[width=0.38\textwidth]{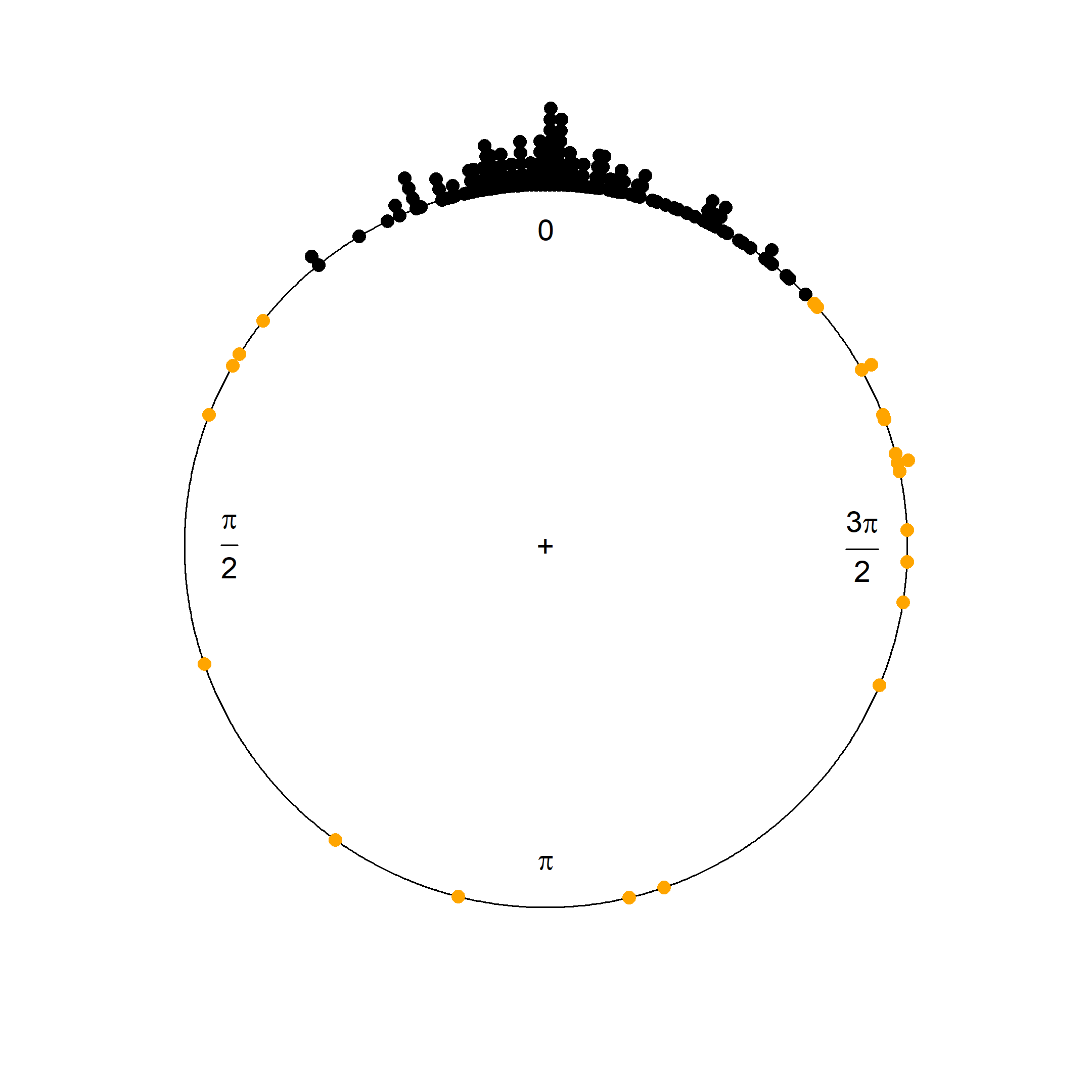}}
\hfill
\subfloat[uWN distribution]{\includegraphics[width=0.38\textwidth]{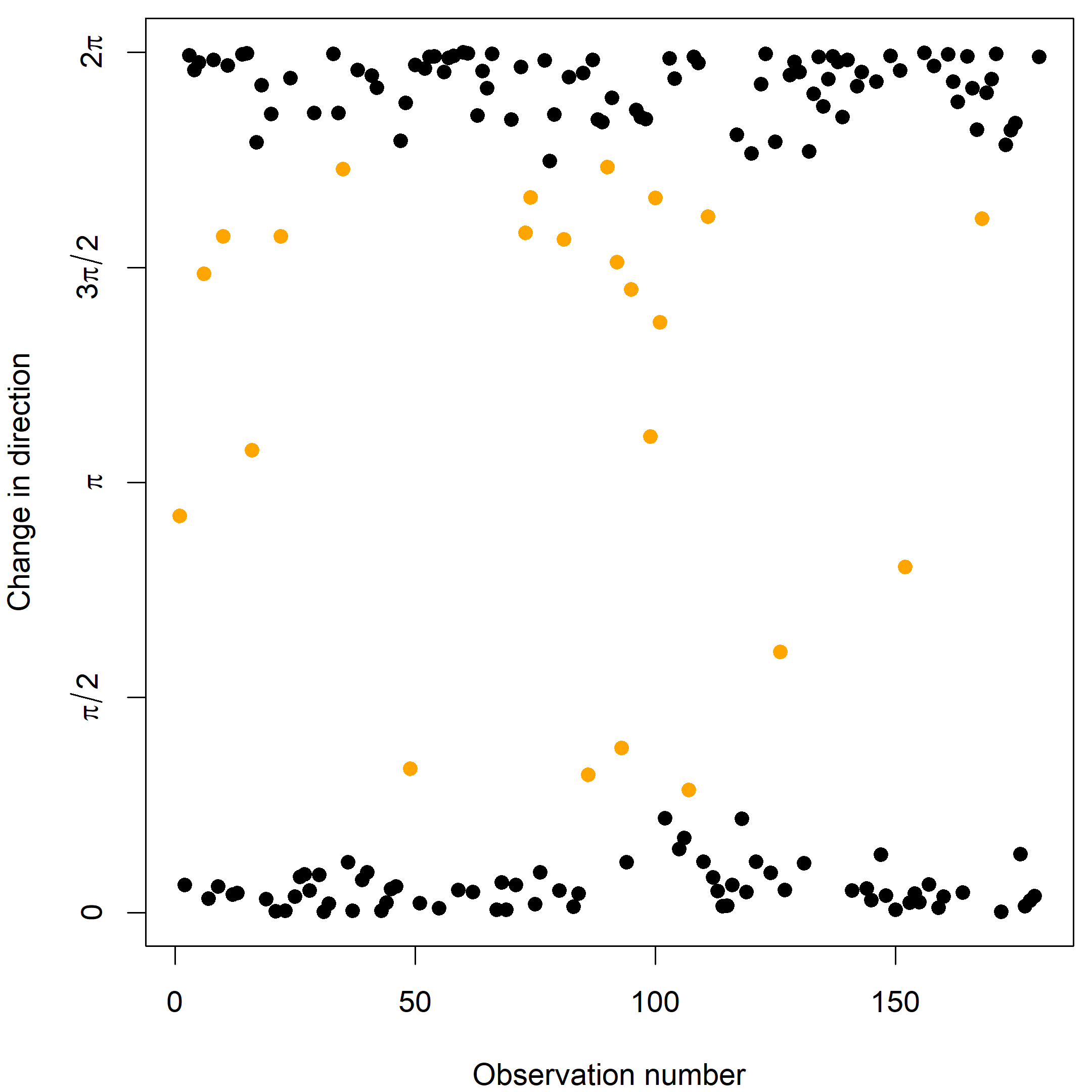}}
 
\subfloat[cWN distribution]{\includegraphics[width=0.38\textwidth]{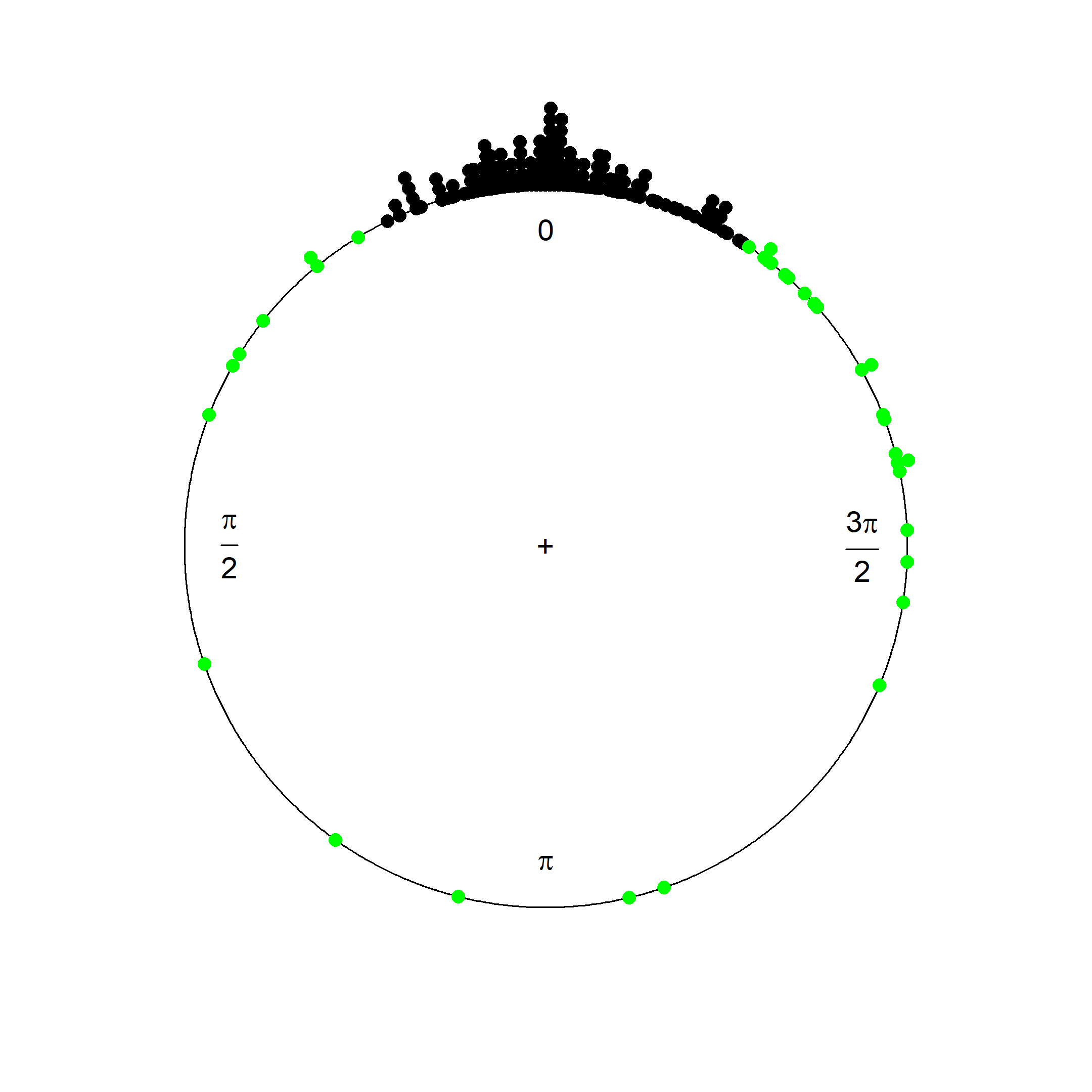}}
\hfill
\subfloat[cWN distribution]{\includegraphics[width=0.38\textwidth]{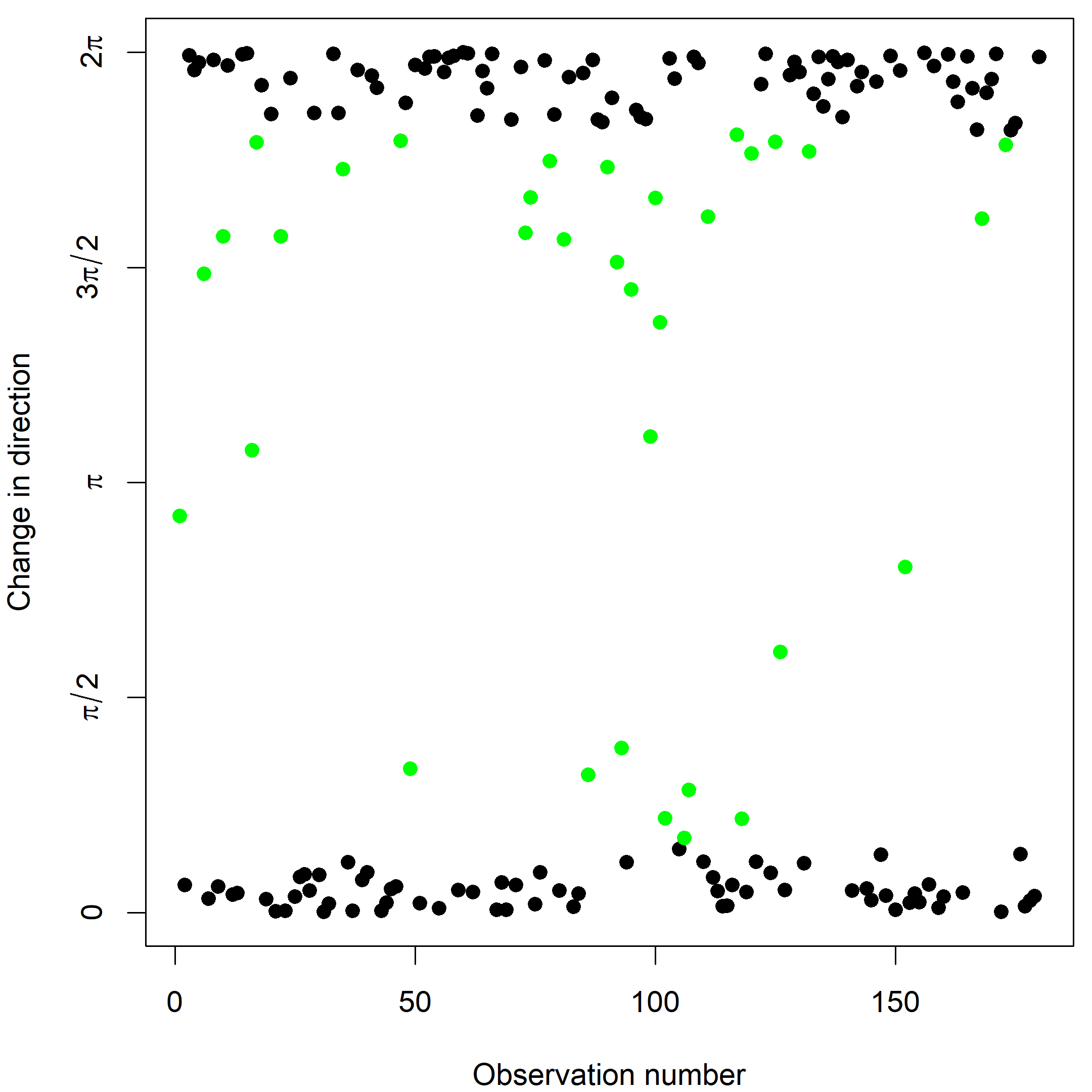}}
 
\subfloat[cuWN distribution]{\includegraphics[width=0.38\textwidth]{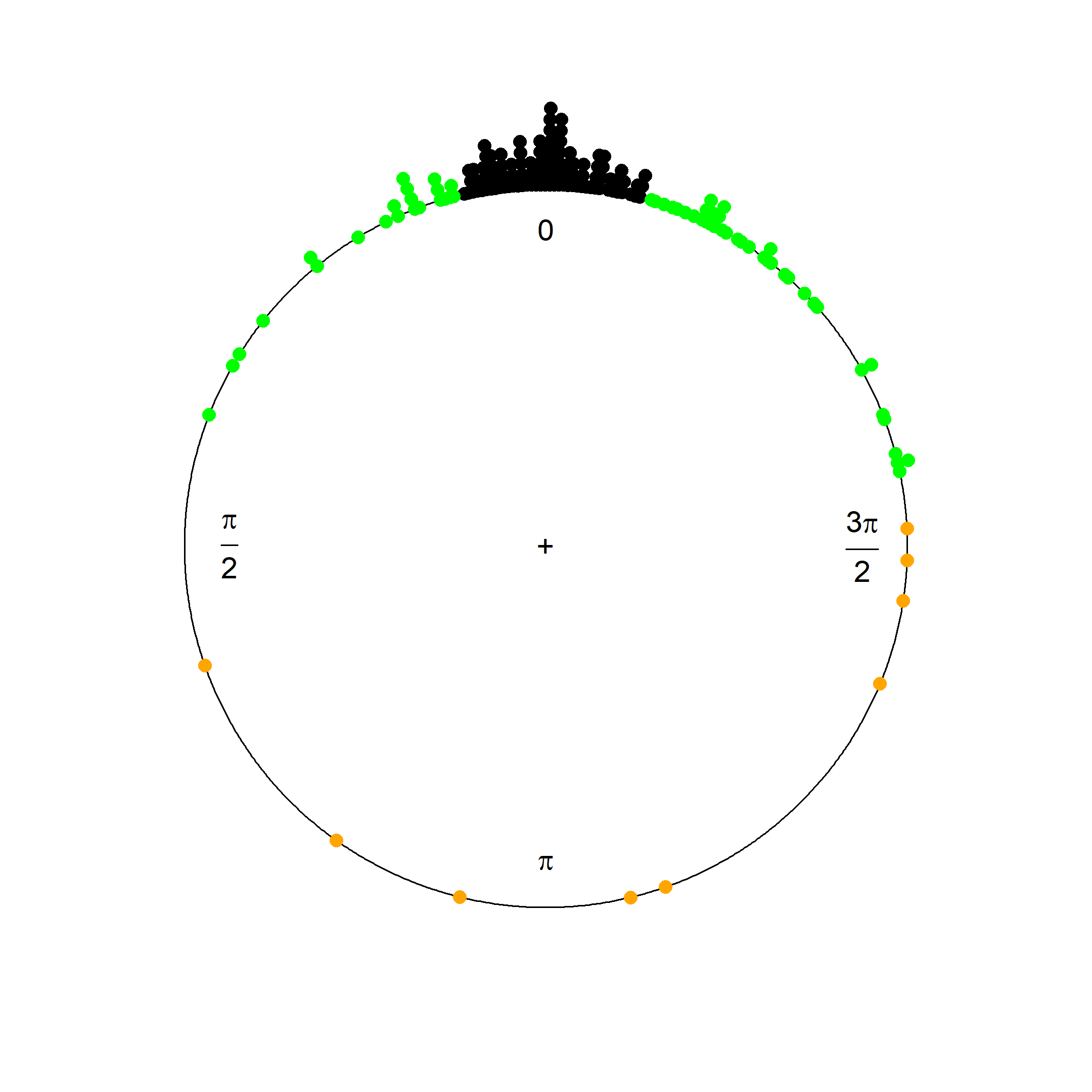}}
\hfill
\subfloat[cuWN distribution]{\includegraphics[width=0.38\textwidth]{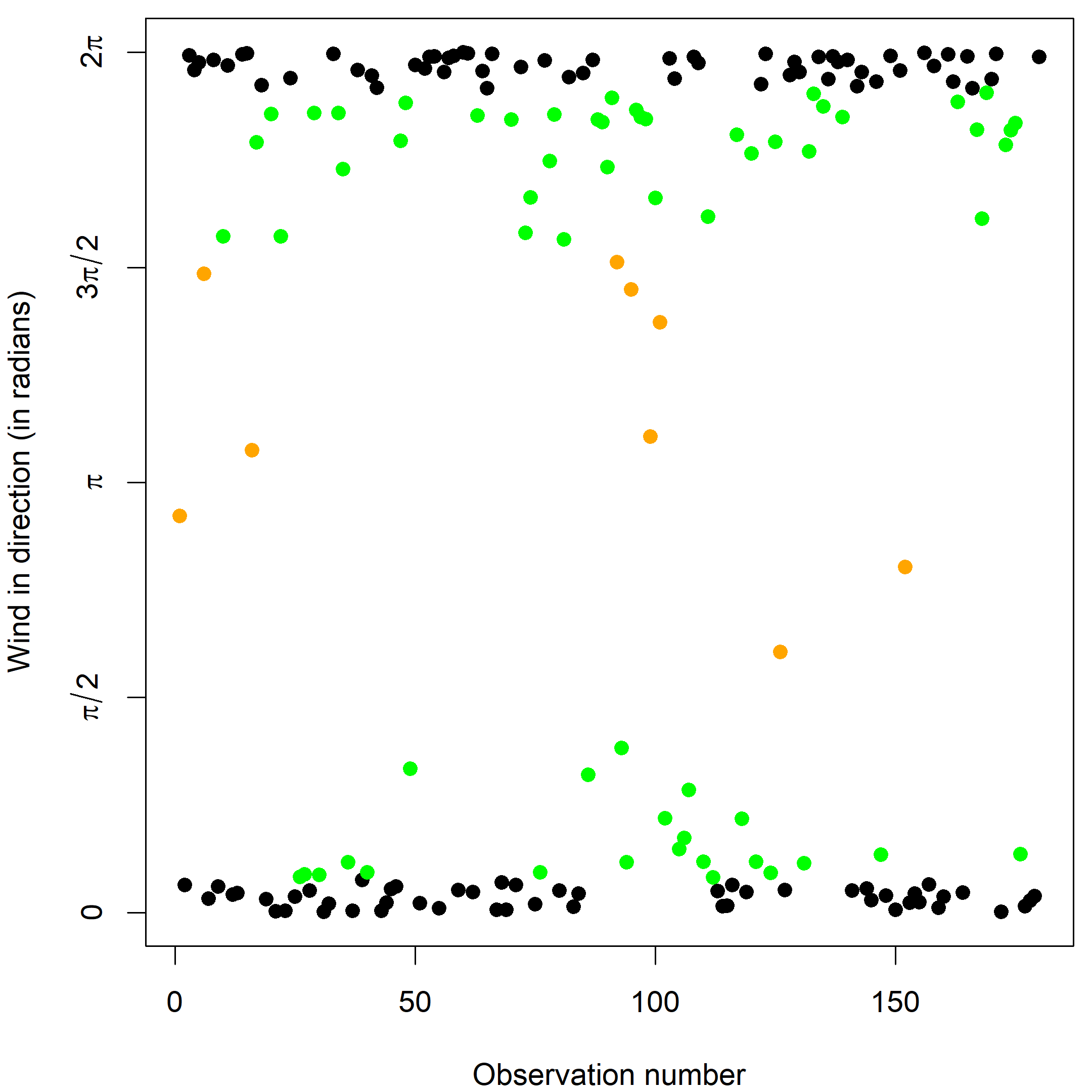}}
\caption{Anomaly detection results for the larva dataset under the uWN, cWN, and cuWN models. Observations assigned to the reference component are shown in black, gross anomalies (uniform component) in orange, and mild anomalies (contaminant component) in green.}
\label{fig: larva detection}
\end{figure}

The parameter estimates are reported in \tablename~\ref{tab:model_parm}. Beyond their statistical meaning, these parameters admit direct biological interpretation.
For the cuWN model, the estimated mean direction is $\hat\mu = 6.280$, which corresponds to a direction very close to $0$ (or equivalently $2\pi$). 
This value identifies the mean direction of the larval movement and coincides with the dominant cluster of observations visible in \figurename~\ref{fig: larva detection}. 
The strength of this directional preference is quantified by the concentration parameter of the reference component ($\hat\kappa = 49.639$). 
This large concentration implies low dispersion within the reference component, reflecting pronounced directional persistence in the larva’s trajectory. 
Thus, movements assigned to the reference component are confined to a more concentrated area around the mean direction, while additional dispersion is captured by the mild and gross contamination components.
The contamination parameters provide additional insight into the variation in larval behavior. The estimate $\hat\delta_{\text{U}} = 0.084$ indicates that approximately 8\% of the observations behave as gross anomalies, that is, movements that do not align with the mean direction and are consistent with a diffuse background mechanism. Biologically, these may correspond to abrupt turns or exploratory movements unrelated to the main trajectory.
Furthermore, $\hat\delta_{\text{C}} = 0.426$ indicates that about 43\% of the observations are classified as mild anomalies. These observations retain the same mean direction but exhibit reduced concentration, as quantified by the attenuation parameter $\hat\eta = 17.161$. Since the contaminant component uses $\kappa/\eta$, this corresponds to a moderate reduction in concentration relative to the main component. 
In practical terms, a substantial portion of the larva’s movements are still directionally aligned but display increased variability, reflecting short-term fluctuations rather than completely random behavior.
 
\begin{table}[!htbp]
\caption{Maximum likelihood estimates of model parameters for the larva dataset.} 
\centering
\begin{tabular}{lrrrrr}
  \toprule
Model & $\hat\mu$ & $\hat\kappa$ & $\hat\delta_{\text{U}}$ & $\hat\delta_{\text{C}}$ & $\hat\eta$ \\ 
  \midrule
WN & 6.165 & 2.387 &  &  &  \\ 
  uWN & 6.233 & 13.217 & 0.166 &  &  \\ 
  cWN & 6.248 & 18.346 &  & 0.263 & 28.110 \\ 
  cuWN & 6.280 & 49.639 & 0.084 & 0.426 & 17.161 \\ 
  VM & 6.181 & 3.668 &  &  &  \\ 
  uVM & 6.234 & 13.527 & 0.164 &  &  \\ 
  cVM & 6.264 & 25.952 &  & 0.360 & 16.749 \\ 
  cuVM & 6.280 & 53.380 & 0.080 & 0.453 & 15.196 \\ 
   \bottomrule
\end{tabular}
\label{tab:model_parm}
\end{table}
 
The VM-based models convey a similar message. 
In particular, the cuVM model yields a large estimate of $\kappa$ (i.e., $\hat\kappa=53.380$), reflecting strong directional persistence in the dominant component, while simultaneously allocating a non-negligible fraction of observations to contamination components. The agreement between WN- and VM-based extensions reinforces the robustness of the findings.
 
For the cuWN model, the estimated mixing proportions of mild ($\hat\delta_{\text{C}} = 0.426$) and gross ($\hat\delta_{\text{U}} = 0.084$) anomalies are broadly consistent with the anomaly classifications obtained using \eqref{eq a posteriori probability contaminated 3} and \eqref{eq a posteriori probability uniform 3}, which identified approximately 5\% of the observations as gross anomalies and 34\% as mild anomalies. 
The slight discrepancy between the estimated mixing proportions and the empirical classification percentages is expected, since the former represent component weights under the fitted mixture model, whereas the latter arise from maximum \emph{a posteriori} allocation of individual observations and therefore involve a deterministic classification rule.
 
\figurename~\ref{fig: larva detection} illustrates these results for the uWN, cWN, and cuWN models. 
The circular plots show how detected anomalies are distributed along the circle, while the dotcharts (ordered by observation index) reveal their temporal structure. 
Notably, mild anomalies (green) tend to cluster around the main direction but exhibit greater dispersion, whereas gross anomalies (orange) appear as isolated departures from the dominant movement pattern. 
This visualization confirms that the cuWN model not only improves model fit according to information criteria but also provides an interpretable classification of larval movement behaviour.
 
\subsection{Col de la Roa wind dataset}
\label{sec: wind data}
 
The Col de la Roa wind dataset is available in the \textbf{circular} package \citep{lund2017package} in \textsf{R} and contains daily wind direction measurements recorded at the Col de la Roa meteorological station in the Italian Alps \citep{agostinelli2007robust}. 
For this dataset, wind direction was recorded daily from January 29 to March 31, 2001, for the period from 3:00 am to 4:00 am included, resulting in $n = 310$ measurements.
 
This dataset provides an environmental application where wind directions are influenced by local topography, nocturnal thermal circulation, and transient atmospheric disturbances. 
In mountainous regions, wind regimes often exhibit a dominant direction due to valley-channeling effects, while occasional synoptic perturbations or turbulence may generate substantial deviations from the prevailing flow. 
As such, this dataset represents an ideal setting for distinguishing between mild directional variability and more diffuse anomalous behavior.
 
Tables~\ref{tab:model_comparison wind} and \ref{tab:model_parm wind} summarize the results of fitting the WN and VM distributions, as well as their extended variants based on our framework. 
According to the AIC, the cuVM model provides the best fit, while the BIC slightly favors the more parsimonious cVM model. 
In contrast, the reference WN and VM models, which do not account for anomalies, rank last under both criteria. 
This clear improvement indicates that the wind direction data cannot be adequately captured by a single unimodal symmetric distribution. 
Instead, the presence of both dispersed yet structured deviations and more diffuse irregular behavior is strongly supported by the data.
 
\begin{table}[!htbp]
\centering
\caption{Model comparison for the Col de la Roa wind dataset based on log-likelihood, AIC, and BIC values, together with the number of parameters (\#par) and relative rankings across WN- and VM-based models.} 
\begin{tabular}{lrrrrrr}
  \toprule
model & \#par & loglik & AIC & rank & BIC & rank \\ 
  \midrule
WN & 2 & -435.733 & 875.465 & 8 & 882.938 & 8 \\ 
  uWN & 3 & -386.096 & 778.192 & 6 & 789.401 & 6 \\ 
  cWN & 4 & -380.846 & 769.693 & 4 & 784.639 & 3 \\ 
  cuWN & 5 & -378.032 & 766.063 & 2 & 784.746 & 4 \\ 
  VM & 2 & -417.071 & 838.141 & 7 & 845.615 & 7 \\ 
  uVM & 3 & -385.570 & 777.140 & 5 & 788.350 & 5 \\ 
  cVM & 4 & -379.926 & 767.851 & 3 & 782.798 & 1 \\ 
  cuVM & 5 & -377.834 & 765.668 & 1 & 784.351 & 2 \\ 
   \bottomrule
\end{tabular}
\label{tab:model_comparison wind}
\end{table}
 
\begin{table}[!htbp]
\centering
\caption{Maximum likelihood estimates of mean direction, concentration, and contamination parameters for the fitted models applied to the Col de la Roa wind dataset.} 
\begin{tabular}{lrrrrr}
  \toprule
model & $\hat\mu$ & $\hat\kappa$ & $\hat\delta_{\text{U}}$ & $\hat\delta_{\text{C}}$ & $\hat\eta$ \\ 
  \midrule
WN & 0.427 & 0.990 &  &  &  \\ 
  uWN & 0.139 & 8.059 & 0.351 &  &  \\ 
  cWN & 0.139 & 14.319 &  & 0.490 & 34.046 \\ 
  cuWN & 0.093 & 54.167 & 0.302 & 0.447 & 15.899 \\ 
  VM & 0.292 & 1.760 &  &  &  \\ 
  uVM & 0.138 & 8.315 & 0.347 &  &  \\ 
  cVM & 0.130 & 18.329 &  & 0.543 & 22.714 \\ 
  cuVM & 0.096 & 55.386 & 0.291 & 0.464 & 14.967 \\ 
   \bottomrule
\end{tabular}
\label{tab:model_parm wind}
\end{table}
 
The parameter estimates offer insightful meteorological interpretation. 
Under the cuWN model, the estimated mean direction $\hat\mu = 0.093$ indicates a dominant prevailing wind direction. 
The high concentration parameter ($\hat\kappa = 54.167$) indicates that wind directions assigned to the reference component (i.e., regular directions) are tightly clustered around the prevailing direction, reflecting low dispersion and a stable nocturnal wind regime.
The contamination parameters reveal additional structure. 
The estimate $\hat\delta_{\text{U}} = 0.302$ indicates that approximately 30\% of the observations behave as gross anomalies, corresponding to directions that do not align with the prevailing wind regime and are consistent with a diffuse background mechanism. 
These may reflect transient atmospheric disturbances, local turbulence, or measurement variability.
Furthermore, $\hat\delta_{\text{C}} = 0.447$ suggests that nearly 45\% of the observations correspond to mild anomalies. 
These retain the same mean direction but exhibit reduced concentration, as quantified by the attenuation parameter $\hat\eta = 15.899$. 
Since the contaminant component uses $\kappa/\eta$, this implies a moderate but systematic increase in dispersion relative to the main component. 
In meteorological terms, this likely represents days where the wind remains aligned with the valley direction but with greater dispersion due to weaker channeling or partial synoptic influence.
 
The VM-based results convey a consistent message. For the cuVM model, the concentration parameter is large too ($\hat\kappa = 55.386$), reflecting strong directional persistence in the dominant component, while $\hat\delta_{\text{U}} = 0.291$ and $\hat\delta_{\text{C}} = 0.464$ confirm a substantial contribution of both gross and mild anomalies. 
The larger $\hat\eta = 14.967$ indicates a pronounced reduction in concentration for the contaminant component relative to the main wind regime.
These estimates are broadly consistent with the anomaly classifications obtained via \emph{a posteriori} probabilities, which identified approximately 44\% of the observations as mild anomalies and 22\% as gross anomalies. As discussed previously, minor discrepancies between estimated mixing proportions and empirical classification percentages are expected due to the probabilistic nature of the mixture model and the deterministic maximum \emph{a posteriori} allocation rule.
 
\begin{figure}[!htbp]
\subfloat[uWN distribution]{\includegraphics[width=0.38\textwidth]{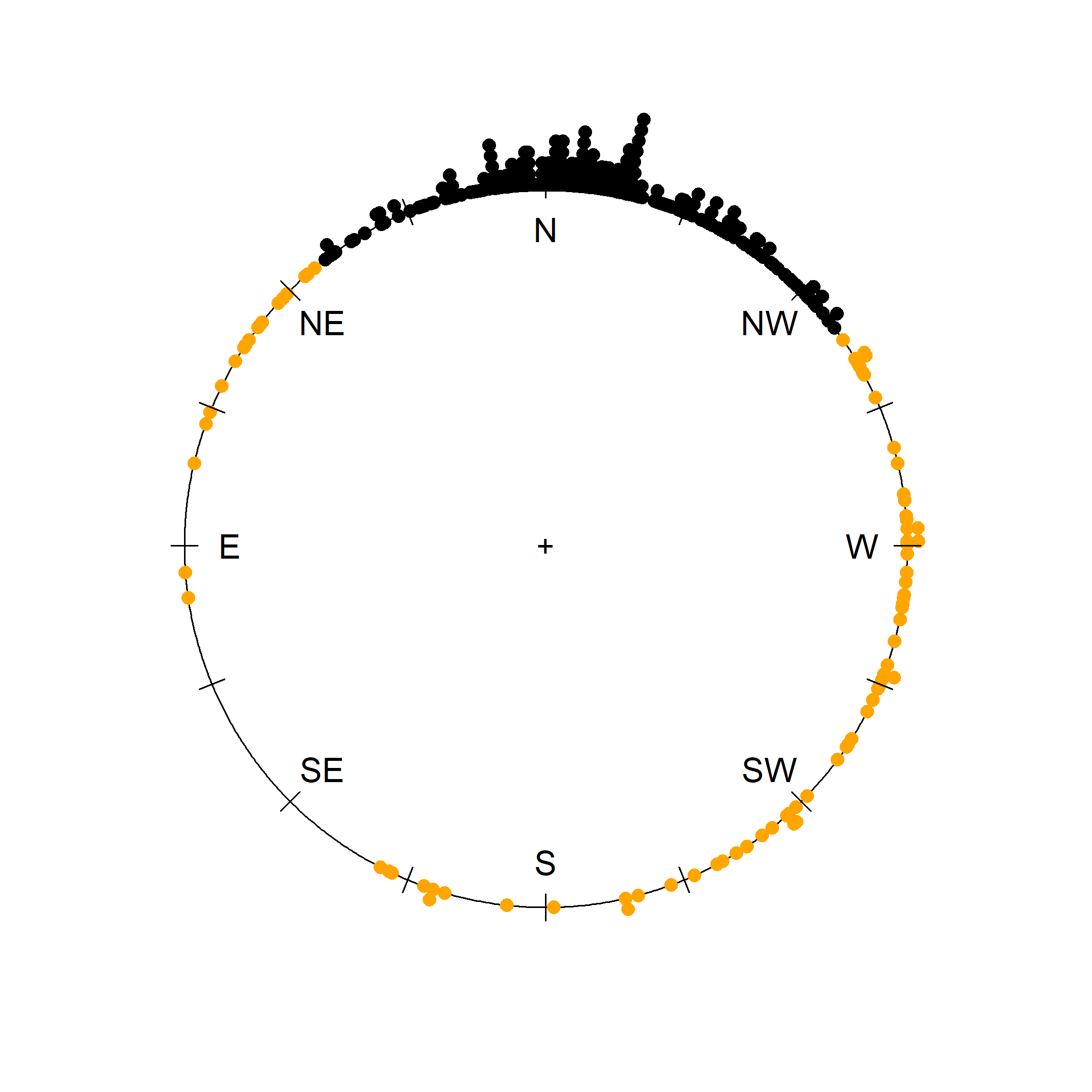}}
\hfill
\subfloat[uWN distribution]{\includegraphics[width=0.38\textwidth]{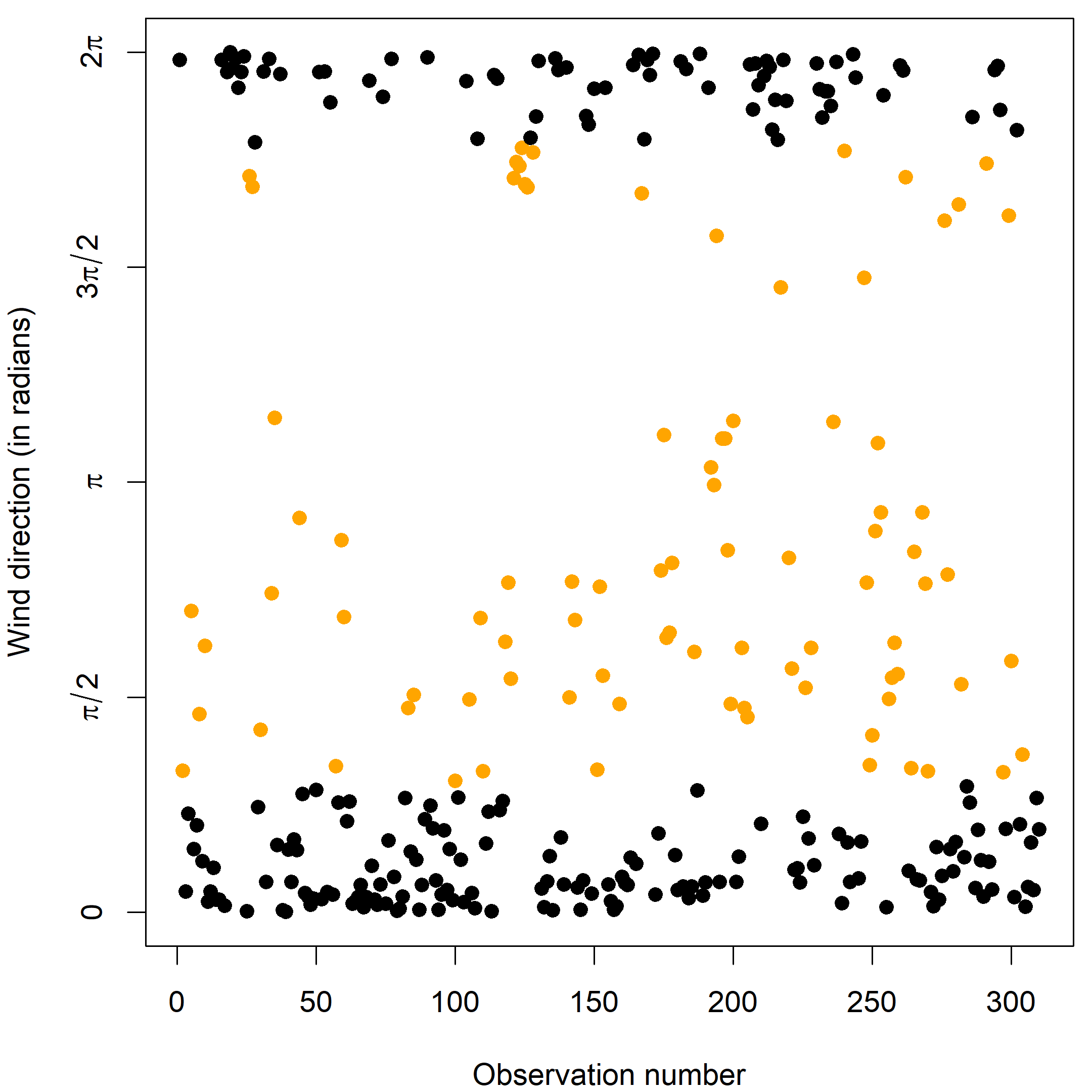}}
 
\subfloat[cWN distribution]{\includegraphics[width=0.38\textwidth]{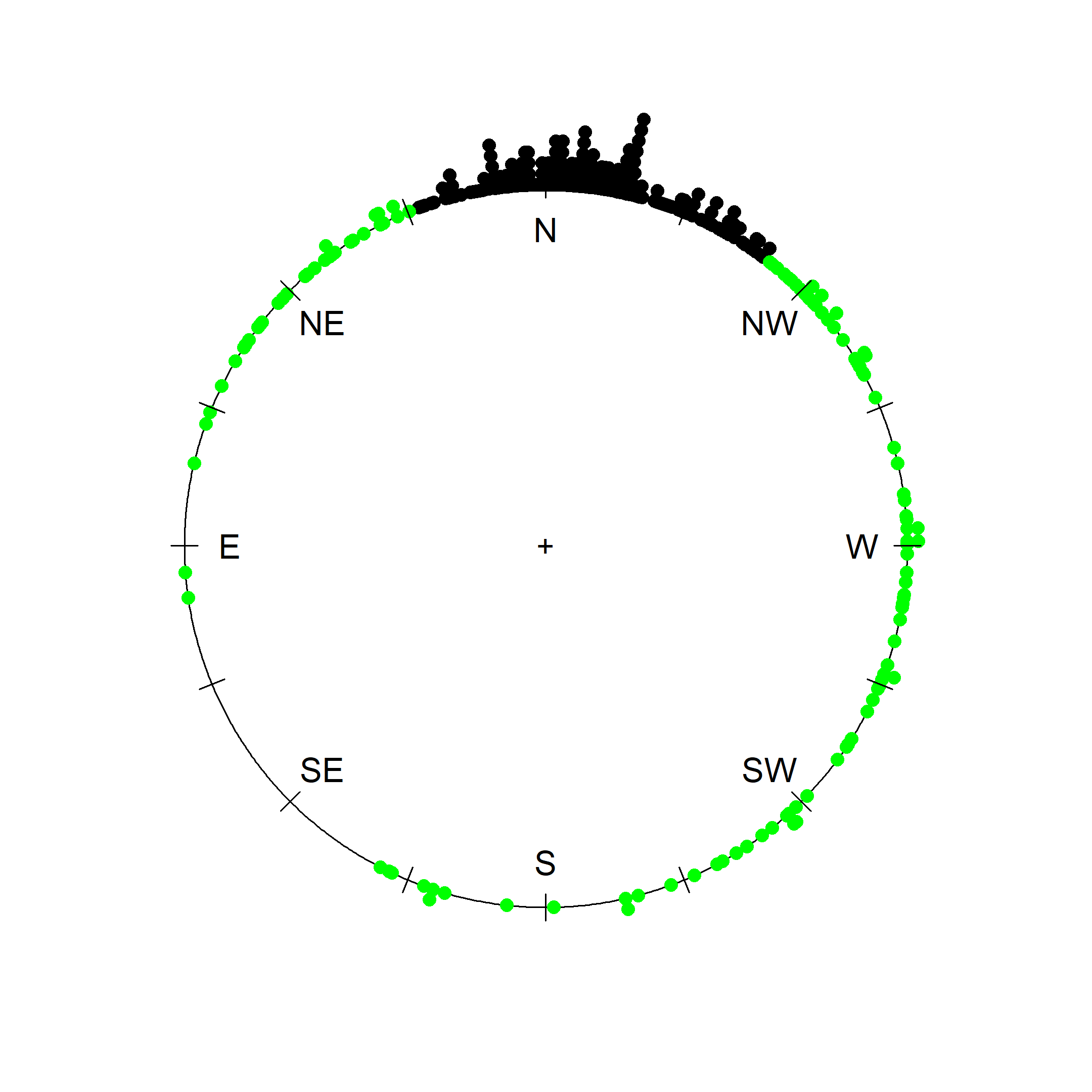}}
\hfill
\subfloat[cWN distribution]{\includegraphics[width=0.38\textwidth]{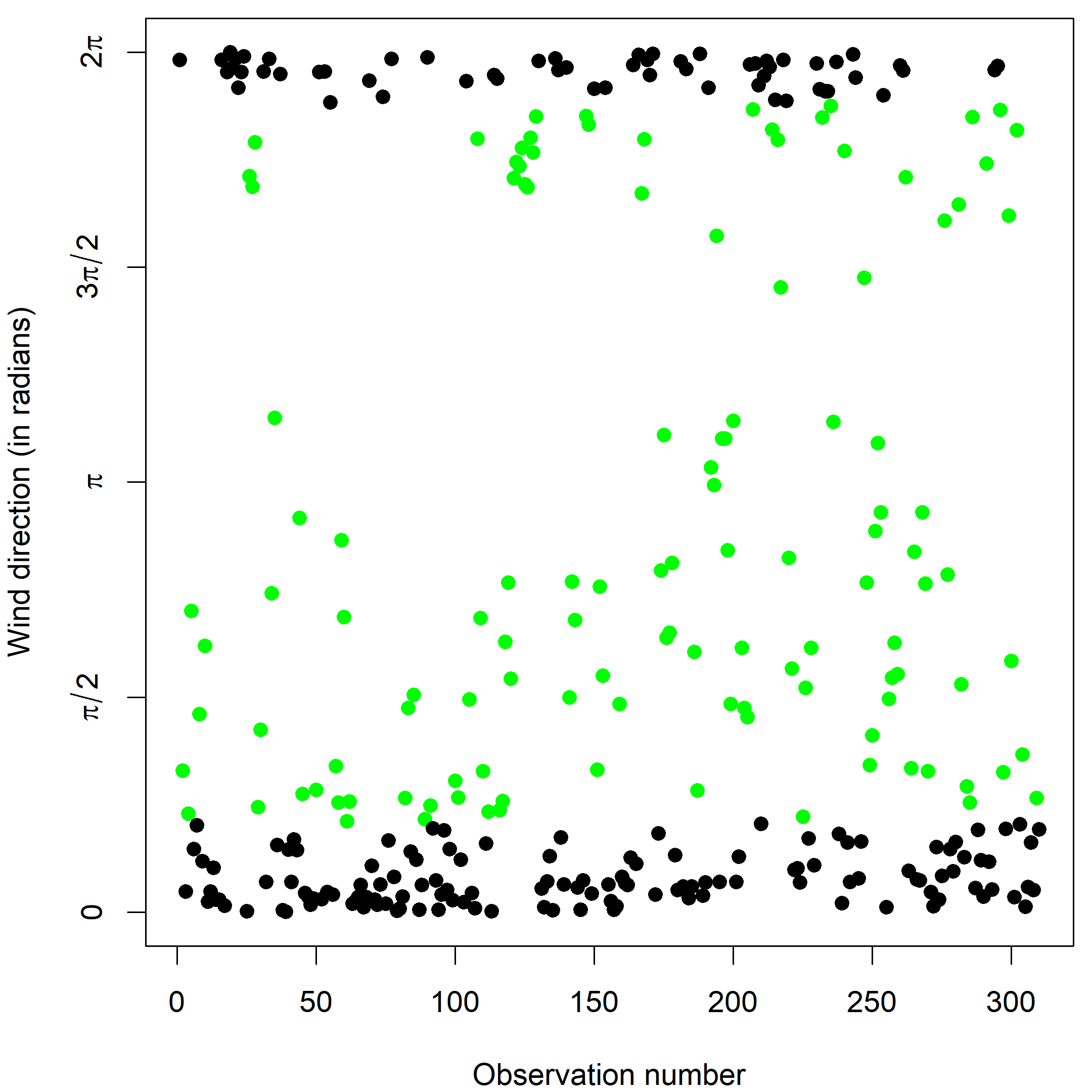}}
 
\subfloat[cuWN distribution]{\includegraphics[width=0.38\textwidth]{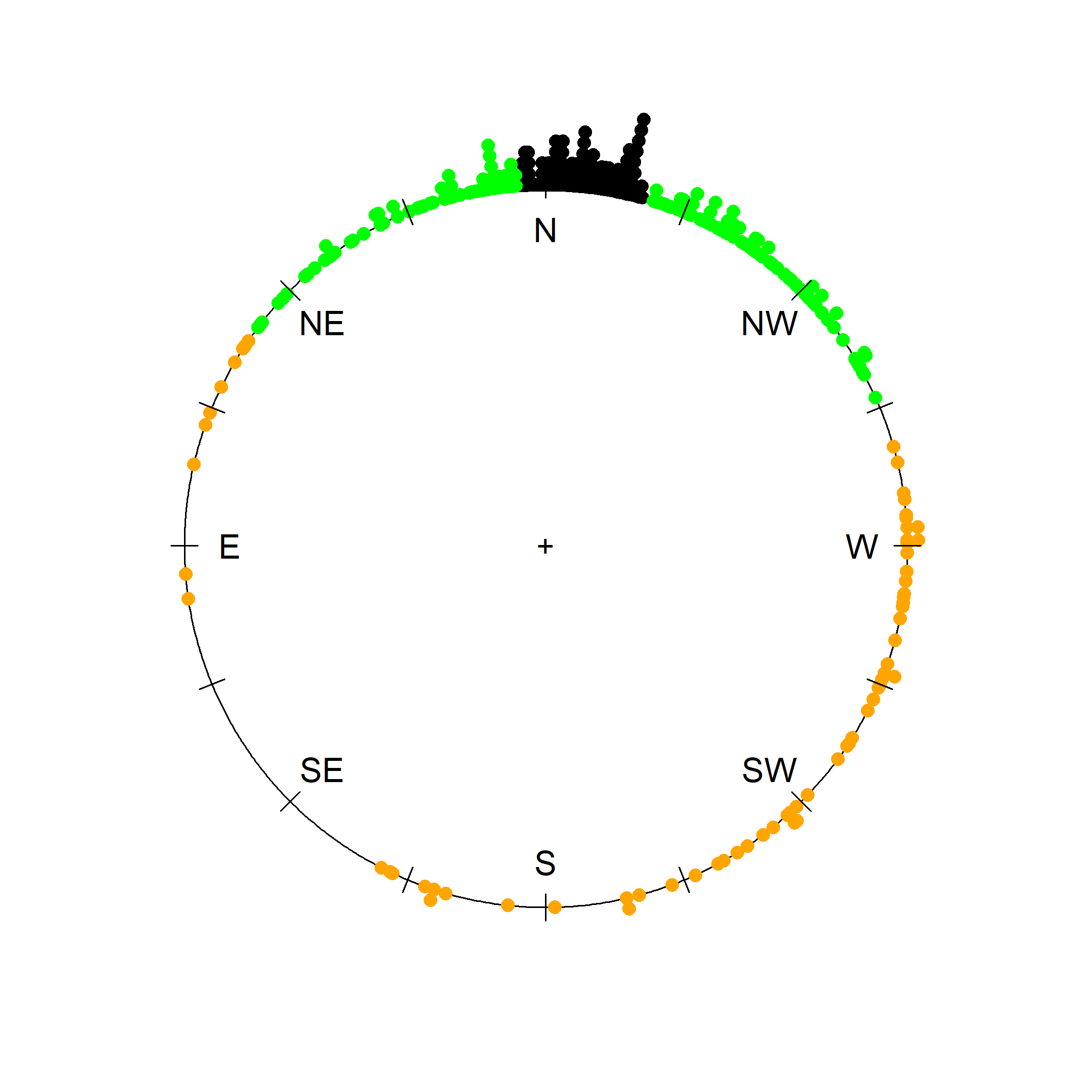}}
\hfill
\subfloat[cuWN distribution]{\includegraphics[width=0.38\textwidth]{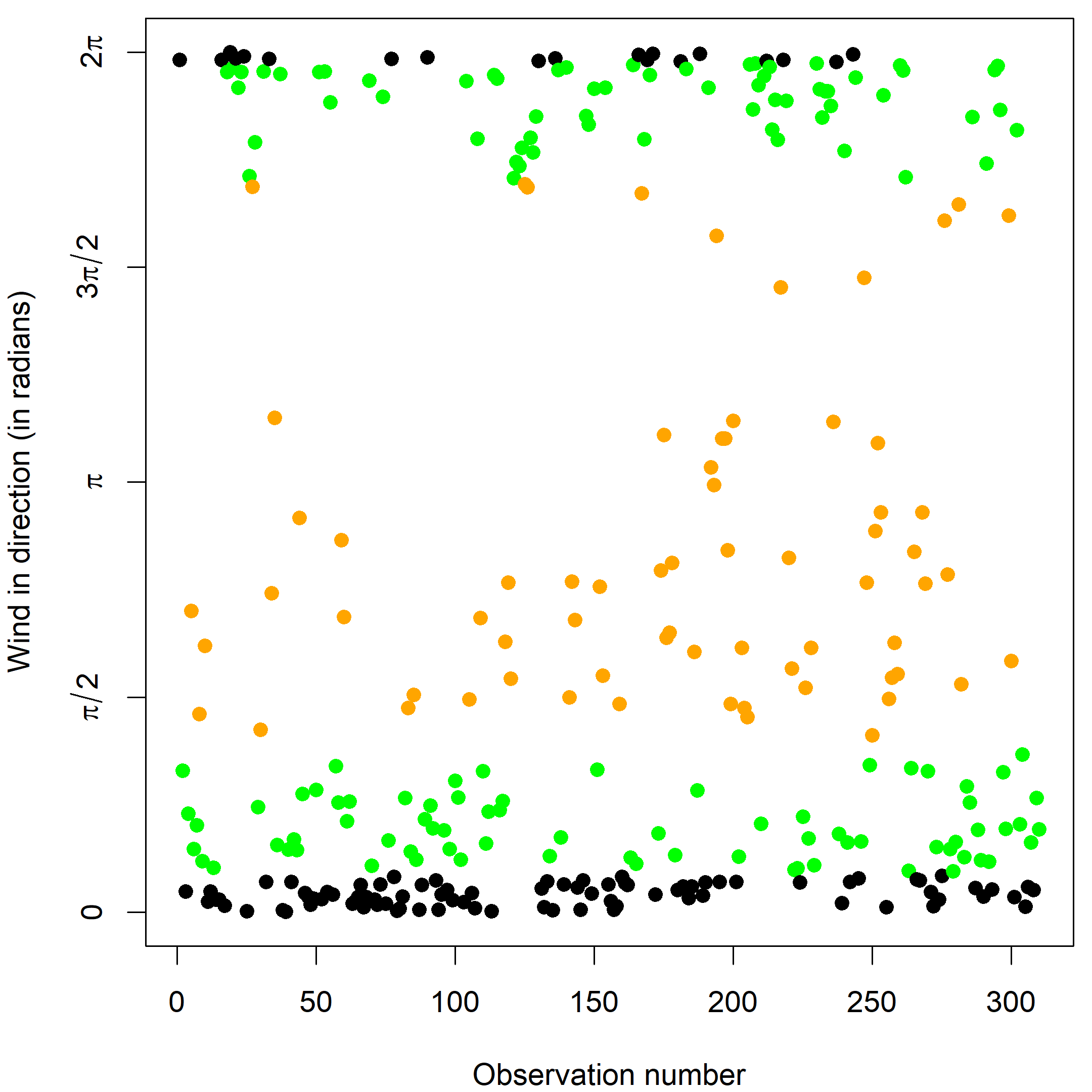}}
\caption{Anomaly detection results for the Col de la Roa wind dataset under the uWN, cWN, and cuWN models. Circular plots (left) and corresponding dotcharts (right) display observations classified into the reference component (black), gross anomalies associated with the uniform component (orange), and mild anomalies associated with the contaminant component (green).}
\label{fig: wind detection}
\end{figure}

\figurename~\ref{fig: wind detection} presents the detection results for the uWN, cWN, and cuWN models. The circular plots display how anomalies are distributed across directions, while the dotcharts (ordered by time index) highlight their temporal structure. Mild anomalies (green) typically remain near the dominant wind direction but show increased dispersion, whereas gross anomalies (orange) appear more scattered across the circle. 
Overall, this visualization confirms that the cuWN and cuVM models not only achieve superior fit according to information criteria, but also provide an interpretable decomposition of the wind regime into dominant flow, moderate directional variability, and diffuse irregular behavior.
 
\section{Conclusion and discussion}
\label{sec: conclussion}
 
Circular data arising in applied sciences often exhibit unimodal and symmetric behavior that can be described through a parametric model defined by a mean direction and concentration parameter (termed as the reference model in this paper). At the same time, some observations may depart from this primary structure due to contamination or other perturbing effects. Forcing all observations into a single homogeneous circular distribution may bias estimation of the mean direction and concentration, thereby compromising both robustness and scientific interpretability.
 
In this paper, we introduced a unified, model-based framework for analysing circular data under two qualitatively different types of departures from the reference structure which we defined in the directional realm as: mild anomalies, which preserve the mean direction while exhibiting inflated dispersion, and gross anomalies, which are consistent with a diffuse background component and are naturally represented by a circular uniform distribution. This framework safeguards inference on the assumed reference model from the influence of anomalous observations as well as serves as an automated mechanism for anomaly detection.
 
The proposed constructions generate a family of models centred on the reference density: the reference model itself (no anomalies), a mild-anomaly model, a gross-anomaly model, and a general specification referred to as the double-contaminated model, accommodating both types of departures simultaneously.
 
For illustrative purposes, we considered two commonly used unimodal and symmetric circular distributions as reference models: the wrapped normal and von Mises distributions.
The framework provides outputs that are directly interpretable in applied terms: mixing proportions quantify the prevalence of mild and gross anomalies, while the attenuation parameter measures the severity of dispersion inflation relative to the reference model. Importantly, posterior component probabilities provide an automatic probabilistic classification of observations, avoiding ad hoc deletion rules or subjective cutoffs and enabling practitioners to localise regions of the circle where each type of behaviour is most likely.
 
The practical relevance of these features was illustrated by an extensive simulation study and by two real-data applications in biology (larval movement directions) and environmental science (alpine wind directions). 
In both datasets, the anomaly-aware specifications substantially improved information-criterion fit over classical wrapped normal and von Mises models, while providing an informative decomposition into a dominant directional pattern, a more dispersed but analogous pattern, and diffuse irregular behaviour. 
This highlights how mixture-based robustness can serve not only anomaly detection, but also scientific interpretation and exploratory structure discovery in directional data.
 
Several extensions follow naturally.
First, the same mixture logic can be coupled with skew (e.g., \citealp{abe2011sine}) or otherwise flexible circular reference models 
(e.g., \citealp{kato2015tractable}) to accommodate asymmetric patterns 
while retaining the mild/gross anomaly interpretation (see also \citealp{kato2010family,kato2013extended}).
Second, the hierarchical concentration-mixture representation suggests a direct route to generalisation beyond the circle, including directional data on spheres and other manifolds, where analogous uniform (or diffuse) components and concentration-attenuation mechanisms can be defined.
Finally, future work may address regression settings with directional responses and/or predictors.
 
\section*{Declarations}
 
\subsection*{Funding}
This work was based upon research supported in part by the National Research Foundation (NRF) of South Africa (SA), grant RA231117164450, the Centre of Excellence in Mathematical and Statistical Sciences, based at the University of the Witwatersrand (SA). The opinions expressed and conclusions arrived at are those of the authors and are not necessarily to be attributed to the NRF.
Antonio Punzo acknowledges the support by the Italian Ministry of University and Research (MUR) under the PRIN 2022 grant number 2022XRHT8R (CUP: E53D23005950006), as part of `The SMILE Project: Statistical Modelling and Inference to Live the Environment', funded by the European Union -- Next Generation EU.
Priyanka Nagar acknowledges the support of the National Graduate Academy for Mathematical and Statistical Sciences, based at the University of Pretoria (SA).
 
\subsection*{Competing interests}
The authors declare that they have no competing interests.
 
\subsection*{Data availability}
The data that support the findings of this study are openly available in the \textbf{circular} package in \textsf{R} software at \url{https://cran.r-project.org/web/packages/circular/index.html}.

\bibliographystyle{chicago}
\bibliography{database}

@article{abe2011sine,
  title={Sine-skewed circular distributions},
  author={Abe, Toshihiro and Pewsey, Arthur},
  journal={Statistical Papers},
  volume={52},
  number={3},
  pages={683--707},
  year={2011},
  publisher={Springer}
}

@article{kato2015tractable,
  title={A tractable and interpretable four-parameter family of unimodal distributions on the circle},
  author={Kato, Shogo and Jones, MC},
  journal={Biometrika},
  volume={102},
  number={1},
  pages={181--190},
  year={2015},
  publisher={Oxford University Press}
}

@article{kato2013extended,
  title={An extended family of circular distributions related to wrapped Cauchy distributions via Brownian motion},
  author={Kato, Shogo and Jones, MC},
  journal={Bernoulli},
  pages={154--171},
  year={2013},
  publisher={JSTOR}
}

@article{kato2010family,
  title={A family of distributions on the circle with links to, and applications arising from, M{\"o}bius transformation},
  author={Kato, Shogo and Jones, MC},
  journal={Journal of the American Statistical Association},
  volume={105},
  number={489},
  pages={249--262},
  year={2010},
  publisher={Taylor \& Francis}
}

@article{abe2012circular,
  title={Circular distributions of fallen logs as an indicator of forest disturbance regimes},
  author={Abe, Toshihiro and Kubota, Yasuhiro and Shimatani, Kenichiro and Aakala, Tuomas and Kuuluvainen, Timo},
  journal={Ecological Indicators},
  volume={18},
  pages={559--566},
  year={2012},
  publisher={Elsevier}
}

@book{ley2017modern,
  title={Modern Directional Statistics},
  author={Ley, C. and Verdebout, T.},
  isbn={9781351645782},
  series={Chapman \& Hall/CRC Interdisciplinary Statistics},
  url={https://books.google.com/books?id=XlsvDwAAQBAJ},
  year={2017},
  publisher={CRC Press}
}

@book{pewsey2013circular,
  title={Circular Statistics in \textsf{R}},
  author={Pewsey, A. and Neuh{\"a}user, M. and Ruxton, G.D.},
  isbn={9780191650765},
  lccn={2013940576},
  series={Ebrary online},
  url={https://books.google.com/books?id=6lhoAgAAQBAJ},
  year={2013},
  publisher={OUP Oxford}
}

@book{Agga:Outl:2013,
  title={Outlier Analysis},
  author={Aggarwal, C. C.},
  year={2013},
  publisher={Springer},
  address={New York}
}

@book{Ritt:Robu:2015,
  title={Robust Cluster Analysis and Variable Selection},
  author={Ritter, G.},
  series={Chapman \& Hall/CRC Monographs on Statistics \& Applied Probability},
  year={2015},
  volume={137},
  publisher={CRC Press}
}

@article{davies1993identification,
  title={The identification of multiple outliers},
  author={Davies, Laurie and Gather, Ursula},
  journal={Journal of the American Statistical Association},
  volume={88},
  number={423},
  pages={782--792},
  year={1993},
  publisher={Taylor \& Francis}
}

@book{Hawk:Iden:2013,
  title={Identification of Outliers},
  author={Hawkins, D.},
  series={Monographs on Statistics and Applied Probability},
  year={2013},
  publisher={Springer},
  address={The Netherlands}
}

@article{tomarchio2020dichotomous,
  title={Dichotomous unimodal compound models: application to the distribution of insurance losses},
  author={Tomarchio, Salvatore D and Punzo, Antonio},
  journal={Journal of Applied Statistics},
  volume={47},
  number={13-15},
  pages={2328--2353},
  year={2020},
  publisher={Taylor \& Francis}
}

@book{huber2011robust,
  title={Robust Statistics},
  author={Huber, P.J. and Ronchetti, E.M.},
  isbn={9781118210338},
  series={Wiley Series in Probability and Statistics},
  url={https://books.google.com/books?id=j1OhquR_j88C},
  year={2011},
  publisher={Wiley}
}

@article{melnykov2025contaminated,
  author  = {Melnykov, Yana},
  title   = {On the Use of Contaminated Normal Distributions for Modeling Data Groups with Heavy Tails and Outliers},
  journal = {Journal of Classification},
  year    = {2025},
  doi     = {10.1007/s00357-025-09518-1}
}

@article{lim2025heckmanCN,
  author  = {Lim, Heeju and Ordoñez, José Alejandro and Punzo, Antonio and Lachos, Victor H.},
  title   = {Heckman Selection--Contaminated Normal Model},
  journal = {Journal of Computational and Graphical Statistics},
  year    = {2025},
  pages = {1--13}
}

@book{fruhwirth2006finite,
  author    = {Fr\"uhwirth-Schnatter, Sylvia},
  title     = {Finite Mixture and Markov Switching Models},
  publisher = {Springer},
  address = {New York},
  year = {2006}
}

@book{mardia2009directional,
  title={Directional Statistics},
  author={Mardia, Kanti. V. and Jupp, Peter E.},
  isbn={9780470317815},
  series={Wiley Series in Probability and Statistics},
  url={https://books.google.com/books?id=PTNiCm4Q-M0C},
  year={2009},
  publisher={Wiley}
}

@article{pewsey2004large,
  title={The large-sample joint distribution of key circular statistics},
  author={Pewsey, Arthur},
  journal={Metrika},
  volume={60},
  number={1},
  pages={25--32},
  year={2004},
  publisher={Springer}
}

@article{aitkin1980mixture,
  title={Mixture models, outliers, and the {EM} algorithm},
  author={Aitkin, Murray and Wilson, Granville Tunnicliffe},
  journal={Technometrics},
  volume={22},
  number={3},
  pages={325--331},
  year={1980},
  publisher={Taylor \& Francis}
}

@article{collett1980outliers,
  title={Outliers in circular data},
  author={Collett, D},
  journal={Journal of the Royal Statistical Society: Series C (Applied Statistics)},
  volume={29},
  number={1},
  pages={50--57},
  year={1980},
  publisher={Wiley Online Library}
}

@article{ko1992robust,
  title={Robust estimation of the concentration parameter of the von {M}ises-{F}isher distribution},
  author={Ko, Daijin},
  journal={The Annals of Statistics},
  pages={917--928},
  year={1992},
  publisher={JSTOR}
}

@incollection{Demni2023anomaly,
  title={Anomaly detection in Circular Data},
  author={Demni, Houyem and Porzio, Giovanni, C.},
  booktitle={Book of the Short Papers-SIS 2023 Ancona},
  pages={63--68},
  year={2023},
  publisher={Pearson}
}

@article{abuzaid2012statistics,
  title={Statistics for a new test of discordance in circular data},
  author={Abuzaid, Ali Hassan and Hussin, Abdul G and Rambli, A and Mohamed, I},
  journal={Communications in Statistics-Simulation and Computation},
  volume={41},
  number={10},
  pages={1882--1890},
  year={2012},
  publisher={Taylor \& Francis}
}

@article{mahmood2017detection,
  title={Detection of outliers in univariate circular data using robust circular distance},
  author={Mahmood, Ehab A and Rana, Sohel and Midi, Habshah and Hussin, Abdul Ghapor},
  journal={Journal of Modern Applied Statistical Methods},
  volume={16},
  number={},
  pages={418--438},
  year={2017}
}

@article{abuzaid2009new,
  title={A new test of discordancy in circular data},
  author={Abuzaid, Ali Hassan and Mohamed, Ibrahim B and Hussin, Abdul G},
  journal={Communications in Statistics-Simulation and Computation},
  volume={38},
  number={4},
  pages={682--691},
  year={2009},
  publisher={Taylor \& Francis}
}

@article{mohamed2016new,
  title={A new discordancy test in circular data using spacings theory},
  author={Mohamed, Ibrahim B and Rambli, ADZHAR and Khaliddin, NURLIZA and Ibrahim, Adriana Irawati Nur},
  journal={Communications in Statistics-Simulation and Computation},
  volume={45},
  number={8},
  pages={2904--2916},
  year={2016},
  publisher={Taylor \& Francis}
}

@article{lund2017package,
  title={Package ‘circular’},
  author={Lund, Ulric and Agostinelli, Claudio and Agostinelli, Maintainer Claudio},
  journal={Repository CRAN},
  volume={775},
  number={5},
  pages={20--135},
  year={2017}
}

@book{jammalamadaka2001topics,
  title={Topics in Circular Statistics},
  author={Jammalamadaka, S Rao and Sengupta, Ambar},
  volume={5},
  year={2001},
  publisher={{W}orld {S}cientific Publishing}
}

@article{agostinelli2007robust,
  title={Robust estimation for circular data},
  author={Agostinelli, Claudio},
  journal={Computational Statistics \& Data Analysis},
  volume={51},
  number={12},
  pages={5867--5875},
  year={2007},
  publisher={Elsevier}
}

@article{jones2012inverse,
  title={Inverse {B}atschelet distributions for circular data},
  author={Jones, MC and Pewsey, Arthur},
  journal={Biometrics},
  volume={68},
  number={1},
  pages={183--193},
  year={2012},
  publisher={Oxford University Press}
}

@article{kent1983identifiability,
  title={Identifiability of finite mixtures for directional data},
  author={Kent, John T},
  journal={The Annals of Statistics},
  pages={984--988},
  year={1983},
  publisher={JSTOR}
}

@article{holzmann2004identifiability,
  title={Identifiability of finite mixtures-with applications to circular distributions},
  author={Holzmann, Hajo and Munk, Axel and Stratmann, Bernd},
  journal={Sankhy{\=a}: The Indian Journal of Statistics},
  pages={440--449},
  year={2004},
  publisher={JSTOR}
}

@article{r2020r,
  title={R: A language and environment for statistical computing. {R} {F}oundation for {S}tatistical {C}omputing, {V}ienna, {A}ustria},
  author={{R Core Team}},
  journal={\href{http://www.R-project.org/}{http://www.R-project.org/}},
  year={2020}
}

@article{biernacki2003choosing,
  title={Choosing starting values for the {EM} algorithm for getting the highest likelihood in multivariate {G}aussian mixture models},
  author={Biernacki, Christophe and Celeux, Gilles and Govaert, G{\'e}rard},
  journal={Computational Statistics \& Data Analysis},
  volume={41},
  number={3-4},
  pages={561--575},
  year={2003},
  publisher={Elsevier}
}

@article{punzo2016parsimonious,
  title={Parsimonious mixtures of multivariate contaminated normal distributions},
  author={Punzo, Antonio and McNicholas, Paul D},
  journal={Biometrical Journal},
  volume={58},
  number={6},
  pages={1506--1537},
  year={2016},
  publisher={Wiley Online Library}
}

@article{otto2026modeling,
  title={Modeling Bounded Count Environmental Data Using a Contaminated Beta-Binomial Regression Model},
  author={Otto, Arnoldus F and Punzo, Antonio and Ferreira, Johannes T and Bekker, Andri{\"e}tte and Tomarchio, Salvatore D and Tortora, Cristina},
  journal={Environmetrics},
  volume={37},
  number={1},
  pages={e70067},
  year={2026},
  publisher={Wiley Online Library}
}

@article{akaike1974new,
	title={A new look at the statistical model identification},
	author={Akaike, Hirotugu},
	journal={{IEEE} {T}ransactions on {A}utomatic {C}ontrol},
	volume={19},
	number={6},
	pages={716--723},
	year={1974},
	publisher={Ieee}
}

@article{schwarz1978estimating,
	title={Estimating the dimension of a model},
	author={Schwarz, Gideon},
	journal={The {A}nnals of {S}tatistics},
        volume={6},
        number={2},
	pages={461--464},
	year={1978},
	publisher={JSTOR}
}

@article{zhang2023model,
  title={On model-based clustering of directional data with heavy tails},
  author={Zhang, Yingying and Melnykov, Volodymyr and Melnykov, Igor},
  journal={Journal of Classification},
  volume={40},
  number={3},
  pages={527--551},
  year={2023},
  publisher={Springer}
}

@article{punzo2017robust,
  title={Robust clustering in regression analysis via the contaminated {G}aussian cluster-weighted model},
  author={Punzo, Antonio and McNicholas, Paul D},
  journal={Journal of Classification},
  volume={34},
  number={2},
  pages={249--293},
  year={2017},
  publisher={Springer}
}

@article{otto2025contaminated,
  title={A contaminated regression model for count health data},
  author={Otto, Arnoldus F and Ferreira, Johannes T and Tomarchio, Salvatore Daniele and Bekker, Andri{\"e}tte and Punzo, Antonio},
  journal={Statistical Methods in Medical Research},
  volume={34},
  number={2},
  pages={369--389},
  year={2025},
  publisher={SAGE Publications Sage UK: London, England}
}

@article{dong2024contaminated,
  title={Contaminated Kent mixture model for clustering non-spherical directional data with heavy tails or scatter},
  author={Dong, Aqi and Melnykov, Volodymyr},
  journal={Statistics \& Probability Letters},
  volume={208},
  pages={110058},
  year={2024},
  publisher={Elsevier}
}

@article{goddard2022investigation,
  title={Investigation of ecologically relevant wind patterns on Marion Island using Computational Fluid Dynamics and measured data},
  author={Goddard, KA and Craig, KJ and Schoombie, J and Le Roux, PC},
  journal={Ecological Modelling},
  volume={464},
  pages={109827},
  year={2022},
  publisher={Elsevier}
}

\appendix

\section{Further simulation results}\label{App: estimation}
 
In this section, we present additional results from the simulation study described in Section~\ref{Sec: simulation study}, based on a larger sample size of $n = 500$.

\begin{figure}[!htbp]
   	\centering
		\includegraphics[scale=0.48]{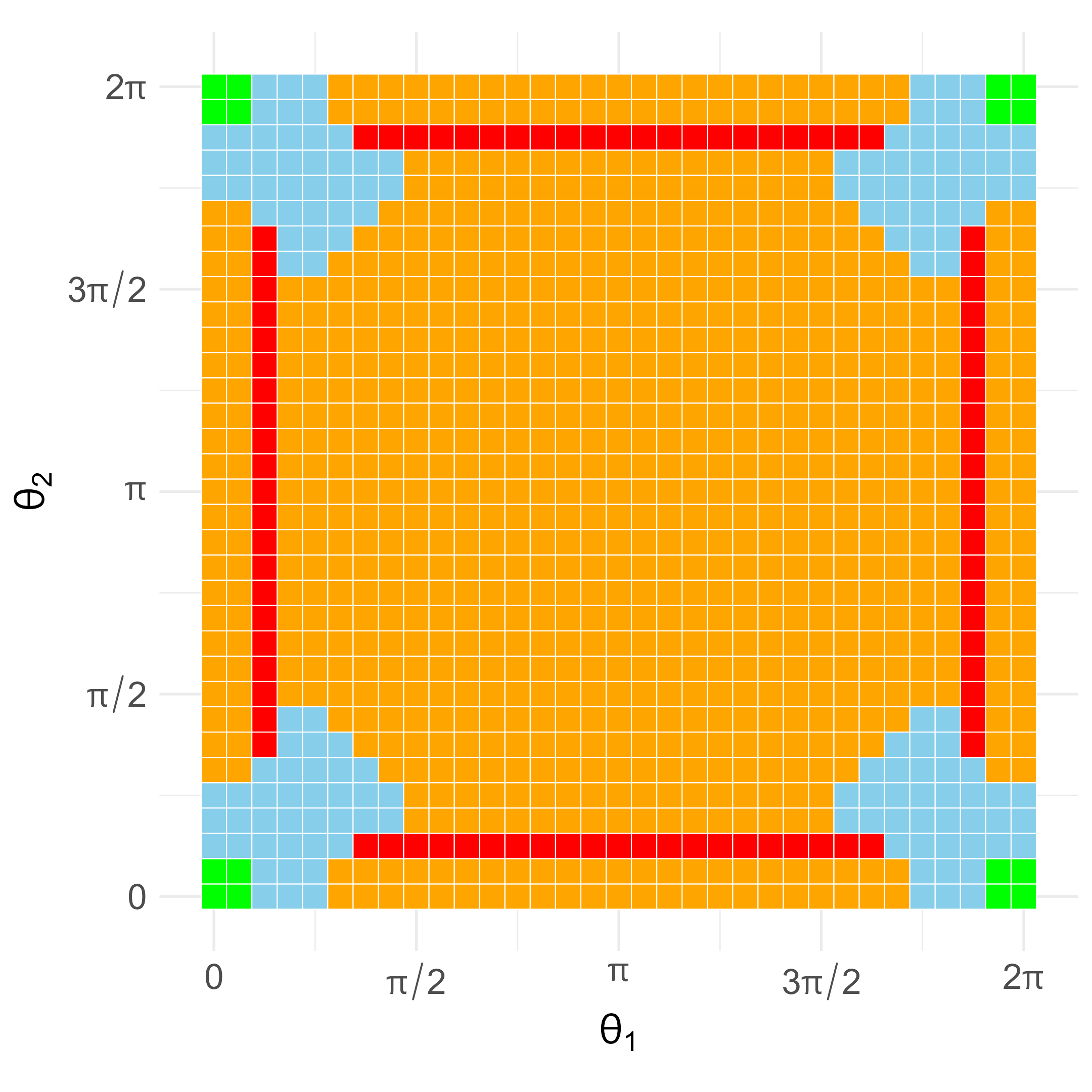}
	\caption{Anomaly detection map for $n=500$, $\kappa=100$, where each cell represents the modal classification across 100 replications for two anomaly points placed at positions $(\theta_1, \theta_2)$. Green: no anomalies detected (WN); orange: gross anomaly only (uWN); sky blue: mild anomaly only (cWN); red: both mild and gross anomalies (cuWN).}
    \label{fig:detection_n500_sigma01}
\end{figure}
 
\begin{figure}[!htbp]
\centering 
\subfloat[$\mu$, WN model]{\includegraphics[width=0.45\textwidth]{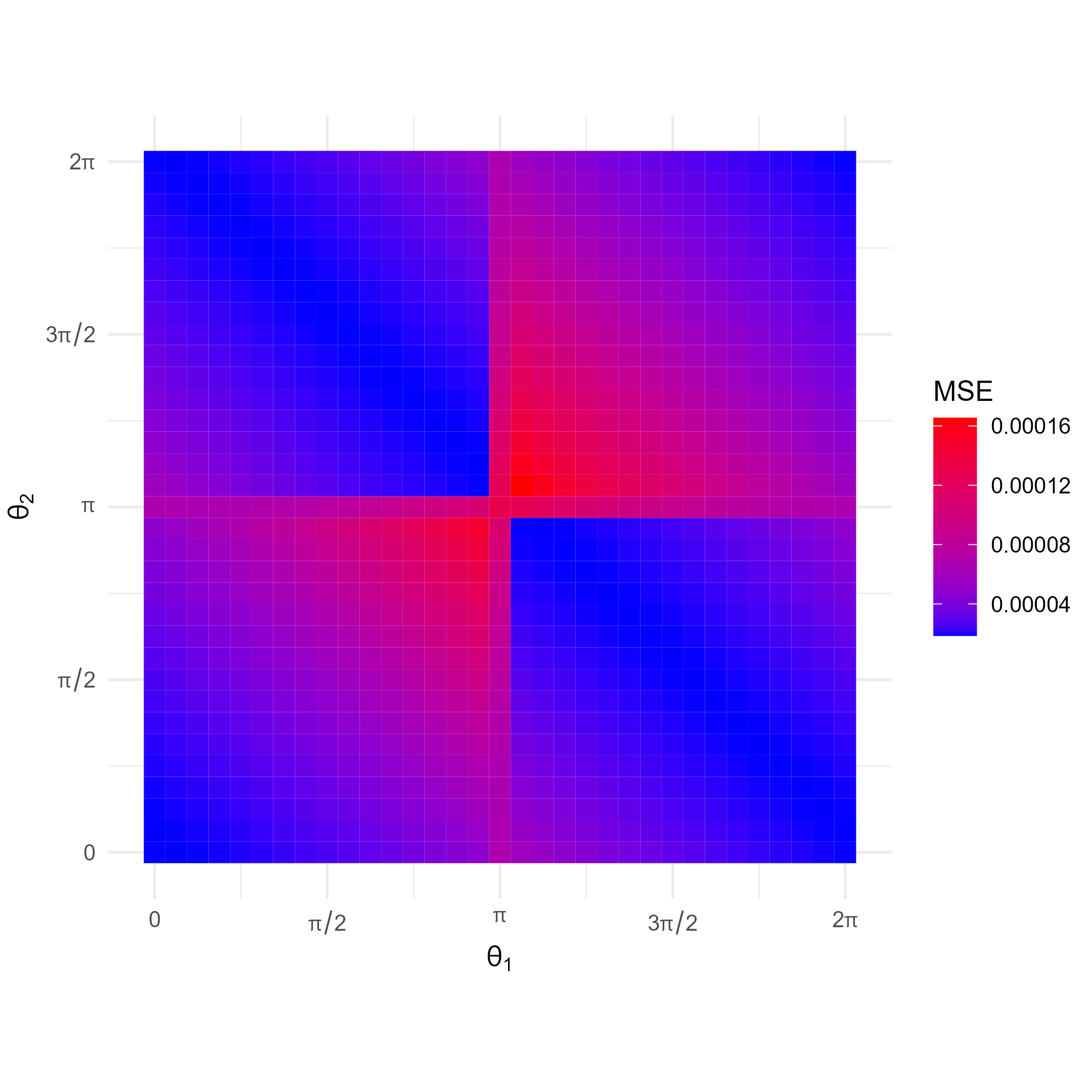}}
\hfill
\subfloat[$\mu$, cuWN model]{\includegraphics[width=0.45\textwidth]{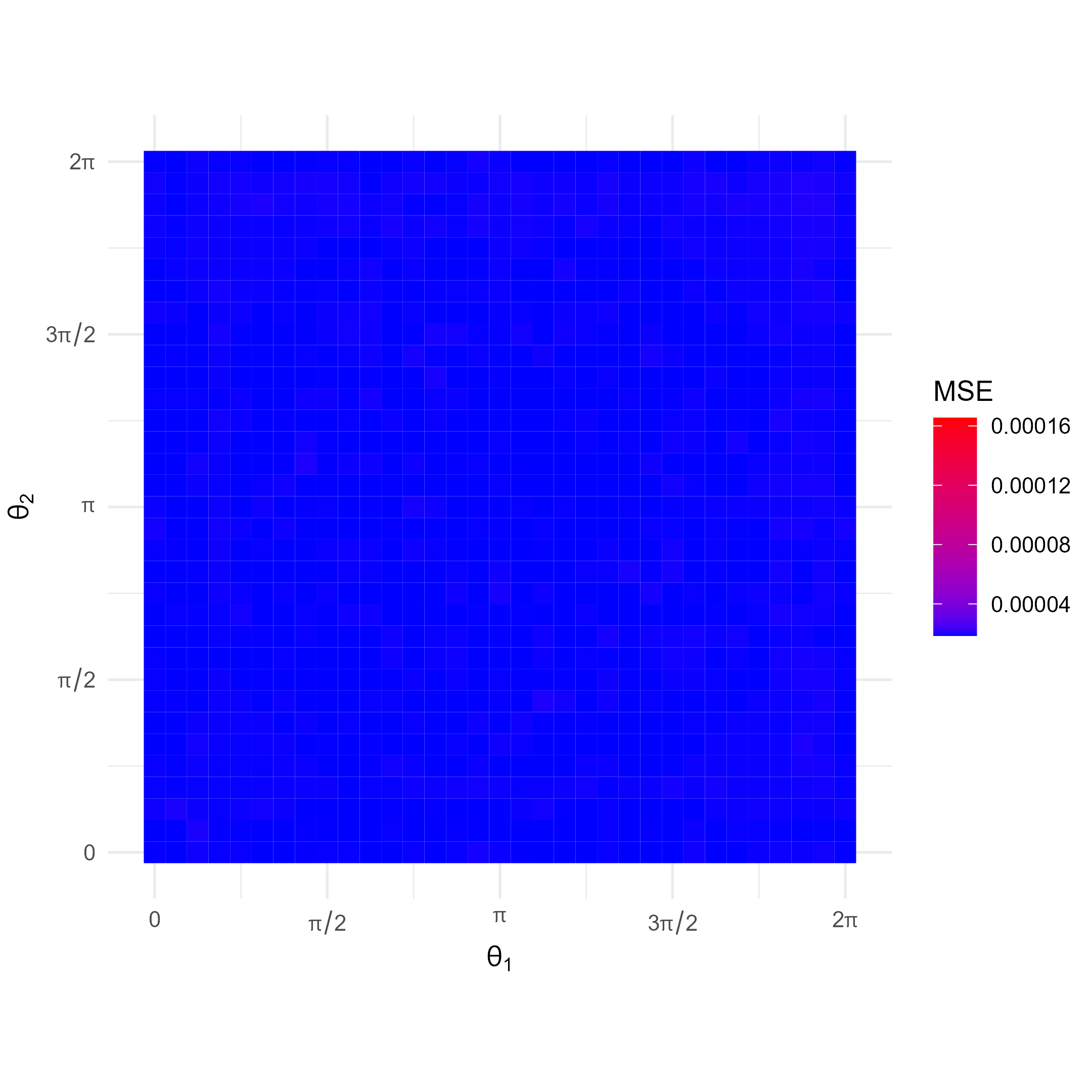}}
 
\subfloat[$\kappa$, WN model]{\includegraphics[width=0.45\textwidth]{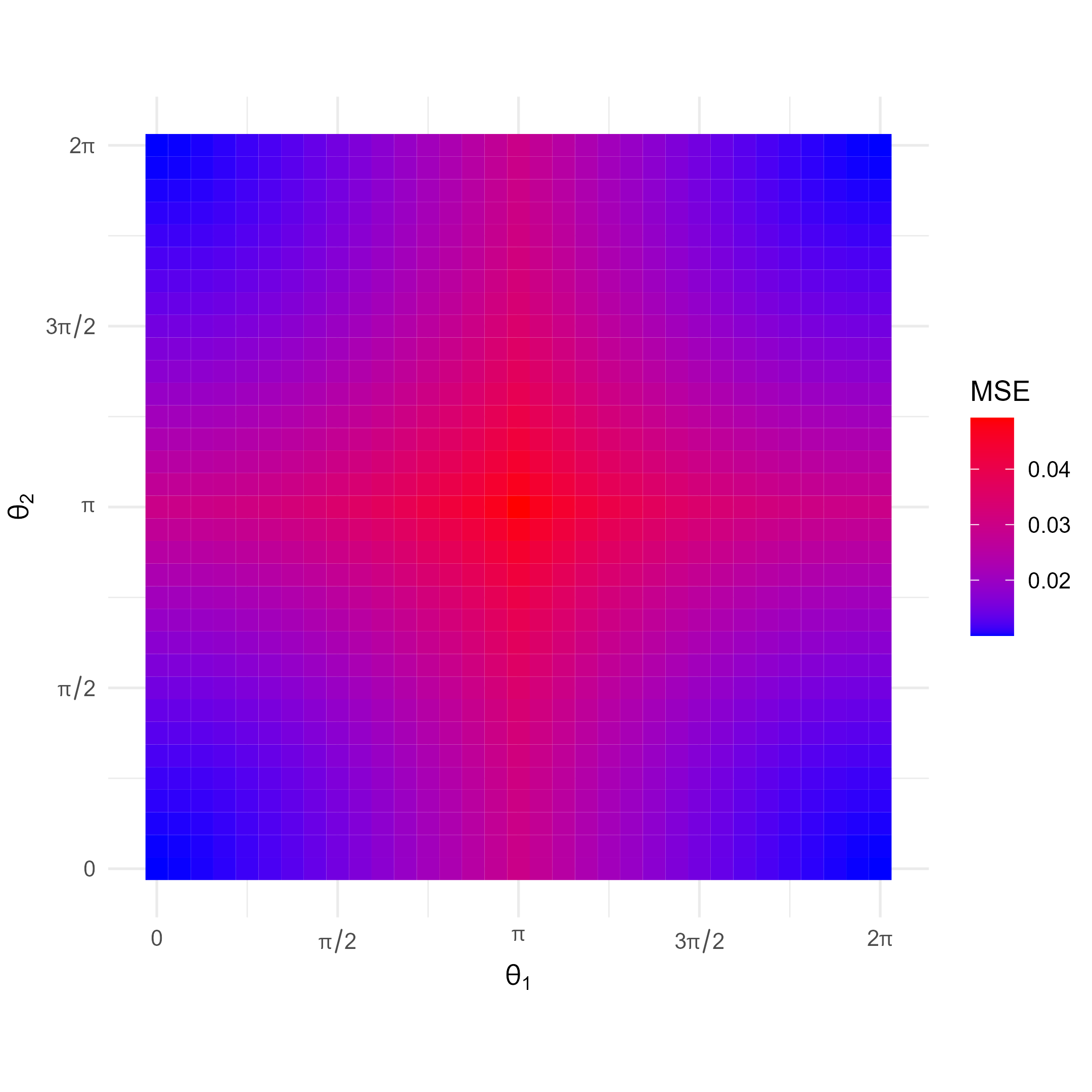}}
\hfill
\subfloat[$\kappa$, cuWN model]{\includegraphics[width=0.45\textwidth]{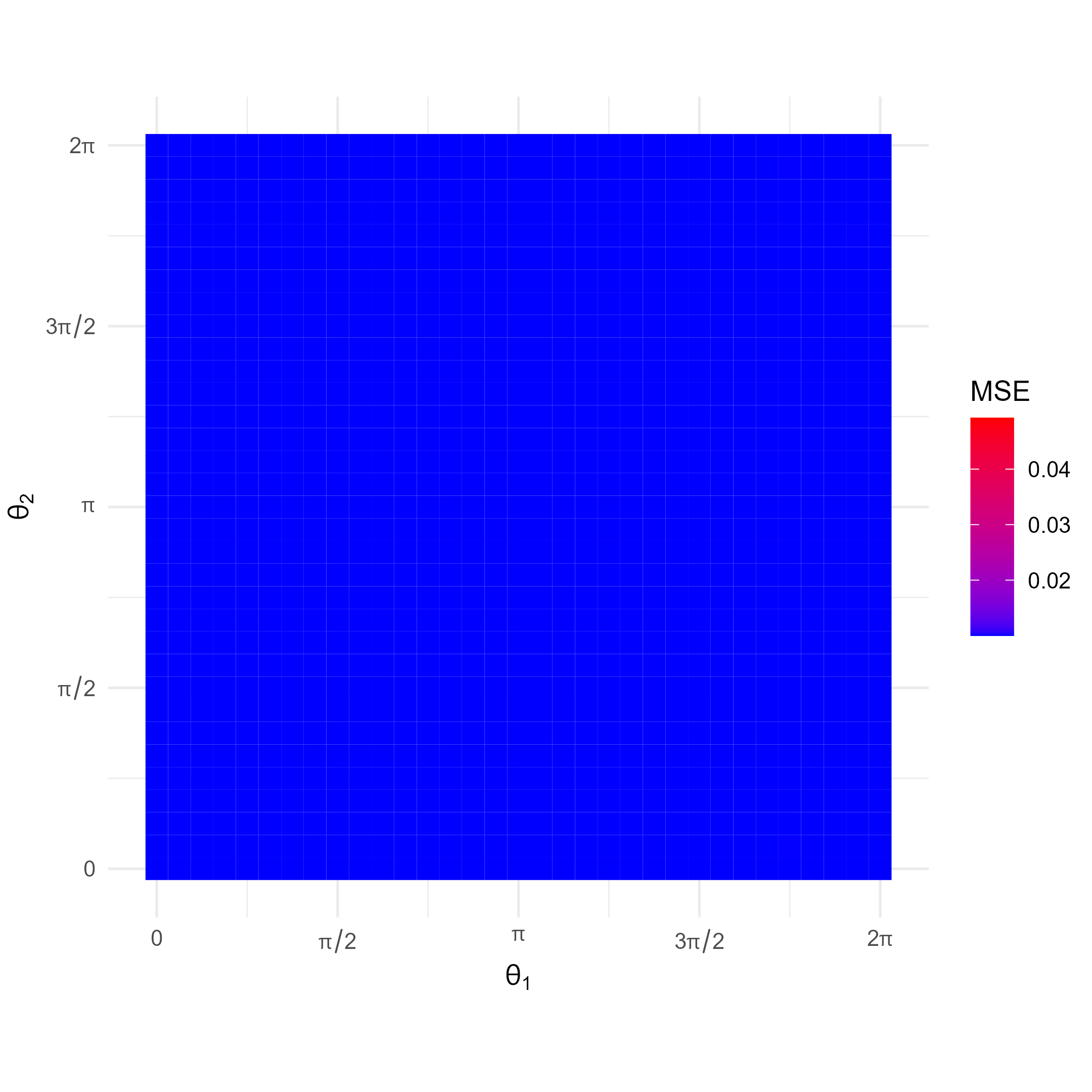} }
\caption{Mean squared error (MSE) for estimating $\mu$ (top row) and $\kappa$ (bottom row) with $n = 500$ and $\kappa = 100$. The left column displays results under the WN model, while the right column corresponds to the cuWN model. }
\label{fig:mse_n500_sigma01}
\end{figure}
 
\begin{figure}[!htbp]
   	\centering
		\includegraphics[scale=0.48]{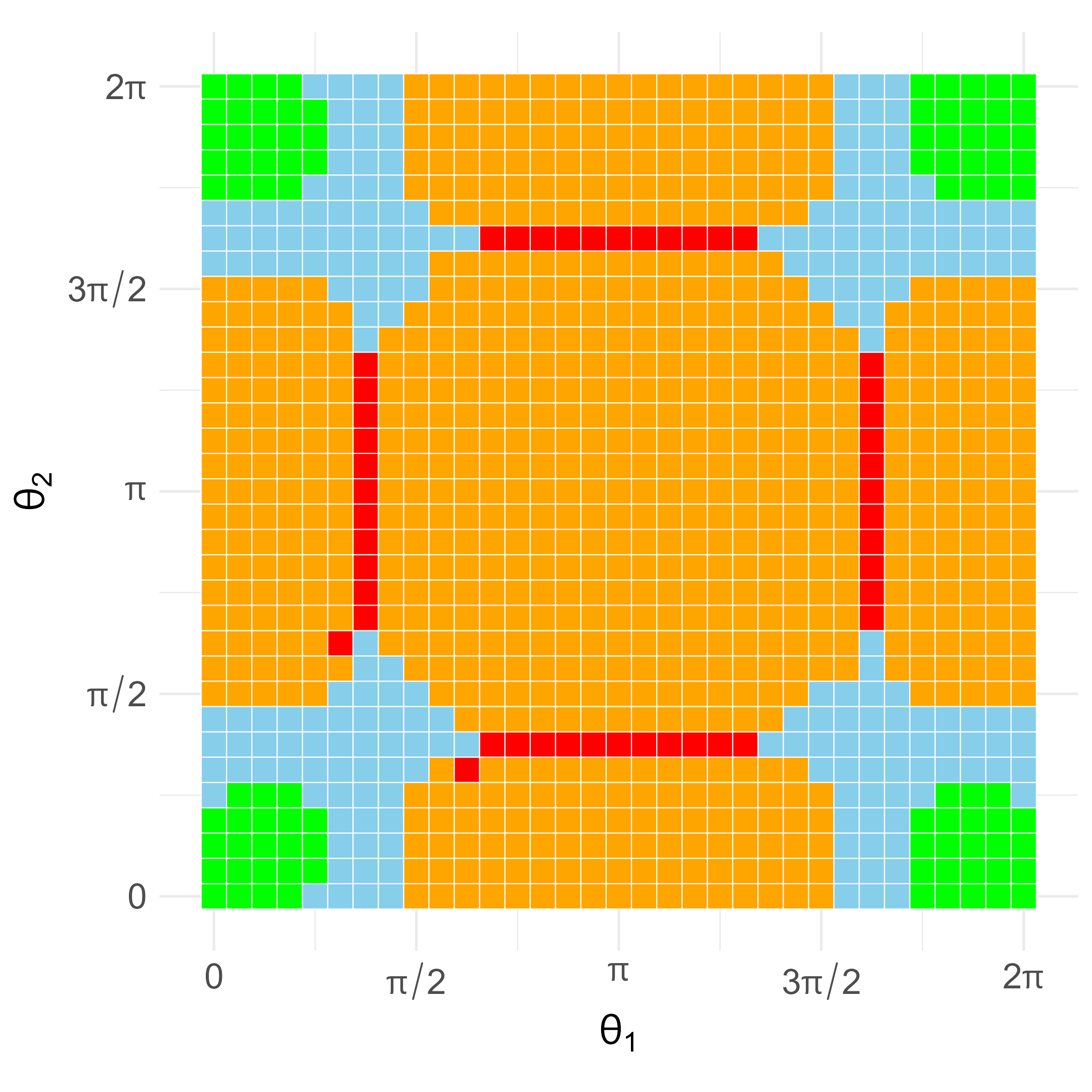}
	\caption{Anomaly detection map for $n=500$, $\kappa=10$, where each cell represents the modal classification across 100 replications for two anomaly points placed at positions $(\theta_1, \theta_2)$. Green: no anomalies detected (WN); orange: gross anomaly only (uWN); sky blue: mild anomaly only (cWN); red: both mild and gross anomalies (cuWN).}
    \label{fig:detection_n500_sigma03}
\end{figure}
 
\begin{figure}[!htbp]
\centering 
\subfloat[$\mu$, WN model]{\includegraphics[width=0.45\textwidth]{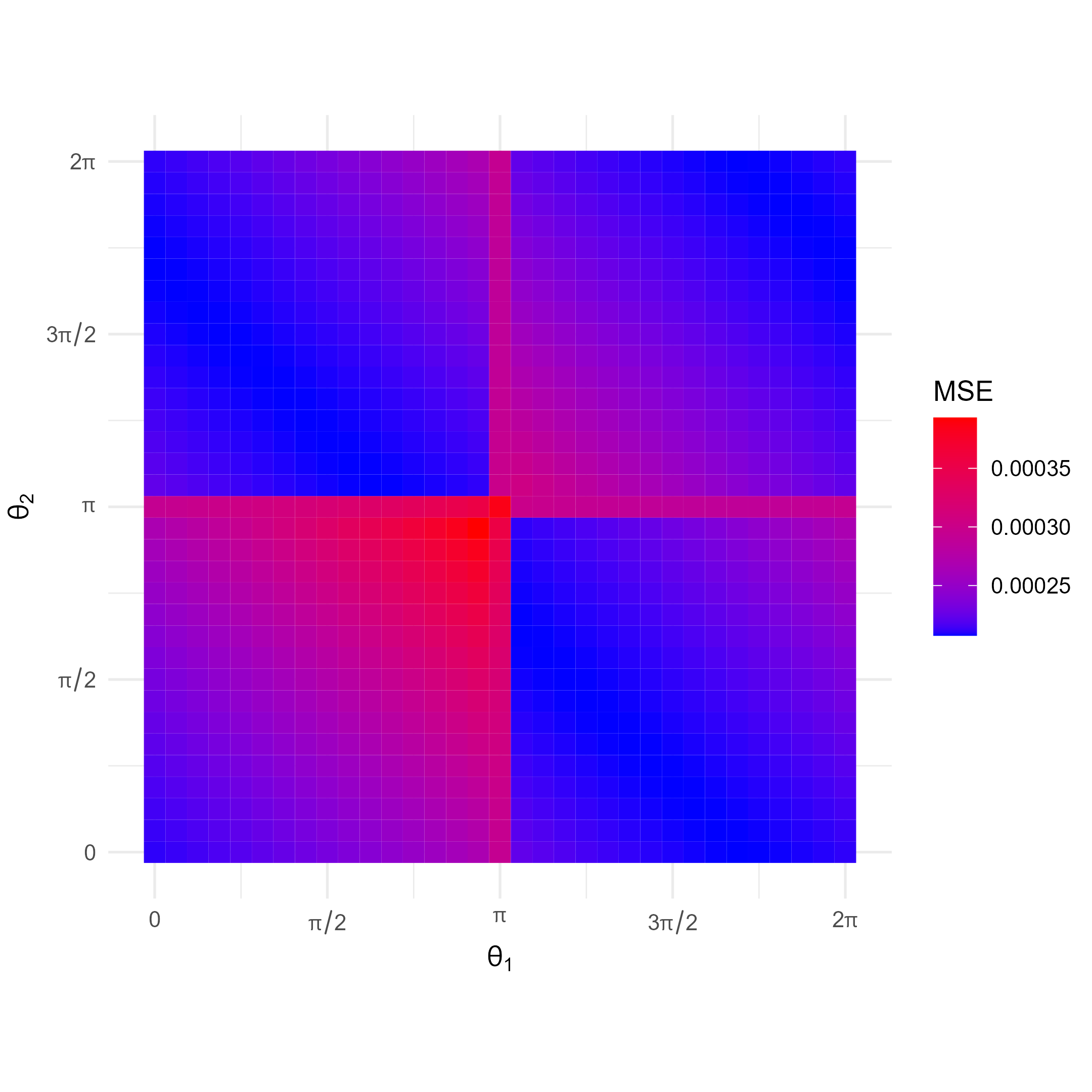}}
\hfill
\subfloat[$\mu$, cuWN model]{\includegraphics[width=0.45\textwidth]{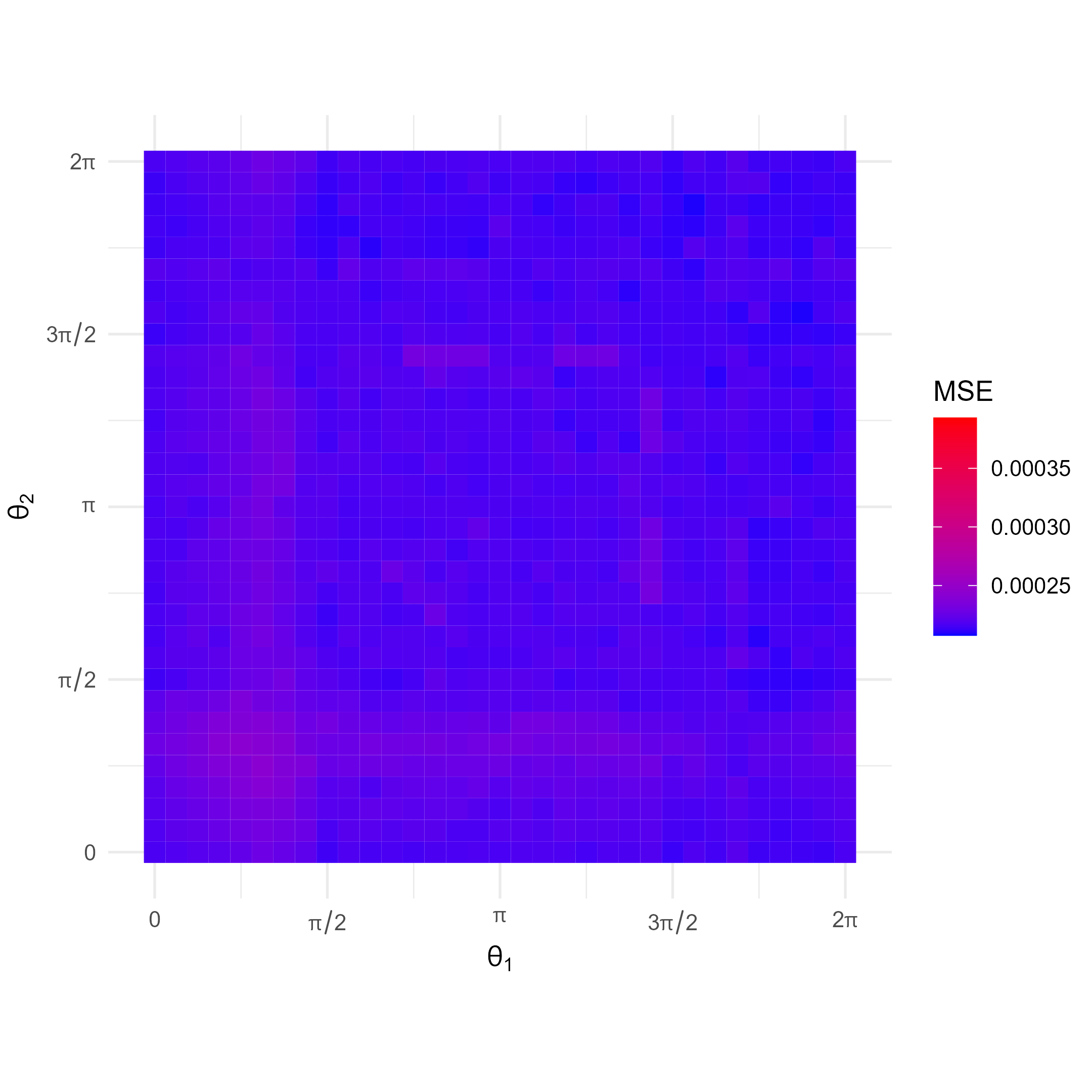}}
 
\subfloat[$\kappa$, WN model]{\includegraphics[width=0.45\textwidth]{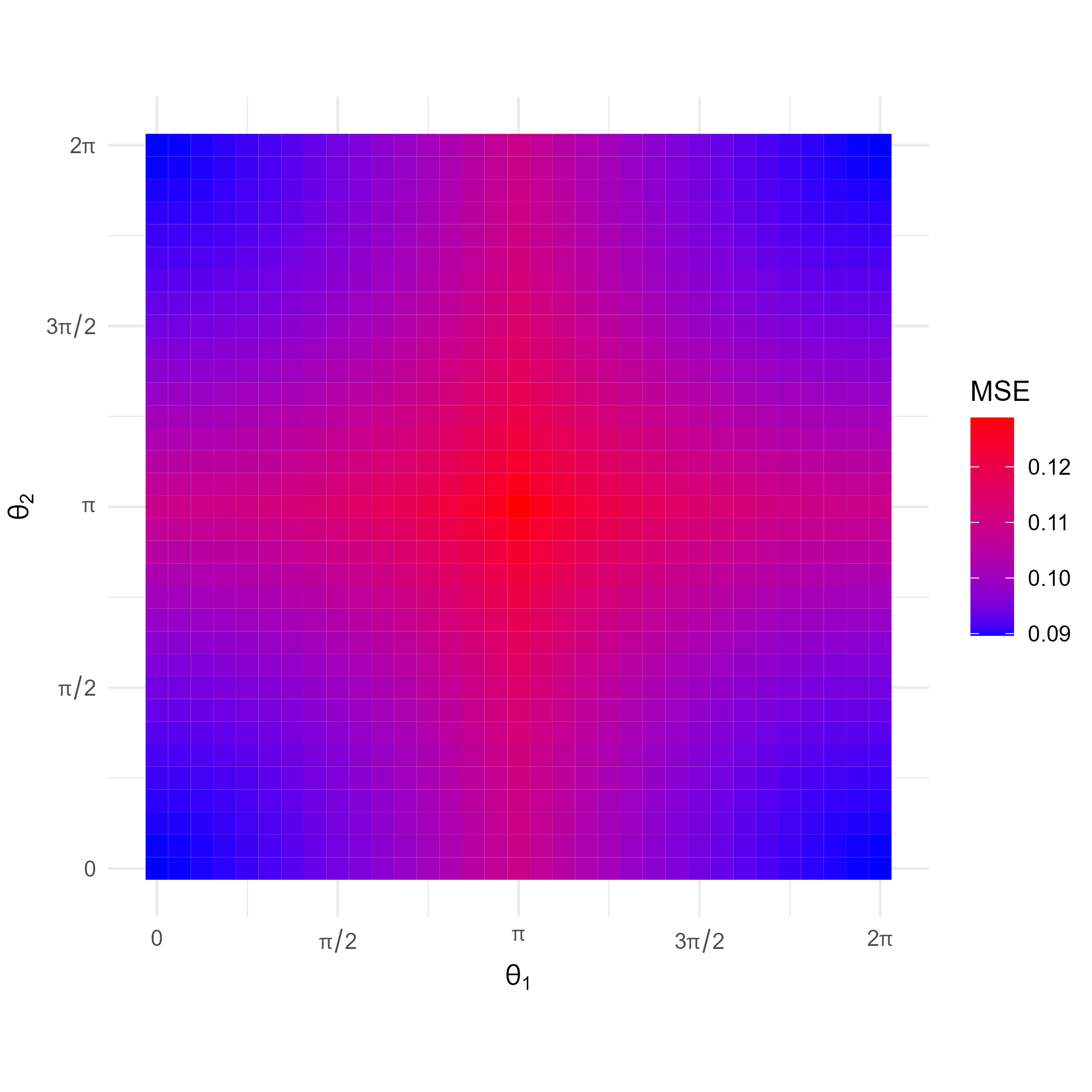}}
\hfill
\subfloat[$\kappa$, cuWN model]{\includegraphics[width=0.45\textwidth]{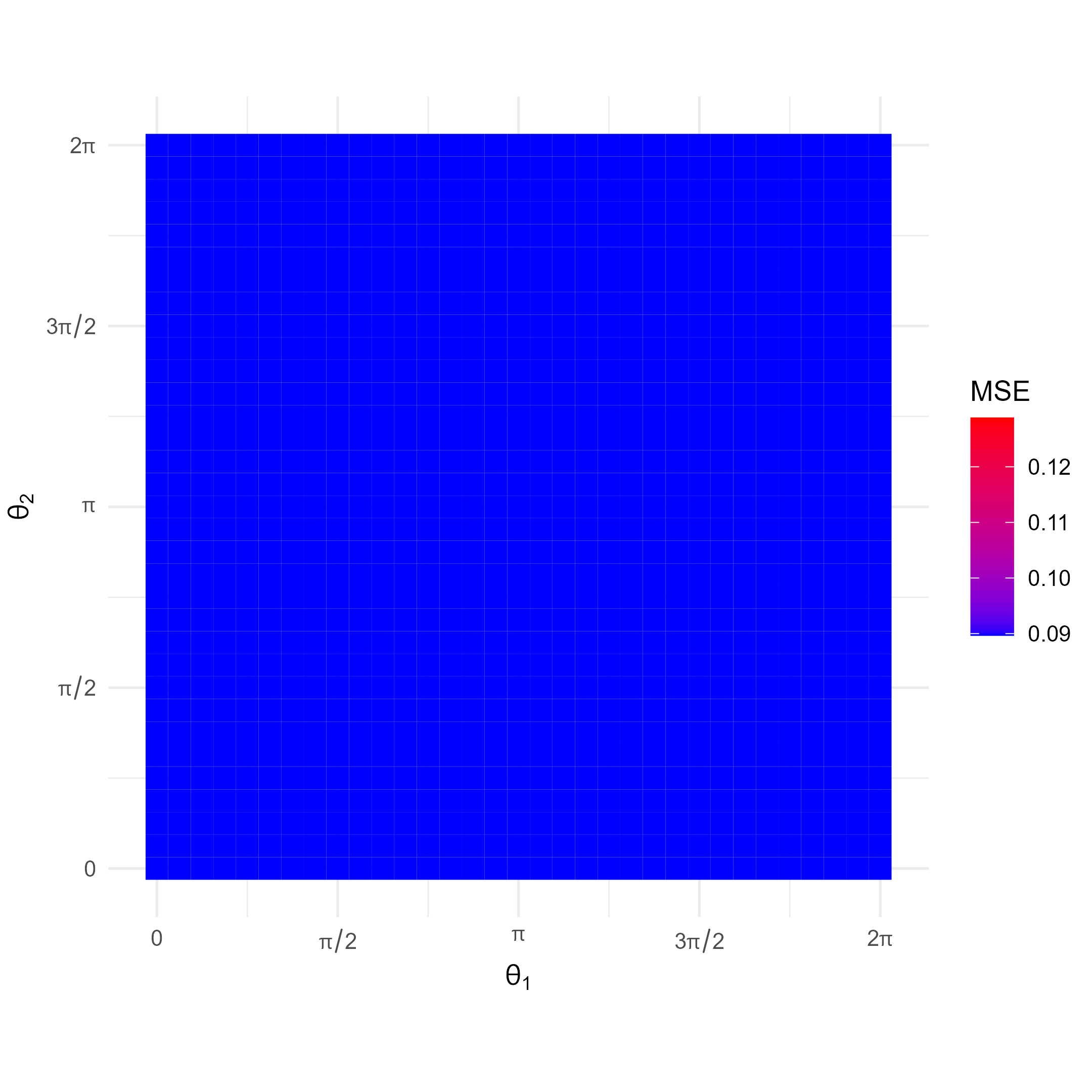} }
\caption{Mean squared error (MSE) for estimating $\mu$ (top row) and $\kappa$ (bottom row) with $n = 500$ and $\kappa = 10$. The left column displays results under the WN model, while the right column corresponds to the cuWN model.}
\label{fig:mse_n500_sigma03}
\end{figure}

\end{document}